\documentclass[aps,prl,12pt,showkeys]{revtex4}

\usepackage{amsmath}
\usepackage{amssymb}
\usepackage{times}
\usepackage{mathptmx} 

\begin{document}
\title{An Introduction to Quantum Game Theory}
\author{J. Orlin Grabbe}
\email[Email: ]{quantum@orlingrabbe.com}
\date{April 19, 2005}

\begin{abstract}
This essay gives a self-contained introduction to quantum game theory, and is primarily
oriented to economists with little or no acquaintance with quantum mechanics. It assumes
little more than a basic knowledge of vector algebra.  Quantum mechanical notation and
results are introduced as needed.  It is also shown that some fundamental problems of quantum 
mechanics can be formulated as games.
\end{abstract}


\maketitle

Quantum game theory is an important development in quantum computation, and has implications
both for classical economic game theory and for quantum mechanics.  Unfortunately, the quantum
mechanical and quantum computational knowledge assumed in the literature presents a serious communication
barrier for most economists.  In the other direction, quantum game theory does not always seem to be
cognizant of many traditional results in classical economic game theory.  This essay is an attempt to
bridge the gap somewhat, by providing economists with a self-contained introduction to quantum games.
The essay assumes, for the most part, little more than a knowledge of vector algebra as mathematical
background, and introduces apparatus and results from quantum mechanics and quantum computation
as needed.  Key concepts such as Grover's search algorithm, Shor's factoring algorithm, and the quantum 
teleportation and pseudo-telepathy protocols based on entanglement are presented in detail, along with 
12 quantum games that illustrate the differences between quantum and classical game theory.  Along the
way we will see that many of the classical issues in quantum mechanics can be given a game theoretic
formulation.

\subsection{Some background history}

Game theory traditionally began in 1944 with \emph{The Theory of Games and Economic
Behavior}, by \emph{John von Neumann} and \emph{Oscar Morgenstern}.  But it had
antecedents stemming from the Hungarian mathematician von Neumann's earlier simultaneous
interest in game theory and the foundations of quantum mechanics. Since we are interested in
quantum games, we will describe the development briefly as follows.  In 1900 \emph{Max Planck}, 
attempting to get rid of the infinite energy implied in the then current formula for 
black body radiation, proposed a solution in which electromagnetic radiation energy was 
only emitted or absorbed in discrete energy units or \emph{quanta}, multiples of a fundamental
unit $h$: $h\nu, 2h\nu, 3h\nu \cdots$, where $\nu$ is the frequency of the radiating oscillator, and
$h$ is now known as Planck's constant.  In 1905 \emph{Albert Einstein} used Planck's quantum as an explanation for the photoelectric effect, whereby metals required incident light of a minimum frequency before they would release electrons. Incident light of frequency $\nu$ appeared to behave as a collection of particles (`photons'), each with energy
$E = h\nu$. \emph{Niels Bohr} then developed a useful, if unsatisfactory, model of the
atom as a nucleus surrounded by planetary electrons whose orbits assumed only
discrete values for the angular momentum, corresponding to multiples of Planck's quantum of energy:
$\frac{h}{2\pi}, \frac{2h}{2\pi}, \frac{3h}{2\pi}, \cdots$.  In 1924 \emph{Louis de Broglie} helped clarify the picture by associating with matter a wave, and noting that waves in closed loops, such as the electron `circling' the nucleus, were required to fit evenly around the loop---i.e. to have \emph{whole number} cycles.
The whole numbers $1,2,3,\cdots$ were thus associated with Planck's quanta (times a constant $a$): $1ah, 2ah, 3ah, \cdots$.  This was the \emph{old} quantum theory.  

The \emph{new} quantum theory began in 1925 when \emph{Werner
Heisenberg} conceived of representing physcial quantities by sets of time-dependent
\emph{complex} numbers. Heisenberg's \emph{matrix mechanics} essentially involved
$N \times N$ input-output matrices $H$, representing transitions between states of matter.  
If we denote by $\psi$ the state of the system we are interested in at time $t\ $(we
will for the moment set $t$ to zero), where $\psi$ is a
$N \times 1$ vector, then Heisenberg was working with the eigenvector-eigenvalue system
\begin{equation}
 H \psi = E \psi 
\end{equation}
where $E$, a scalar, represents some quantized energy level.  Assuming the system of $N$ equations is nondegenerate, there are $N$ solutions for $E$, say $E_n,\mbox{ }n = 1,2,\ldots,N$.  The
$E_n$ eigenvalues, or energy levels, are associated with an $N$-eigenvector-basis for the
state space of $\psi$.

The following year \emph{Erwin Schr\"{o}dinger}, looking for an electromagnetic interpretation
of the same phenomena, published his famous wave equation
\begin{equation}
 i\hbar\frac{\partial \psi}{\partial t}
  = \frac{-\hbar^2}{2m} \left(
    \frac{\partial^2}{\partial x^2}
    + \frac{\partial^2}{\partial y^2}
    + \frac{\partial^2}{\partial z^2}
  \right) \psi + V \psi,
  \end{equation}
where $i = \sqrt -1$, $\hbar$ is Planck's quantum of energy $h$ divided by $2\pi$, and $V$ is potential energy.
To Schr\"{o}dinger's delight, he discovered that his approach and Heisenberg's matrix
mechanics were mathematically equivalent, one form of this equivalence being suggested by
the equation $ i\hbar\frac{\partial \psi}{\partial t} = H \psi$.  If we, for example, set
$\psi = A exp^{(-i\frac{E}{\hbar}t)}$ in Schr\"{o}dinger's equation (2), and let $H =
\frac{-\hbar^2}{2m} \left(
    \frac{\partial^2}{\partial x^2}
    + \frac{\partial^2}{\partial y^2}
    + \frac{\partial^2}{\partial z^2}
  \right) + V$, then we obtain $E \psi = H \psi$, which is Heisenberg's equation (1).
  
\begin{sloppypar}
A few years later \emph{John von Neumann}, whose interest in quantum mechanics was
inspired by Heisenberg, `showed that quantum mechanics can be formalized
as a calculus of Hermitian operators in Hilbert space and that the theories of
Heisenberg and Schr\"{o}dinger are merely particular representations of this calculus.' \cite[p.22]{Jammer}
Recall that a \emph{Hermitian matrix} is one that is its own complex-conjugate transpose.  For example,
consider the matrix $\sigma_y = \left( \begin{array}{cc}
0 & -i \\
i & 0 \end{array} \right)$. The transpose of this matrix is $\sigma_y^T = \left( \begin{array}{cc}
0 & i \\
-i & 0 \end{array} \right)$.  Then if we take the complex conjugate, by changing the signs of the
imaginary parts, $i \rightarrow -i, -i \rightarrow i$, we again obtain the matrix $\sigma_y$.  So
$\sigma_y$ is Hermitian.  A Hermitian matrix may be considered an \emph{operator} on a vector in
Hilbert space.  Recall that Hilbert space is simply a vector space defined over the complex numbers $\mathbf C$,
with a defined \emph{norm} or \emph{length} or \emph{inner product}.  For the vector $\psi$ the
norm is $||\psi|| = \sqrt {\psi^\dagger \psi}$, where $\psi^\dagger$ is the complex conjugate transpose
of $\psi$.  Hilbert spaces may be infinite dimensional, but we will only consider finite dimensional spaces
in this essay.
\end{sloppypar}

It was during this heady period that game theory arose.  The name `game' was
introduced in 1921 by the French mathematician Emil Borel, who was preoccupied with bluffing
in poker and initiated `la th\'{e}orie du jeu'. In his 1928 paper \cite{vN28}, written for Karl
Menger's Vienna Colloquium,  von Neumann defined, and completely solved, two-person zero-sum games.
He speculated on $N$-person games, which were more complicated due to the possibility of coalitions:
with three people or more, some people could benefit from cooperation. Later, in a famous paper delivered to the Princeton economics club in 1932, the same year
his book on the foundations of quantum mechanics was published, von Neuman laid out the whole
apparatus of linear programming and the foundations of his later game theory book with Morgenstern.
(This paper was not published util 1937 \cite{vN37}.)   

Central to many results was the linear programming problem and its dual \cite{DGale}.  The linear programming
problem is this:  given an $m \times n$ matrix $A$, an $n \times 1$-vector $b$, and an $m \times 1$-vector $c$, find a
non-negative $m \times 1$-vector $x$ such that
\begin{equation}
x^T c\mbox{ is a maximum }
\end{equation}
subject to
\begin{equation}
x^T A \leq b^T .
\end{equation}
The dual problem is that of finding a non-negative $n \times 1$-vector $y$ such that
\begin{equation}
y^T b \mbox{ is a minimum }
\end{equation}
subject to
\begin{equation}
Ay \geq c .
\end{equation}
The only major game theoretic result missing from von Neumann-Morgenstern (and indeed one missing from the
quantum game theory literature) is the theory of the core   \cite[chapter 8]{LR}.  The \emph{core} arises in 
$N$-person game theory.  In $N$-person game theory players' interests are not necessarily opposed,  Some 
players may improve their (expected) payoffs by forming coalitions with other players.  A maximum value can 
be determined for each subset of players, which gives rise to the \emph{characteristic function} of the game. 
Let $S$ be a member of the set of subsets of $N$.  The characteristic function $v(S)$ is a mapping from the set of subsets (i.e. coalitions) of players to an (expected) payoff value in the set of real numbers $R$:
\begin{equation}
v(S): S \rightarrow R .
\end{equation}
The value $v(S)$ is determined as the maximum value obtainable by $S$ in the two-person game between 
the coalition $S$ and the coalition of all remaining players $N - S$.
An \emph{imputation} is a set of numbers (allocations or payoffs) $\{\pi_i\}$ assigned to each player $i$ in $N$.  The core $C_x$ is the set of imputations $C_x = \{\{\pi_i\}_x\}$ such that
\begin{equation}
v(S) \leq \sum_{i \in S} \pi_i\mbox{ for every subset } S \mbox{ in } N ,\mbox{ and } \sum_{i \in N} \pi_i = v(N) .
\end{equation}
The core (it may be empty) is critical to economic equilibrium. The core restricts the value of any coalition to
be not greater than the sum of the imputed payoffs to each member of the coalition individually.  Debreu and
Scarf \cite{DeSc} showed that in a replicated market game the core shrinks down to a set of imputations which
can be interpreted in terms of a price system emerging as its limit.

Meanwhile, in quantum mechanics, the reactionary forces of determinism were at work.  In a 1935 paper  \cite{EPR} 
\emph{Einstein-Podolsky-Rosen} (EPR) attempted to prove the \emph{incompleteness} of quantum
mechanics by considering entangled pairs of particles which go off in different directions.  The particles
may become separated by light-years.  Nevertheless a measurement of one particle will instantly affect
the state of the other particle, an example of quantum mechanics' `spooky action at a distance'.  (We
will discuss entanglement later, in the body of this essay, but essentially two particles are entangled
if their wave functions cannot be written as tensor products.)  This instantaneous effect is sometimes 
called the `EPR channel', though properly speaking it should be called the \emph{Bohr channel} because Bohr argued for its existence, while EPR argued against it.  \emph{John Bell}  \cite{JSB} formulated a set of inequalities that would distinguish experimentally whether quantum mechanics was incomplete, or whether physics is \emph{non-local}, permitting
instantaneous propagation of some effects of some causes.  Fortunately Bohr was right and EPR were wrong, as
experimental evidence has decisively demonstrated.\cite{NG2005}  The Bohr channel is now the basis of quantum teleportation, and, indeed, every quantum computer is in some sense a demonstration of the Bohr effect.

As it stands today, quantum game theory can probably be viewed as a subbranch of quantum computation.
With respect to the latter development, it was apparently \emph{Richard Feynman}  \cite{RF} who first foresaw 
the unusual power of quantum computers, noting that simulation of quantum evolution in a classical 
computer would invole an exponential slowdown in time. Once again there is a direct line from von
Neumann \cite{vN56} (with Stan Ulam \cite{SU}): `In the nineteen fifties, Ulam and von Neumann began to discuss computational models known as cellular automata, in which simple rules of computation applied to systems with many
degrees of freedom could produce complex patterns of behavior.  By the nineteen eighties, Friedkin, Feynman,
Minsky and others were speculating on the possibility of describing the laws of physics and the universe
in terms of cellular automata and computation.  Underlying their ideas was a dissatisfaction with the
conventional description of physics based on continuous space and time.'  \cite{JR}

\emph{David Deutsch}  \cite{DD} suggested that quantum \emph{superposition} might allow the parallel performance of many classical computations.  Indeed, we shall see that superposition is the key new ingredient that makes quantum games different from classical games, whether or not the superposed states are \emph{entangled}.  For dynamic games,
superposition suffices, though static games generally require entanglement also.  (Superposition is the ability
of a quantum observable to be in a linear combination of two or more states at the same time.)

The `killer app' that created a storm of interest in quantum computation came when \emph{Peter Shor}  \cite{PS} showed that a quantum mechanical algorithm could factor numbers in polynomial time. This was an exponential speed-up over factoring algorithms available to classical computers. Shor's algorithm relies mainly on superposition and an
ingenious application of the quantum Fourier transform.  Another result was obtained
by \emph{Lov Grover}  \cite{LG}, who showed a quantum mechanical way to speed up the search for items in an $N$-item database from $O(N)$ steps to $O(\sqrt N)$ steps. Grover's result is based upon the rotation of quantum states 
(vectors) in Hilbert space.

Quantum game theory seems to have crystallized when \emph{David Meyer} gave a talk on the subject at Microsoft
Corporation (see \cite{DM} for an account).  Of the twelve quantum games considered in this essay, three are
due to Meyer (the Spin Flip game, and Guess a Number games I and II).

As von Neumann and Morgenstern noted  \cite{vNM}, `In order to elucidate the conceptions which we are applying to
economics, we have given and may give again some illustrations from physics.  There are many social scientists
who object to the drawing of such parallels on various grounds, among which is generally found the assertion
that economic theory cannot be modeled after physics since it is a science of social, of human phenomena, has to
take psychology into account, etc.  Such statements are at least premature.' One may conversely note that some
may similarly object to mixing economic concepts with those of quantum mechanics, but such objections are at least premature.  Indeed, the human brain is arguably a quantum computer  \cite{HS1}  \cite{HS2}  \cite{RP}  \cite{DD89} \cite{DD2002}, though the mind may be more than that, so to ignore quantum mechanics in questions of psychology, much less economics, is folly indeed.  In the reverse direction, the role of the human mind in the quantum measurement problem has been a subject of contention \cite{JJ} since it was first clearly delineated by von Neumann.  In any event, quantum games may have lessons both for economics and quantum mechanics.

\subsection{Preliminary mathematical pieces}

Before defining a game, we are going to give an example of one.  This example, the
Spin Flip Game in the next section, will highlight some of the differences between traditional game theory and quantum game theory.  In order to explain how the Spin Flip Game
works, we will need some modest mathematical preliminaries, involving $2\times1$
vectors and $2\times2$ matrices.
  
The following simple vectors will prove quite useful for our purposes:
\begin{equation}
u = \left( \begin{array}{c}
1 \\
0 \end{array} \right),
\ \ d = \left( \begin{array}{c}
0 \\
1 \end{array} \right).
\end{equation}
These are, of course, \emph{basis} vectors for 2-dimensional (complex) space, as any
point can be expressed in the form of $au+bd$ (where, in general, it is assumed that $a$
and $b$ are complex scalars, $a, b \in \mathbf C$). But $u$ and $d$ can also represent many
'spaces' or states outside geometry: Yes or No responses, Up or Down spin states of an electron (with
spin measured in the $z$ direction), Heads or Tails in a probability sequence, Success or Failure 
of a bidding process or an electronic device, and so on.  A choice of $u$ or $d$ can also represent 
player moves in a game, and we can represent a sequence of such moves by the \emph{bits} in a binary number, or the quantum equivalent \emph{qubits}.  Bits and qubits differ by the fact that a bit $b$ is a single number, $b \in \{0,1\}$,
while a qubit $q$ is a vector in a two-dimensional Hilbert space, $q \in \{au+bd\}$.  (Later we will introduce
the Dirac notation $|0\rangle$, $|1\rangle$, and in this essay there is the correspondence $u \leftrightarrow$ 
$|u\rangle \leftrightarrow
\left( \begin{array}{c}
1 \\
0 \end{array} \right) \leftrightarrow |0\rangle \leftrightarrow \mbox{ bit } 0$, and the similar correspondence
$d \leftrightarrow$ $|d\rangle \leftrightarrow
\left( \begin{array}{c}
0 \\
1 \end{array} \right) \leftrightarrow |1\rangle \leftrightarrow \mbox{ bit } 1$.
For example, to foreshadow what is to come, the 5-qubit register or sequence $|10011\rangle$ could represent the 
tensor product of vectors as well as the number $19\mbox{ } (= 2^4+2^1+2^0)$:
\begin{equation}
|10011\rangle =  \left( \begin{array}{c}
0 \\
1 \end{array} \right) \otimes \left( \begin{array}{c}
1 \\
0 \end{array} \right) \otimes \left( \begin{array}{c}
1 \\
0 \end{array} \right)
\otimes \left( \begin{array}{c}
0 \\
1 \end{array} \right) \otimes \left( \begin{array}{c}
0 \\
1 \end{array} \right)
\end{equation}
$ = (0,0,0,0,0,0,0,0,0,0,0,0,0,0,0,0,0,0,0,1,0,0,0,0,0,0,0,0,0,0,0,0)^T$. In the latter vector, the $1$ is in
the 20th slot, not the 19th, because we start counting from $0$, which occupies the first slot.
The same sequence could have also been written $duudd$.)

Next we need some way to transform one state into another. For a two-state system,
it is useful to do this with the Pauli spin matrices.  The three 
$2 \times 2$ \emph{Pauli spin matrices} are 
\begin{equation}
\sigma_x = \left( \begin{array}{cc}
0 & 1 \\
1 & 0 \end{array} \right),
\ \sigma_y = \left( \begin{array}{cc}
0 & -i \\
i & 0 \end{array} \right),
\ \sigma_z = \left( \begin{array}{cc}
1 & 0 \\
0 & -1 \end{array} \right).
\end{equation}
These three matrices, along with the following unit matrix $\mathbf 1$,
\begin{equation}
\mathbf 1 = \left( \begin{array}{cc}
1 & 0 \\
0 & 1 \end{array} \right),
\end{equation}
span $2 \times 2$ Hermitian matrix space (recall that a Hermitian matrix has diagonal elements that are real, and mirror-image off-diagonal elements that are complex
conjugates of each other).
Each of the spin matrices has a simple effect on the base states $u$ and $d$.  In
particular,
\begin{equation}
\mathbf 1 u = u,\mbox{   }\mathbf 1 d = d
\end{equation}
\begin{equation}
\sigma_x u = d, \mbox{   }\sigma_x d = u
\end{equation}
\begin{equation}
\sigma_z u = u, \mbox{   }\sigma_z d = -d .
\end{equation}
Table 1 summarizes some matrix properties of the Pauli spin matrices:
\begin{table}[ht]
\begin{tabular}{|c|}
\hline
$\sigma_x^2 = \mathbf 1$\\
$\sigma_y^2 = \mathbf 1$\\
$\sigma_z^2 = \mathbf 1$\\
$\sigma_x \sigma_y = -\sigma_y \sigma_x = i \sigma_z$\\
$\sigma_y \sigma_z = -\sigma_z \sigma_y = i \sigma_x$\\
$\sigma_z \sigma_x = -\sigma_x \sigma_z = i \sigma_y$\\
\hline
\end{tabular}
\caption{Products of Pauli spin matrices}
\end{table}

\subsection{The spin flip game}

Electrons have two spin states: spin up and spin down.
Let us consider a simple game of electron spin flip played between Alice and Bob.
Alice first prepares the electron in spin up state $u$.  After this initial step, Bob
applies either the $\sigma_x$ or the $\mathbf 1$ matrix to $u$, resulting in either
\begin{equation}
\sigma_x u = d \mbox{ or } \mathbf 1 u = u .
\end{equation}
Then Alice (not knowing Bob's action or the state of the electron) takes a turn, also applying either $\sigma_x$ or $\mathbf 1$ to the electron spin.  Then Bob (not knowing Alice's action
or the state of the electron) takes another turn.  Finally, the electron spin state is measured. If it is in the $u$ state, Bob wins \$1, and Alice loses \$1.  If it is in the $d$ state, Alice wins \$1, while Bob loses the same amount.

The sequence of possible choices by Bob (\emph{columns}) and Alice (\emph{rows}) are summarized in Table II.
\begin{table}[ht]
\begin{tabular}{|r|r|r|r|r|}
\hline
$Alice\backslash Bob$ & $\mathbf 1$,$\mathbf 1$ & $\mathbf 1$,$\sigma_x$ & $\sigma_x$,$\mathbf 1$ & $\sigma_x$,$\sigma_x$\\
\hline
$\mathbf 1$ & $\mathbf 1$,$\mathbf 1$,$\mathbf 1$ & $\mathbf 1$,$\mathbf 1$,$\sigma_x$ & $\sigma_x$,$\mathbf 1$,$\mathbf 1$ & $\sigma_x$,$\mathbf 1$,$\sigma_x$\\
$\sigma_x$ & $\mathbf 1$,$\sigma_x$,$\mathbf 1$ & $\mathbf 1$,$\sigma_x$,$\sigma_x$ & $\sigma_x$,$\sigma_x$,$\mathbf 1$ & $\sigma_x$,$\sigma_x$,$\sigma_x$\\
\hline
\end{tabular}
\caption{Sequence of player moves}
\end{table}
Note that Alice's move is the middle one in each sequence of three, reading from right to
left.  For example $\mathbf 1,\mathbf 1,\sigma_x$ means that Bob played $\sigma_x$,
followed by Alice's play of $\mathbf 1$, followed by Bob's play of $\mathbf 1$.  The net
result is $\mathbf 1 \mathbf 1 \sigma_x u = d$.  Thus Alice wins \$1.  The sequence of spin states after each move, starting from the initial $u$ state are shown in Table III.  Again,
each sequence of three should be read from right to left.
\begin{table}[ht]
\begin{tabular}{|r|r|r|r|r|}
\hline
$Alice\backslash Bob$ & $\mathbf 1$,$\mathbf 1$ & $\mathbf 1$,$\sigma_x$ & $\sigma_x$,$\mathbf 1$ & $\sigma_x$,$\sigma_x$\\
\hline
$\mathbf 1$ & $u$,$u$,$u$ & $d$,$d$,$d$ & $d$,$u$,$u$ & $u$,$d$,$d$\\
$\sigma_x$ & $d$,$d$,$u$ & $u$,$u$,$d$ & $u$,$d$,$u$ & $d$,$u$,$d$\\
\hline
\end{tabular}
\caption{Sequence of spin states}
\end{table}

Finally, Table IV shows the payoff to \emph{Alice}, positive if the final spin is in the
$d$ state, negative if it is in the $u$ state.
\begin{table}[ht]
\begin{tabular}{|r|r|r|r|r|}
\hline
$Alice\backslash Bob$ & $\mathbf 1$,$\mathbf 1$ & $\mathbf 1$,$\sigma_x$ & $\sigma_x$,$\mathbf 1$ & $\sigma_x$,$\sigma_x$\\
\hline
$\mathbf 1$ & $-1$ & $+1$ & $+1$ & $-1$\\
$\sigma_x$ & $+1$ & $-1$ & $-1$ & $+1$\\
\hline
\end{tabular}
\caption{Payoffs to Alice}
\end{table}

This is the basic Spin Flip Game, which we are going to extend in two directions:
first, by considering probabilistic moves, and, second, by considering \emph{quantum
superposition} (without \emph{quantum entanglement}) of states.  But before doing this, let's
consider some basic game theory terminology.

\subsection{First game definitions and strategies}
As is implicit in the previous section, a \emph{game} $\Gamma$ may be defined as a set 
$\Gamma = \Gamma$(players,moves or actions,outcomes,payoffs).  In the Spin Flip Game, the \emph{players} were Alice and Bob, the \emph{moves} were the application of the matrices $\sigma_x$ or $\mathbf 1$, the \emph{outcomes} were the spin states $u$ or $d$, and the
\emph{payoffs} to Alice were either +1 or -1, according to whether the final state
was $d$ or $u$, respectively.  Since this was a \emph{two-person, zero-sum} game,
the payoffs to Bob were the exact opposite of those to Alice.

Omitted thus far in the account of the game is any explanation how Alice and Bob determined their
moves---how they decided whether to play $\sigma_x$ or $\mathbf 1$.  A \emph{strategy}
is a rule for determining a move at any stage of a game.  That is, in our example, a \emph{move} is a member
of the set $\{\mathbf 1, \sigma_x\}$, while a \emph{strategy} is a \emph{function} $f$ mapping the
state of the game to the set of moves: $f:\mbox{ game state } \rightarrow \{\mathbf 1, \sigma_x\}$.
(There seems to be confusion on this point in the quantum game theory literature.) This is not quite a good definition, since the `state of the game' may not be known to a player; a player may know little more than his or her move.  So let's revise this to:  a \emph{strategy} for Alice is a mapping $f_A:\{\mbox{ Alice's information }\} \rightarrow \{\mbox{ Alice's moves }\}$.  Similarly for Bob. In the Spin Flip Game Alice, after initial preparation of the electron, 
has only one opportunity to choose a move, so she has a single strategy at the second,
or middle, step of the sequence of three moves. Bob has strategies for the first and last
steps. Thus, associated with a sequence of moves is a sequence of strategies.  In economics, strategies are highly dependent on a player's \emph{information}.  Of particular interest is \emph{asymmetric} information, where one player has some information advantage over another, or where the information sets of the players are not the same. If Bob
can make quantum moves that Alice cannot, then clearly Bob has an information advantage in at least that
respect. Strategies are endogenous to a game, given the game's allowed moves and payoffs, so strategies are not
properly part of the game's definition.  Rather, solving a game essentially means determining the optimal
strategies for the players.

The concept of information set is important.  In the Spin Flip Game we said that neither
Bob nor Alice could know the other person's moves.  Suppose we relaxed this assumption.
Then Alice would know Bob's first move, and could choose her move accordingly, but it would make no difference.  Bob, seeing Alice's move (and knowing his own first move), could always
choose a final move that would leave the electron in a spin up state $u$.  He would win
100 percent of the time.  It would not be a `game', but rather a racket.  So in this case
we must limit the information sets of Alice and Bob in order to make it a game in the
first place.

Now, as an example let us consider the following strategies, $f_A$ and $f_B$, for Alice and Bob,
respectively.  These will be called \emph{mixed} strategies because they involve selection of a move with some probability mechanism.
\begin{eqnarray}
f_A & = & \mbox{ play } \mathbf 1 \mbox{ with probability } p = \frac{1}{2}, \mbox{ play } \sigma_x \mbox{ with probability } q = \frac{1}{2} \\
f_B & = & \mbox{ play } \mathbf 1 \mbox{ with probability } p = \frac{1}{2}, \mbox{ play } \sigma_x \mbox{ with probability } q = \frac{1}{2}.
\end{eqnarray}
Then, looking at the columns of Table IV, we see that Alice's \emph{expected payoff} 
$\overline{\pi}_A$, no matter what Bob does, is always
\begin{equation}
\overline{\pi}_A = \frac{1}{2} (+1) + \frac{1}{2} (-1) = 0
\end{equation}
while, looking at the rows of Table IV, Bob's expected payoff is always
\begin{equation}
\overline{\pi}_B = \frac{1}{4} (+1) + \frac{1}{4} (-1) + \frac{1}{4} (-1) + \frac{1}{4} (+1) = 0 .
\end{equation}
Of course, for the concept of \emph{mixed} strategies and \emph{expected} payoffs to make
much sense, we should consider a sequence of $N$ games
\begin{equation}
\Gamma_N \Gamma_{N-1} \Gamma_{N-2} \cdots \Gamma_3 \Gamma_2 \Gamma_1 .
\end{equation}
The \emph{actual} payoff to Alice, letting $x$ stand for the number of wins
in $N$ games, will be a member of the \emph{payoff set}
\begin{equation}
\Pi = \{f(x;N)\} = \{2x-N, \mbox{ for } x = 0,1,\cdots,N\}
\end{equation} 
while the probability of these payoffs are
\begin{equation}
P(\Pi) = \{f(x;N,p)\} = \{\left( \begin{array}{c}
N \\
x \end{array} \right) p^x q^{N-x}, \mbox{ for } x = 0,1,\cdots,N\}.
\end{equation} 
For example, with $N = 3$, the possible payoffs to Alice are $\{-3,-1,1,3\}$, and
if $p = \frac{1}{2}$ these have respective probabilities $\{\frac{1}{8},\frac{3}{8},\frac{3}{8},\frac{1}{8}\}$.  Alice's expected payoff
$\overline{\pi}_A$ is $0$, but if $N$ is odd, her actual payoff will never be $0$.

Physicists will recognize equation (22) as giving the possible outcome states when a massive particle of spin $\frac{N\hbar}{2}$ is measured. The spin in this case defines an $(N+1)$-state quantum system, with possible
outcomes for the spin values (in terms of the fundamental unit $\frac{\hbar}{2})$ given by equation (22).  Thus
the \emph{measured} spin states of the massive particle may be thought of as being determined by $N$ Spin Flip games between Alice and Bob.

In the matrix of payoffs analogous to Table IV, for a general two-person, zero-sum game, let Alice's moves be
represented by the mixed strategy (the set of probabilities over moves) $P_A = \{a_1,a_2,\cdots,a_m\}$, while
the mixed strategy of Bob is represented by $P_B = \{b_1,b_2,\cdots,b_n\}$.  Let the payoffs to Alice be
represented by the $m \times n$ matrix $[\pi_{ij}]$.  Then the \emph{expected payoff} to Alice is
\begin{equation}
\overline{\pi}_A = \sum_{j=1}^n \sum_{i=1}^m \pi_{ij} a_i b_j .
\end{equation}
In this context, we should mention the \emph{minimax theorem} which says that for every finite two-person, zero-sum
game
\begin{equation}
max_{\ _{P_A}}(min_{\ _{P_B}} \overline{\pi}_A) = min_{\ _{P_B}}(max_{\ _{P_A}} \overline{\pi}_A).
\end{equation}
That is, Alice chooses probable moves to maximize her expected payoff, while Bob choses probable moves to minimize
Alice's expected payoff.  The minimax theorem says the payoff to Alice's maximizing set of probabilities given Bob's minimizing set of probabilites is equal to the payoff to Bob's minimizing set of probabilities given Alice's
maximizing set of probabilities.  

\subsection{Amplitudes and superpositions and his cheatin' heart}

Let's consider a quantum state (a vector) $\psi$ of the following form, where $a$ and $b$ may
be complex scalars:
\begin{equation}
\psi = a u + b d
\end{equation}
In quantum computation, this superimposed two-dimensional state is known as a \emph{qubit}, which we will discuss in detail later.  Here $a$ and $b$ are \emph{amplitudes}, and a (von Neumann) measurement of
$\psi$ will obtain the base state $u$ with probability $|a|^2$, while the measurement will yield base state $d$ with probability $|b|^2$, where $|a|^2 + |b|^2 = 1$.
(Recall that for a complex number $a$, and its complex conjugate $a^*$, we have
$aa^* = a^*a = |a|^2$.)  

This raises the possibility of games, including variants
of the Spin Flip Game, for which there is no classical analog. 
For example, set $a = b = \frac{1}{\sqrt 2}$.  Then the probability of either $u$
or $d$ is $|\frac{1}{\sqrt 2}|^2 = \frac{1}{2}$.    Thus probability is built
into measurements of the state vector, irrespective of whether a mixed strategy is chosen by
either Bob or Alice.
 
Here $u$ and $d$ are orthonormal (that is, the inner product of $u$ with $d$ is 0, and the inner product of
either $u$ or $d$ with itself is 1), so we may obtain $a$ as the inner product
\begin{equation}
\langle\psi,u\rangle = a \langle u,u\rangle + b \langle d,u\rangle = a(1) + b(0) = a.
\end{equation}
A similar computation will yield $b$.  
 
$\mathbf{Alice\mbox{ }Cheats.}$ Now let us consider a variation of the Spin Glip Game---let's call it \emph{Alice Cheats}---in which Alice has a way of cheating in the initial preparation of
the spin state of the electron.  First, suppose she initially prepares the electron
in spin state $d$, knowing that Bob thinks it will be in spin state $u$.  Otherwise
the game is exactly as before: both Bob and Alice play either $\mathbf 1$ or $\sigma_x$. 
It is easy to see that the arrangement of spin states changes in Table III, and the arrangemnt of 
payoffs to Alice changes in Table
IV, but the set of payoffs $\Pi$ is still the same, and the corresponding payoff probabilities $P(\Pi)$ to Alice are unchanged.  Thus Alice has cheated to no avail. She simply changed the initial state from $u$ to $d$, and it had no impact on the outcome of the game.
Where she previously got +1, she now gets -1, and vice-versa.

So Alice tries something else.  She choses the initial state to be 
$\frac{1}{\sqrt 2} (u + d)$.  Then whether Bob plays $\mathbf 1$ or $\sigma_x$,
his move leaves the state of the game unchanged:
\begin{equation}
\mathbf 1 [\frac{1}{\sqrt 2} (u + d)]  = \frac{1}{\sqrt 2} ( \mathbf 1 u +  \mathbf 1 d) = \frac{1}{\sqrt 2} (u + d),
\end{equation}
\begin{equation}
\sigma_x [\frac{1}{\sqrt 2} (u + d)]  = \frac{1}{\sqrt 2} ( \sigma_x u +  \sigma_x d) = \frac{1}{\sqrt 2} (d + u).
\end{equation}
Since $u+d = d+u$, the state is unchanged by the play of either $\mathbf 1$ or
$\sigma_x$.  However, when the final measurement of the (unchanged) state of the
 electron is taken, Alice discovers to her frustration that she once more wins or loses a dollar with equal probability,
because a measurement of the final superposed state yields $u$ or $d$ with equal probability.
For a single game, the payoff set $\Pi$ and corresponding probabilities $P(\Pi)$ are:
\begin{equation}
\Pi = \{-1,+1\}
\end{equation}
\begin{equation}
P(\Pi) = \{(\frac{1}{\sqrt 2})^2, (\frac{1}{\sqrt 2})^2\} = \{\frac{1}{2}, \frac{1}{2}\} .
\end{equation}

$\mathbf {Bob\mbox{ }Cheats}$. Let's return to our basic Spin Flip Game, where a repentent
Alice prepares the electron in an initial $u$ state, with the added detail that she follows
a mixed strategy, and choses $\mathbf 1$ or $\sigma_x$ each with probability $p = \frac{1}{2}$.
But now we allow Bob to cheat.  Since Bob does not prepare the initial electron state, Bob's method of cheating will differ from Alice's.  What dastardly things can Bob do?  Bob has some extra Pauli spin matrices up his sleeve, namely $\sigma_y$ and $\sigma_z$, as well as linear combinations of these. In addition, Bob has the final move.
Let's suppose that Bob plays the so-called Hadamard operator $H = \frac{1}{\sqrt 2}(\sigma_x+\sigma_z)$:
\begin{equation}
H = \frac{1}{\sqrt 2} \left( \begin{array}{cc}
1 & 1 \\
1 & -1 \end{array} \right).
\end{equation}
After Bob's first move, the spin state would be
\begin{equation}
H u = \frac{1}{\sqrt 2} \left( \begin{array}{cc}
1 & 1 \\
1 & -1 \end{array} \right)\left( \begin{array}{c}
1 \\
0 \end{array} \right) = \frac{1}{\sqrt 2} \left( \begin{array}{c}
1 \\
1 \end{array} \right) = \frac{1}{\sqrt 2} (u + d) .
\end{equation}
As we saw in equations (28-29), Alice's mixed strategy will not change this state.  Then Bob plays $H$ again to obtain:
\begin{equation}
H (Hu) = \frac{1}{\sqrt 2} \left( \begin{array}{cc}
1 & 1 \\
1 & -1 \end{array} \right) \frac{1}{\sqrt 2} \left( \begin{array}{c}
1 \\
1 \end{array} \right) = \frac{1}{2} \left( \begin{array}{c}
2 \\
0 \end{array} \right) = u.
\end{equation}
Bob will always win.  This results from Bob's ability to create a \emph{superposition} of
states (and his having the final move). Like Schr\"{o}dinger's cat that is simultaneously both alive and dead, the electron spin is simultaneously both $u$ and $d$ after Bob applies the Hadamard matrix $H$ to $u$. Alice cannot 
alter the outcome by playing a classical
mixed strategy that choses a play of $\mathbf 1$ with probability $p$ and $\sigma_x$ with probability
$1-p$.   

\subsection{Guess a number games}

To understand the \emph{Guess a Number Game}, we will first need to introduce some more concepts, including
\emph{qubits}, the \emph{Walsh-Hadamard transformation} (the $n$-bit analogue of the Hadamard transformation) and some elements of the \emph{Grover search algorithm}  \cite{LG}. The Grover search algorithm is one of the fundamental techniques of quantum computation, so it is not surprising it shows up in quantum game theory.

$\mathbf {\mbox{ } Dirac \mbox { } notation.\mbox{ } }$ For convenience, we are going to alter our designations for $u$ and $d$ into forms that will denote each $2 \times 1$ vector and also its $1 \times 2$ \emph{complex conjugate} transpose: 
\begin{equation}
|u\rangle = \left( \begin{array}{c}
1 \\
0 \end{array} \right), \langle u| = (1,0),
\ \ |d\rangle = \left( \begin{array}{c}
0 \\
1 \end{array} \right), \langle d| = (0,1).
\end{equation}
Note that if $|x\rangle = \left( \begin{array}{c}
1 \\
-i \end{array} \right), \mbox{ then } \langle x| = (1,i)$. This is the \emph{Dirac bracket notation}, where $\langle x|$ is the \emph{bra} and $|x\rangle$ is the \emph{ket}.  The bras are horizontal, and the kets are vertical.  Notice that we
may then use the form $|u\rangle \langle d|$:
\begin{equation}
|u\rangle \langle d| = \left( \begin{array}{c}
1 \\
0 \end{array} \right) (0,1) = \left( \begin{array}{cc}
0 & 1 \\
0 & 0 \end{array} \right)
\end{equation}
where $|u\rangle \langle d|$ turns a $|d\rangle$ into an $|u\rangle$; namely, $|u\rangle \langle d|d\rangle = |u\rangle$; and an $|u\rangle$ into a $2 \times 1$ zero vector, namely $|u\rangle \langle d|u\rangle = \left( \begin{array}{c}
0 \\
0 \end{array} \right)$.

$\mathbf { Qubits. }$ Consider an $n$-bit binary number $x$:
\begin{equation}
x = b_{n-1} b_{n-2}\cdots b_2 b_1 b_0 ,
\end{equation}
where each $b_i$ is either $0$ or $1$, $b_i \in \{0,1\}$.
Note that the decimal equivalent of $x$ is
\begin{equation}
x = b_{n-1}2^{n-1}+b_{n-2}2^{n-2}+\cdots+b_{2}2^{2}+b_{1}2^{1}+b_{0}2^{0}.
\end{equation}
In a quantum computer, each $b_i$ may be represented by $|u\rangle$ or $|d\rangle$, respectively. We make the
correspondence $|u\rangle \rightarrow |0\rangle, |d\rangle \rightarrow |1\rangle$, and call $\{|0\rangle,
|1\rangle\}$ the \emph{computational basis}. The latter representation, however, makes them quantum bits or 
\emph{qubits}---vectors in a two-dimensional Hilbert space.  Each qubit can be any linear combination $a |0\rangle + c |1\rangle$, where $|a|^2 + |c|^2 = 1$. For example, consider the 3-qubit state 
\begin{eqnarray}
|\psi\rangle = |q_2\rangle \otimes |q_1\rangle \otimes |q_0\rangle \mbox{ where }\\
|q_2\rangle = \frac{1}{\sqrt 2}(|0\rangle + |1\rangle)\\
|q_1\rangle = |1\rangle\\
|q_0\rangle = |1\rangle .
\end{eqnarray}
Then the quantum register is the superposition of $|3\rangle$ and $|7\rangle$:
\begin{eqnarray}
|\psi\rangle = \frac{1}{\sqrt 2}(|0\rangle + |1\rangle) \otimes |1\rangle \otimes |1\rangle\\
= \frac{1}{\sqrt 2} (|011\rangle + |111\rangle)\\
= \frac{1}{\sqrt 2} (|3\rangle + |7\rangle).
\end{eqnarray}
This calculation will be further clarified below.

A collection
of $n$ qubits is called a \emph{quantum register} of size $n$. There are $N = 2^n$ such numbers or quantum
register states $x$ in terms of the computational basis $b_i$, $b_i \in \{|0\rangle, |1\rangle\}$; hence $x \in S = \{0,1,2,\cdots ,N-1\}$.  So our Hilbert space has dimension $N = 2^n$. That is, a classical computer with $n$
bits has a total of $2^n$ possible states.  By contrast, a quantum computer with $n$ qubits can be in any
superposition of these $2^n$ states, which results in an arbitrary state or vector in $2^n$-dimensional Hilbert
space. A superposition $|\psi_s\rangle$ of \emph{all} the computational basis states, letting $a_x$ be the probability amplitude associated with the number or state $x$, would be designated
\begin{equation}
|\psi_s\rangle = \sum_{x=0}^{2^n-1} a_x |x\rangle .
\end{equation}
If all amplitudes $a_x$ are equal, then this superposition is designated
\begin{equation}
|\psi_s\rangle = \frac{1}{\sqrt {2^n}} \sum_{x=0}^{2^n-1} |x\rangle .
\end{equation}
Note that in the summation in equation (47), $|x\rangle$ runs through all basis states or numbers, and all the basis states are orthogonal to each other.  Hence for a given number or state $|z\rangle$, we have that the amplitude
for $|z\rangle$ is the inner product
\begin{equation}
\langle z|\psi_s\rangle = \frac{1}{\sqrt {2^n}}.
\end{equation}
A measurement of $|\psi_s\rangle$ will thus yield $|z\rangle$ with probability
\begin{equation}
|\langle z|\psi_s\rangle|^2 = \frac{1}{2^n}.
\end{equation}

Now, when we have a \emph{many-state} system of $|u\rangle$s and $|d\rangle$s (i.e., $|0\rangle$s and $|1\rangle$s)
like this, each in a Hilbert space
$\mathbf H_2$ of $2$ dimensions, we simply place the states side by side.  Two such states side
by side form a Hilbert space of $\mathbf H_4 = \mathbf H_2 \otimes \mathbf H_2$ dimensions.  Basis vectors in a $2$-qubit quantum register could thus be represented
\begin{equation}
|0\rangle |0\rangle =
|u\rangle \otimes |u\rangle = \left( \begin{array}{c}
1 \\
0 \end{array} \right) |u\rangle = \left( \begin{array}{c}
u \\
\mathbf 0 \end{array} \right) = \left( \begin{array}{c}
1 \\
0 \\
0 \\
0 \end{array} \right).
\end{equation}

\begin{equation}
|0\rangle |1\rangle =
|u\rangle \otimes |d\rangle = \left( \begin{array}{c}
1 \\
0 \end{array} \right) |d\rangle = \left( \begin{array}{c}
d \\
\mathbf 0 \end{array} \right) = \left( \begin{array}{c}
0 \\
1 \\
0 \\
0 \end{array} \right).
\end{equation}

\begin{equation}
|1\rangle |0\rangle =
|d\rangle \otimes |u\rangle = \left( \begin{array}{c}
0 \\
1 \end{array} \right) |u\rangle = \left( \begin{array}{c}
\mathbf 0 \\
u \end{array} \right) = \left( \begin{array}{c}
0 \\
0 \\
1 \\
0 \end{array} \right).
\end{equation}

\begin{equation}
|1\rangle |1\rangle =
|d\rangle \otimes |d\rangle = \left( \begin{array}{c}
0 \\
1 \end{array} \right) |d\rangle = \left( \begin{array}{c}
\mathbf 0 \\
d \end{array} \right) = \left( \begin{array}{c}
0 \\
0 \\
0 \\
1 \end{array} \right).
\end{equation}
Physicists, who get bored with the excessive notation, usually compress the tensor product of qubits as
\begin{equation}
|u\rangle \otimes |u\rangle \otimes \cdots \otimes |u\rangle \rightarrow |u\rangle |u\rangle \cdots |u\rangle .
\end{equation}
And then often compress it again:
\begin{equation}
|u\rangle |u\rangle \cdots |u\rangle \rightarrow |uu \cdots u\rangle .
\end{equation}
All these different ways of writing multiple states mean the same thing.  Thus, numbers represented as
$n$-qubit vectors lie in a space of dimension $2^n$, and may be written as $1 \times 2^n$ column vectors (each
of the $2^n$ slots in the column vector determined by the state of $n$-qubits), as
illustrated for $\mathbf H_2 \otimes \mathbf H_2$ above.  
We now introduce a matrix, $W_{2^n}$, that operates on these vectors.

$\mathbf {The\mbox{ }Walsh-Hadamard\mbox{ } Transformation.\mbox{ } }$ The \emph{Walsh-Hadamard transformation}, $W_{2^n}$, is defined recursively in the following way. Set
\begin{equation}
W_2 = H = \frac{1}{\sqrt 2} \left( \begin{array}{cc}
1 & 1 \\
1 & -1 \end{array} \right),
\end{equation}
\begin{equation}
W_{2^n} = \frac{1}{\sqrt {2^n}} \left( \begin{array}{cc}
W_{2^{n-1}} & W_{2^{n-1}} \\
W_{2^{n-1}} & -W_{2^{n-1}} \end{array} \right), \mbox{ for } n>1.
\end{equation}
Note that $W_4$ is
\begin{equation}
W_4 = W_2 \otimes W_2 = \frac{1}{2} \left( \begin{array}{cc}
1W_2 & 1W_2 \\
1W_2 & -1W_2 \end{array} \right) = \frac{1}{2}\left( \begin{array}{cccc}
1 & 1 & 1 & 1 \\
1 & -1 & 1 & -1 \\
1 & 1 & -1 & -1 \\
1 & -1 & -1 & 1  \end{array} \right).
\end{equation}
Thus, for example
\begin{equation}
W_{4} |uu\rangle = \frac{1}{2}\left( \begin{array}{cccc}
1 & 1 & 1 & 1 \\
1 & -1 & 1 & -1 \\
1 & 1 & -1 & -1 \\
1 & -1 & -1 & 1  \end{array} \right) \left( \begin{array}{c}
1 \\
0 \\
0 \\
0 \end{array} \right) = \frac{1}{2}\left( \begin{array}{c}
1 \\
1 \\
1 \\
1 \end{array} \right).
\end{equation}
We can rearrange the output, and see that it is a superposition of the elements of $S = \{0,1,2,3\}$:
\begin{eqnarray}
\frac{1}{2}\left( \begin{array}{c}
1 \\
1 \\
1 \\
1 \end{array} \right)
= \frac{1}{2}[\left( \begin{array}{c}
1 \\
0 \\
0 \\
0 \end{array} \right)
+ \left( \begin{array}{c}
0 \\
1 \\
0 \\
0 \end{array} \right)
+
\left( \begin{array}{c}
0 \\
0 \\
1 \\
0 \end{array} \right)
+\left( \begin{array}{c}
0 \\
0 \\
0 \\
1 \end{array} \right)]
 =  \frac{1}{2}[|00\rangle + |01\rangle + |10\rangle + |11\rangle]\\ 
 = \frac{1}{2}[|0\rangle + |1\rangle + |2\rangle + |3\rangle ] = \frac{1}{\sqrt {2^n}} \sum_{x=0}^{2^n-1} |x\rangle
\end{eqnarray}
where here $n = 2$, and we have mapped the binary numbers to their decimal equivalents.  Thus, if $|\psi\rangle =
W_4 |uu\rangle$ and we take a measurement of $|\psi\rangle$, we will find a given number $y$, $y \in S$, with probability $[\frac{1}{2}]^2 = \frac{1}{4}$.  We may take the vectors $|x\rangle$ as basis vectors
for our Hilbert space $\mathbf H_4$.  Applying $W_{2^n}$ to $n$-bits, all in state $|0\rangle$, results in an equally weighted superposition of all states (numbers) in $S = \{0,1,\cdots,2^n-1\}$:
\begin{equation}
W_{2^n} |00 \cdots 000\rangle = \frac{1}{\sqrt {2^n}} \sum_{x=0}^{2^n-1} |x\rangle .
\end{equation}

What happens if the qubits in the initial state of the quantum register are not all $|0\rangle$ (not all 
$|u\rangle$)?  Define the \emph{bit-wise inner product, or dot product, $x\cdot y$}, for $x = x_{n-1} x_{n-2} \cdots x_2 x_1 x_0$,  $y = y_{n-1} y_{n-2} \cdots y_2 y_1 y_0$, as $x\cdot y = x_{n-1} y_{n-1}+ x_{n-2} y_{n-2}+\cdots + x_2 y_2 + x_1 y_1 + x_0 y_0$  mod 2.  (In the present example, taking the result mod 2 is redundant.)
Then if the register was initially in state $|y\rangle$, the transformation is
\begin{equation}
|\psi\rangle  = W_{2^n} |y\rangle = \sum_{x=0}^{2^n-1} (-1)^{x\cdot y} |x\rangle .
\end{equation}

\begin{table}[ht]
\begin{tabular}{|r|r|r|r|}
\hline
$|y\rangle$ & $|x\rangle$ & $x\cdot y$ & $(-1)^{x \cdot y}$\\
\hline
$|110\rangle$ & $|000\rangle$ & $0$ & $1$\\
$|110\rangle$ & $|001\rangle$ & $0$ & $1$\\
$|110\rangle$ & $|010\rangle$ & $1$ & $-1$\\
$|110\rangle$ & $|011\rangle$ & $1$ & $-1$\\
$|110\rangle$ & $|100\rangle$ & $1$ & $-1$\\
$|110\rangle$ & $|101\rangle$ & $1$ & $-1$\\
$|110\rangle$ & $|110\rangle$ & $2$ & $1$\\
$|110\rangle$ & $|111\rangle$ & $2$ & $1$\\
\hline
\end{tabular}
\caption{Walsh transform with intitial qubit $|110\rangle$}
\end{table}

For example, suppose $|y\rangle$ is the 3-qubit state $|110\rangle$.  Then the bit-wise dot products and signs
are shown in Table V.  Thus we may write the output state $|\psi\rangle$ as
\begin{eqnarray}
|\psi\rangle = W_{2^n} |y\rangle = \frac{1}{\sqrt {2^3}} (|000\rangle + |001\rangle - |010\rangle - |011\rangle - |100\rangle - |101\rangle + |110\rangle + |111\rangle)\\
= \frac{1}{\sqrt {2^3}} (|0\rangle + |1\rangle - |2\rangle - |3\rangle - |4\rangle
- |5\rangle + |6\rangle + |7\rangle).
\end{eqnarray}

The transformation of qubits must be \emph{unitary}.  Recall that a matrix $U$ is unitary if its inverse is
equal to its complex conjugate transpose: $U^{-1}$ = $U^\dagger$. Thus $U^\dagger U = \mathbf 1$. (For a Hermitian matrix $M$, $M^\dagger = M$, so a Hermitian matrix is unitary provided $M^2 = \mathbf 1$.) The Pauli spin matrices, the Hadamard matrix $H$, and the Walsh matrix $W_{2^n}$ are all unitary.  A unitary transformation conserves lengths of vectors.  This
can be seen if we compare the squared length of $|\psi\rangle$ and $U|\psi\rangle$: 
\begin{eqnarray}
\langle\psi |\psi\rangle = |\psi|^2\\
\langle\psi |U^\dagger U |\psi\rangle = \langle\psi| \mathbf 1 |\psi\rangle = |\psi|^2 .
\end{eqnarray}

One more unitary transformation we will need is the following:
\begin{equation}
U_f|x\rangle|y\rangle = |x\rangle|y +_2 f(x)\rangle,
\end{equation}
where $f:\{0,1\} \rightarrow \{0,1\}$, and $+_2$ means addition modulo $2$.  Note that $U_f$ operates on two qubits
at once, $|x\rangle|y\rangle$. In this case, the $|x\rangle$ qubit is considered the \emph{control} qubit and
does not change in the operation; $|y\rangle$ is the data or \emph{target} qubit, and changes according to whether $f(x) = 0$ or $f(x)=1$.  If $f(x) = x$, then $U_f$ here is called the \emph{c-NOT} or \emph{XOR} gate,
often denoted by the negation symbol $\neg$.  It takes the control and target qubits as inputs, and replaces the target qubit with the sum of the two inputs modulo 2:
\begin{equation} 
\neg |x\rangle|y\rangle = |x\rangle|y +_2 x\rangle .
\end{equation}

Note for future reference with respect to the Grover search algorithm the effect of $U_f$ when 
$|y\rangle = |0\rangle - |1\rangle$:
\begin{equation}
U_f|x\rangle \otimes (|0\rangle - |1\rangle) = |x\rangle \otimes [(|0\rangle - |1\rangle) +_2 f(x)].
\end{equation}
For $f(x) = 0$ we have
\begin{equation}
|x\rangle \otimes [(|0\rangle - |1\rangle) +_2 f(x)] = |x\rangle \otimes [|0\rangle - |1\rangle] = |x\rangle \otimes (-1)^{f(x)}(|0\rangle - |1\rangle).
\end{equation}
For $f(x) = 1$ we have
\begin{equation}
|x\rangle \otimes [(|0\rangle - |1\rangle) +_2 f(x)] = |x\rangle \otimes [|1\rangle - |0\rangle] = |x\rangle \otimes (-1)^{f(x)}(|0\rangle - |1\rangle).
\end{equation}
So, in summary,
\begin{equation}
U_f |x\rangle \otimes (|0\rangle - |1\rangle) = |x\rangle \otimes (-1)^{f(x)}(|0\rangle - |1\rangle).
\end{equation}
Note that if we modify the definition of $f(x)$ so that it is defined on the whole domain of $S = \{0,1,2,\cdots,2^n-1\}$,
$f(x): x \in S \rightarrow \{0,1\}$, then we can use $f(x)$ as an \emph{indicator} or \emph{characteristic} function,
by letting $f(a) = 1$ for some $a \in S$ and $f(x) = 0$ for all $x \not= a$.  Denote this version of $f(x)$ as $f_a(x)$,
and the associated unitary transformation as $U_{f_a}|x\rangle|y\rangle = |x\rangle|y +_2 f_a(x)\rangle$.  Then, as before, we have
\begin{equation}
U_{f_a} |x\rangle \otimes (|0\rangle - |1\rangle) = |x\rangle \otimes (-1)^{f_a(x)}(|0\rangle - |1\rangle).
\end{equation}

$\mathbf {The\mbox{ }Grover\mbox{ }Search\mbox{ }Algorithm.\mbox{ } }$ In computer science an \emph{oracle} is a
black box subroutine into which we are not allowed to look.  An example of an oracle is our characteristic
function $f_a(x): x \in S \rightarrow \{0,1\}$.  It sets $f_a(a) = 1$ and otherwise $f_a(x) = 0,\mbox{ } x\not= a$. If
$f_a(x)$ is able to operate without our knowledge of what $a$ is, then $f_a(x)$ is an oracle. The values of $x$
may be an unsorted list---randomized telephone numbers for example (or ones which are sorted alphabetically by
the owner's names).  The objective is to find $a$ by relying on the output of $f_a(x)$.  If you had $N = 2^n$ items, the
expected number of queries to $f_a(x)$ to find $a$ with a probability of 50 percent would be $\frac{N}{2}$.  Grover, however, showed a quantum computer could find the same item with a probability close to 100 percent in about $\frac{\pi}{4}\sqrt N$ searches.

Suppose we are looking for the number $a$, where $a$ is $n$-bits.  We will want to use our indicator function $f_a(x)$ as
an oracle to help find $a$.

\emph{Initial Preparation. } First we prepare a qubit register with $n+1$ states, all of which are $|0\rangle$:
\begin{equation}
|0\rangle |0\rangle \cdots |0\rangle |0\rangle |0\rangle \otimes |0\rangle ,
\end{equation}
where the tensor product has been explicitly written out for the right-most qubit to set it off from the rest.
We apply the Walsh transform $W_{2^n}$ to the left $n$ $|0\rangle$ qubits and the simple transform $H\sigma_x$ to
the last qubit.  As we have seen before,
\begin{eqnarray}
|\psi_s\rangle = W_{2^n} |0\rangle |0\rangle \cdots |0\rangle |0\rangle |0\rangle = \frac{1}{\sqrt {2^n}} \sum_{x=0}^{2^n-1} |x\rangle\\
H\sigma_x |0\rangle = \frac{1}{\sqrt 2}(|0\rangle - |1\rangle),
\end{eqnarray}
so that the state of the entire computer becomes
\begin{equation}
|\psi_s\rangle \otimes H\sigma_x |0\rangle = \frac{1}{\sqrt {2^n}} \sum_{x=0}^{2^n-1} |x\rangle \otimes \frac{1}{\sqrt 2}(|0\rangle - |1\rangle).
\end{equation}

\emph{Step One. } We then apply our unitary transformation $U_{f_a}$
\begin{equation}
U_{f_a} |x\rangle \otimes (|0\rangle - |1\rangle) = |x\rangle \otimes (-1)^{f_a(x)}(|0\rangle - |1\rangle),
\end{equation}
to obtain 
\begin{eqnarray}
U_{f_a}(|\psi_s\rangle \otimes H\sigma_x |0\rangle) = \frac{1}{\sqrt {2^n}} \sum_{x=0}^{2^n-1} |x\rangle \otimes \frac{1}{\sqrt 2}(-1)^{f_a(x)}(|0\rangle - |1\rangle)\\
 = \frac{1}{\sqrt {2^n}}(-1)^{f_a(x)} \sum_{x=0}^{2^n-1} |x\rangle \otimes \frac{1}{\sqrt 2}(|0\rangle - |1\rangle).
\end{eqnarray}
The effect of $U_{f_a}$ is to change the sign on $|x\rangle = |a\rangle$ and to leave all the other superimposed
states unchanged.  You may ask, how did the sign $(-1)^{f_a (x)}$ get transferred from the right-most qubit in equation
(80) to the superposition of qubits in equation (81)?  The answer is that the right-most qubit is allowed to
\emph{decohere}, to interact with the environment and to `collapse' into $|0\rangle$ or $|1\rangle$.  This forces
the parameters that describe the bipartite state into the left $n$-qubit register.

\emph{Step Two. } Apply $W_{2^n}$ again to the left-most $n$ qubits. (Or apply $W_{2^n} \otimes \mathbf 1_2$ to
$n+1$ qubits, where $\mathbf 1_2$ is the $2 \times 2$ identity matrix.)

\emph{Step Three. } Let $f_0(x)$ be the indicator function for the state $|x\rangle =|0\rangle$.  Apply $-U_{f_0}$ to the
current state of the qubit register (note the negation).  This operation changes the sign on all states $|x\rangle$ except for $|x\rangle = |0\rangle$.  That is, $U_{f_0}$ maps  $|0\rangle \rightarrow -|0\rangle$, and the negation
of $U_{f_0}$, $-U_{f_0}$ restores the original sign on $|0\rangle$ , but changes the sign on all other states.

\emph{Step Four. } Apply $W_{2^n}$ again to the left-most $n$ qubits.  

Repeat Steps One to Four $\frac{\pi}{4}\sqrt N$ times.  Then sample the final state (the left-most $n$ qubits) $|\psi_f\rangle$.   With close to probability 1, $|\psi_f\rangle = |a\rangle$.  

That's the Grover search algorithm, but what does it mean?  What do Steps One, Two, Three, and Four do? Short
answer: they rotate the initial superposition $|\psi_s \rangle$ about the origin until it's as close as possible to
$|a\rangle$.  Let's see the details.

Another way to think of $U_{f_a}$, in Step One, is as the matrix $\mathbf 1 - 2 |a\rangle \langle a|$ operating on the left-most $n$ qubits.  Applying this operation to $|x\rangle$ yields
$|x\rangle$ for all basis states $|x\rangle \not= |a\rangle$ but $-|x\rangle$ for $|x\rangle = |a\rangle$.  Similarly, another way to think of  $U_{f_0}$, in Step Three, is as the matrix $\mathbf 1 - 2 |0\rangle \langle 0|$.  Applying this operation to $|x\rangle$ yields
$|x\rangle$ for all basis states $|x\rangle \not= |0\rangle$ but $-|0\rangle$ for $|x\rangle = |0\rangle$.

Step One is, geometrically, a reflection $R_a$ of $|\psi_s\rangle$ about the hyperplane orthogonal to $|a\rangle$ to
a vector $|\psi_s^R\rangle$.  Since $W_{2^n}^2 = \mathbf 1$, Steps Two to Four correspond to $-W_{2^n} U_{f_0} W_{2^n}^{-1}$.  The operation $W_{2^n} U_{f_0} W_{2^n}^{-1}$ would correspond to a further reflection of $|\psi_s^R\rangle$
about the hyperplane orthogonal to the original $|\psi_s\rangle = \frac{1}{\sqrt {2^n}} \sum_{x=0}^{2^n-1} |x\rangle$. 
However, this isn't what we want.  Instead, let $|\psi_s^\perp\rangle$ be a unit vector perpendicular to $|\psi_s\rangle$. The operation $-W_{2^n} U_{f_0} W_{2^n}^{-1}$ corresponds to a further reflection $R_s$ of
$|\psi_s^R\rangle$ about the hyperplane orthogonal to $|\psi_s^\perp \rangle$. Call this furtherly reflected vector
$|\psi_s^{'}\rangle$. The net effect is a rotation $R_s R_a = -W_{2^n} U_{f_0} W_{2^n}^{-1} U_{f_a}$ of $|\psi_s\rangle
\rightarrow |\psi_s^{'}\rangle$ in the plane spanned by $|\psi_s\rangle$ and $|a\rangle$. (By the plane spanned by $|\psi_s\rangle$ and $|a\rangle$ we mean all states of the form $c|\psi_s\rangle + d|a\rangle$, where $c, d \in \mathbf C$.) 

To summarize: Let $\theta$ be the angle between $|\psi_s\rangle$ and the unit vector orthogonal to $|a\rangle$, the latter designated $|a^{\perp}\rangle$. For simplicity we assume a counter-clockwise ordering $|a^{\perp}\rangle$, $|\psi_s\rangle$, $|a\rangle$.  Then the combination $R_s R_a$ is a counter-clockwise rotation of $|\psi_s\rangle$ by $2\theta$, so that the angle between  $|a^{\perp}\rangle$ and $|\psi_s\rangle$ is now $3\theta$.  That is, $R_s R_a$ moves $|\psi_s\rangle$ \emph{away}
from $|a^{\perp}\rangle$, the vector orthogonal to $|a\rangle$, and hence moves $|\psi_s\rangle$ \emph{toward} $|a\rangle$ itself by the angle $2\theta$.

The whole idea of the Grover search algorithm is to rotate the state $|\psi_s\rangle$ about the origin, in the plane spanned by $|\psi_s\rangle$ and $|a\rangle$,
until $|\psi_s\rangle$ is as close as possible to $|a\rangle$. Then a measurement of $|\psi_s\rangle$ will yield
$|a\rangle$ with high probability.
 
How much do we rotate (how many times do we apply $R_s R_a$)?  We don't want to overshoot or undershoot by rotating
too much or too little.  We want to rotate $|\psi_s\rangle$ around to $|a\rangle$ and then stop.  Consider the
vector or state $|\psi_s\rangle$ lying initially in the plane formed by $|a^{\perp}\rangle$ and $|a\rangle$, with
the angle between $|\psi_s\rangle$ and $|a^{\perp}\rangle$ equal to $\theta$.  That means we can write $|\psi_s\rangle$ as the
initial superposition 
\begin{equation}
|\psi_s\rangle = cos \theta |a^{\perp}\rangle + sin \theta |a\rangle .
\end{equation}
After $k$ applications  of $R_s R_a = -W_{2^n} U_{f_0} W_{2^n}^{-1} U_{f_a}$, the state is
\begin{equation}
(R_s R_a)^k |\psi_s\rangle = cos (2k+1)\theta |a^{\perp}\rangle + sin (2k+1)\theta |a\rangle .
\end{equation}
Note that if $(2k+1) \theta = \frac{\pi}{2}$, then $cos (2k+1) \theta = 0$, $sin (2k+1) \theta =1$, so that
\begin{equation}
(R_s R_a)^k |\psi_s\rangle = |a\rangle .
\end{equation}
Now this may not be achievable, because $k$ must be a whole number, but let's solve for the closest integer, 
where $[\cdot]_{nint}$ denotes nearest integer:
\begin{equation}
k = [\frac{\pi}{4 \theta} - \frac{1}{2}]_{nint} .
\end{equation}
Remember that the inner product of two unit vectors gives the cosine of the angle between them, and that the 
\emph{initial} angle between
$|a\rangle$ and $|\psi_s\rangle$ is $\frac{\pi}{2} - \theta$.  Therefore
\begin{equation}
\langle a|\psi_s\rangle = \frac{1}{\sqrt {2^n}} = cos (\frac{\pi}{2} - \theta) = sin (\theta).
\end{equation}
For $N = 2^n$ large, we can set $sin\mbox{ }\theta \approx \theta$.  Thus, substituting $\frac{1}{\sqrt N} = \theta$ into
our equation for $k$, we obtain 
\begin{equation}
k = [\frac{\pi}{4} \sqrt N - \frac{1}{2}]_{nint} .
\end{equation}
This value of $k$, then, obtains $(R_s R_a)^k |\psi_s\rangle = |a\rangle$ with probability close to $1$. 

$\mathbf {Grover\mbox{ }search\mbox{ }example. \mbox{ }}$ Here is an example of Grover search for $n = 3$ qubits, where $N = 2^n = 8$. (We omit reference to qubit $n+1$, which is in state $\frac{1}{\sqrt 2}(|0\rangle - |1\rangle)$ and
does not change. The dimension of the unitary operators for this example is thus $2^n = 8$ also.)  Suppose the unknown number is $|a\rangle = |5\rangle$. The matrix or black box oracle $U_{f_a}$ is then 
\begin{equation}
U_{f_5} =  \left( \begin{array}{cccccccc}
1 & 0 & 0 & 0 & 0 & 0 & 0 & 0\\
0 & 1 & 0 & 0 & 0 & 0 & 0 & 0\\
0 & 0 & 1 & 0 & 0 & 0 & 0 & 0\\
0 & 0 & 0 & 1 & 0 & 0 & 0 & 0\\
0 & 0 & 0 & 0 & 1 & 0 & 0 & 0\\
0 & 0 & 0 & 0 & 0 & -1 & 0 & 0\\
0 & 0 & 0 & 0 & 0 & 0 & 1 & 0\\
0 & 0 & 0 & 0 & 0 & 0 & 0 & 1   \end{array} \right).
\end{equation}
(Remember that numbering starts with 0 and ends with 7, so that the -1 here is in the slot for $|5\rangle$.)
This matrix reverses the sign on state $|5\rangle$, and leaves the other states unchanged.  The Walsh matrix
$W_8$ is
\begin{equation}
W_8 = \frac{1}{\sqrt {2^3}} \left( \begin{array}{cccccccc}
1 & 1 & 1 & 1 & 1 & 1 & 1 & 1\\
1 & -1 & 1 & -1 & 1 & -1 & 1 & -1\\
1 & 1 & -1 & -1 & 1 & 1 & -1 & -1\\
1 & -1 & -1 & 1 & 1 & -1 & -1 & 1\\
1 & 1 & 1 & 1 & -1 & -1 & -1 & -1\\
1 & -1 & 1 & -1 & -1 & 1 & -1 & 1\\
1 & 1 & -1 & -1 & -1 & -1 & 1 & 1\\
1 & -1 & -1 & 1 & -1 & 1 & 1 & -1   \end{array} \right).
\end{equation}
The matrix $-U_{f_0}$ is
\begin{equation}
-U_{f_0} =  \left( \begin{array}{cccccccc}
1 & 0 & 0 & 0 & 0 & 0 & 0 & 0\\
0 & -1 & 0 & 0 & 0 & 0 & 0 & 0\\
0 & 0 & -1 & 0 & 0 & 0 & 0 & 0\\
0 & 0 & 0 & -1 & 0 & 0 & 0 & 0\\
0 & 0 & 0 & 0 & -1 & 0 & 0 & 0\\
0 & 0 & 0 & 0 & 0 & -1 & 0 & 0\\
0 & 0 & 0 & 0 & 0 & 0 & -1 & 0\\
0 & 0 & 0 & 0 & 0 & 0 & 0 & -1   \end{array} \right).
\end{equation}
This matrix changes the sign on all states except $|0\rangle$. Finally, we have the repeated step $R_s R_a$ in the Grover algorithm:
\begin{equation}
R_s R_5 = -W_{8} U_{f_0} W_{8}^{-1} U_{f_5} = \frac{1}{4} \left( \begin{array}{cccccccc}
-3 & 1 & 1 & 1 & 1 & -1 & 1 & 1\\
1 & -3 & 1 & 1 & 1 & -1 & 1 & 1\\
1 & 1 & -3 & 1 & 1 & -1 & 1 & 1\\
1 & 1 & 1 & -3 & 1 & -1 & 1 & 1\\
1 & 1 & 1 & 1 & -3 & -1 & 1 & 1\\
1 & 1 & 1 & 1 &  1 & 3  & 1 & 1\\
1 & 1 & 1 & 1 &  1 & -1 & -3 & 1\\
1 & 1 & 1 & 1 &  1 & -1 & 1 & -3   \end{array} \right).
\end{equation}

The \emph{initial preparation}  is
\begin{equation}
W_8 |0\rangle|0\rangle|0\rangle = \frac{1}{\sqrt {2^3}} \left(\begin{array}{c}
1 \\
1 \\
1 \\
1 \\
1 \\
1 \\
1 \\
1 \end{array} \right) .
\end{equation}
Since $N = 2^3 = 8$ we calculate the number of rotations $k$ as the nearest integer:
\begin{equation}
k = [\frac{\pi}{4} \sqrt 8 -\frac{1}{2}]_{nint} = 2 .
\end{equation}
Thus, after the first rotation, the state becomes
\begin{equation}
R_s R_5 W_8 |0\rangle|0\rangle|0\rangle = \frac{1}{4 \sqrt 2} \left(\begin{array}{c}
1 \\
1 \\
1 \\
1 \\
1 \\
5 \\
1 \\
1 \end{array} \right)
\end{equation}
and, after the second rotation,
\begin{equation}
(R_s R_5)^2 W_8 |0\rangle|0\rangle|0\rangle = \frac{1}{8 \sqrt 2} \left(\begin{array}{c}
-1 \\
-1 \\
-1 \\
-1 \\
-1 \\
11 \\
-1 \\
-1 \end{array} \right) .
\end{equation}
Note that the amplitude for $|5\rangle$ is now $\frac{11}{8 \sqrt 2}$.  A measurement of $(R_s R_5)^2 W_8 |0\rangle|0\rangle|0\rangle$ will thus yield $|5\rangle$ with probability $(\frac{11}{8 \sqrt 2})^2 = .9453$.

$\mathbf {The\mbox{ }guess\mbox{ }a\mbox{ }number\mbox{ }game\mbox{ }I.\mbox{ }}$  Bob challenges Alice to the following
game.  Alice is to chose a number $a$ from $S = \{0, 1,\cdots,N-1\}$, and he is to attempt to guess it, with a certain number of tries $k$. Alice acts as the oracle $U_{f_a}$ after each of Bob's turns. They agree on $N = 2^{30} = 1,073,741,824$.  Alice knows that, classically, Bob will require $\frac{N}{2} = 2^{29} = 536,870,912$ tries to
guess the number with a probability of 50 percent, so she agrees with Bob to allow up to $k = 100,000,000$, believing
that the advantage is all hers.  Bob, however, intends to use the Grover search algorithm, and never intends
to guess more than $k = [\frac{\pi}{4}\sqrt {2^{30}}-\frac{1}{2}]_{nint} = 25,735$ times.

Bob initially sets up $N+1$ qubits as
\begin{equation}
|\psi_s\rangle \otimes H\sigma_x |0\rangle = \frac{1}{\sqrt {2^n}} \sum_{x=0}^{2^n-1} |x\rangle \otimes \frac{1}{\sqrt 2}(|0\rangle - |1\rangle),
\end{equation}
as in equation (78).  He presents the left-most $n$ qubits, $|\psi_s\rangle$, to Alice.  This is
followed by Alice's move of $R_a$, followed by Bob's play of $R_s$, and so on, until
after $k$ moves the state of the $n$-qubit system is:
\begin{equation}
(R_s R_a)^k |\psi_s\rangle = cos (2k+1)\theta |a^{\perp}\rangle + sin (2k+1)\theta |a\rangle .
\end{equation}
The system is then measured and Bob wins with a probability of $|sin (2k+1)\theta|^2$.  To Alice's surprise she
finds that Bob wins repeatedly, despite playing only a small number of his allowed moves. (Bob's probability of
winning is $p \ge 1 - \frac{1}{N}$.)  After a number of games
she realizes Bob always plays the same number of moves $k = 25,735$.  She becomes suspicious that there is
some conspiracy afoot.  

$\mathbf {The\mbox{ }Bernstein-Vazirani\mbox{ }oracle.\mbox{ }}$  Previously we defined the bitwise
inner product $x\cdot y$.  Let's substitute for $y$ a constant vector $a$ of 0s and 1s, and let 
$f_{bv}^a : \{0,1\}^n \rightarrow \{0,1\}$ be defined as
\begin{equation}
f_{bv}^a (x,a) = x\cdot a
\end{equation}
with an associated transform 
\begin{equation}
T_{bv}^a |x\rangle = (-1)^{f_{bv}^a} |x\rangle = (-1)^{x\cdot a}|x\rangle .
\end{equation}
This is the Bernstein-Vazirani oracle.  How many measurements of $f_{bv}^a (x,a)$ would be required to find $a$?
Classically you would have to perform measurements for all possible values of $x$, and then solve a set
of linear equations for $a$.  But quantum mechanically solving for $a$ only takes one step.

To see why, refer back to equation (63) and the calculation in Table V for the Walsh transform of an initial state $|y\rangle \ne |0\rangle$.  Now compare the effect of the transform $T_{bv}^a$ on an equal superposition of all states:
\begin{equation}
T_{bv}^a |\psi_s\rangle = \frac{1}{\sqrt {2^n}} \sum_{x=0}^{2^n-1} T_{bv}^a |x\rangle
= \frac{1}{\sqrt {2^n}} \sum_{x=0}^{2^n-1} (-1)^{x\cdot a}|x\rangle .
\end{equation}
This is just the Walsh transform of an initial state $|a\rangle$!  Therefore we can find $|a\rangle$ with another
application of the Walsh transform (which is its own inverse):
\begin{equation}
W_{2^n} T_{bv}^a |\psi_s\rangle = |a\rangle .
\end{equation}

$\mathbf {The\mbox{ }guess\mbox{ }a\mbox{ }number\mbox{ }game\mbox{ }II.\mbox{ }}$  Alice says to Bob,
you are getting too many guesses.  Either change the game or I won't play anymore.  Bob says:  I don't
know why you are complaining.  I'm only making a tiny fraction of the number of guesses we agreed on.  But 
I'll tell you what.  I will make only \emph{two} guesses--a preliminary guess, you will give me some feedback 
information, and then I will make a second and final guess of the number.  The feedback I need is $T_{bv}^a$ 
applied as an oracle to my initial guess. (Of course Bob plans to submit $|\psi_s \rangle$ as his initial
guess.)  

Alice agrees, and the game proceeds as follows:

Bob: $prepares\mbox{ } |\psi_s\rangle = W_{2^n} |0 \cdots 00\rangle = \frac{1}{\sqrt {2^n}} \sum_{x=0}^{2^n-1} |x\rangle$

Alice: $ T_{bv}^a |\psi_s\rangle = \frac{1}{\sqrt {2^n}} \sum_{x=0}^{2^n-1} (-1)^{x\cdot a}|x\rangle $

Bob: $ W_{2^n} T_{bv}^a |\psi_s\rangle = |a\rangle$ .

Bob wins.  Again, the key feature was the ability to present a superposition of states to Alice's oracle.

\subsection{Shor's factoring algorithm}

Shor's algorithm is a key result in quantum computation, so we want to look at it in some modest detail.  It will
form the basis of the RSA game. We will need as preliminaries Euler's theorem and the quantum Fourier transform $F$.

\begin{sloppypar}
$\mathbf {Euler's\mbox{ }theorem. }$  Let $N$ be an integer, and let $a$ be an integer less than $N$ and
relatively prime to $N$.  Euler's theorem  \cite[chap. 12]{OO} says that
\begin{equation}
a^\phi = 1\mbox{ } mod\mbox{ } N .
\end{equation}
Here $\phi$ is Euler's totient function, and is the total number of integers less than $N$ that are relatively
prime to $N$.  Example:  Let N = 77.  In this case $\phi = 60$, so $23^{60} = 1$ mod $77$, $39^{60} = 1$ mod $77$, etc.
Euler's theorem implies that the powers of any number relatively prime to $N$ cycle mod $N$:
\begin{equation}
a, a^2, a^3, \cdots, a^{\phi -1}, a^\phi = 1, a, a^2, a^3, \cdots  .
\end{equation}
Thus $\phi$ is the maximum length of a cycle or period. Of course, for a given $a$, there may be a smaller $s<\phi$ such that
$a^s = 1$ mod $N$.  But in that case it is clear $s$ divides $\phi$. The smallest value of $s$ such that
$a^s$ = 1 mod N is called the \emph{order} of $a$, which in the Shor algorithm below we denote by $r$. 
Given knowledge of $\phi$, or any $s$ or $r$ for a given $a$, we can factor $N$.  Since
$a^\phi = 1$ mod $N$, we have, for even $\phi$, $(a^{\frac{\phi}{2}}+1)(a^{\frac{\phi}{2}}-1) = 0$ mod $N$. Let $gcd(x,y)$ 
denote the greatest common divisor of $x$ and $y$. We then check $gcd(N,a^{\frac{\phi}{2}}+1)$ and $gcd(N,a^{\frac{\phi}{2}}-1)$ for a factor. If we don't get a factor, we divide $\phi$ again by two (if the previous division left an even exponent), or else try another value for $a$. 
Example: Let $N = 77$, and $a = 2$.  We find that $2^{60} = 1$ mod $77$, and upon division of $\phi$ by $2$, also
$2^{30} = 1$ mod $77$.  Hence we look at $2^{15}$ mod $N = 43$.  We find that $gcd(77,44) = 11$ and $gcd(77,42) = 7$.  These are the two factors of 77. Obviously, this is not the best way to factor a number, normally, but it is
ideally suited for a quantum algorithm.
\end{sloppypar}

$\mathbf {Quantum\mbox{ }Fourier\mbox{ }transform. }$ The quantum Fourier transform looks a lot like the discrete
Fourier transform.  For a given state $|y\rangle$ the quantum Fourier transform is the unitary transformation
\begin{equation}
F |y\rangle = \frac{1}{\sqrt {2^n}} \sum_{x=0}^{2^n-1} e^{2\pi i xy/{2^n}} |x\rangle .
\end{equation}
In this definition, the term $xy$ denotes ordinary multiplication.  It is \emph{not} the bitwise dot product
$x\cdot y$.  Rather, if $|x\rangle = 7$ and $|y\rangle = 6$, then $xy = 42$. (By contrast, the dot product is
$x\cdot y = 7\cdot 6$ mod $2 = 111 \cdot 110$ mod $2 = 2$ mod $2 = 0$.)
$F |y\rangle$ is periodic in $xy$ with period $2^n$. The Hadamard matrix $H$ we saw previously is simply
the Fourier transform for $n = 1$.  To see this, let x, y each be 0 or 1 in the term
\begin{equation}
\frac{1}{\sqrt {2^n}} e^{2\pi i xy/{2^n}}
\end{equation}
where $n = 1$. We obtain the matrix
\begin{equation}
\frac{1}{\sqrt 2}
\left( \begin{array}{cc}
e^0 & e^0 \\
e^0 & e^{\pi i} \end{array} \right) =
\frac{1}{\sqrt 2}
\left( \begin{array}{cc}
1 & 1 \\
1 & -1 \end{array} \right),
\end{equation}
remembering that $e^{\pi i} = cos(\pi)+ i\mbox{ }sin(\pi) = -1 + 0 = -1$.

The inverse quantum Fourier transform $F^{-1}$ simply reverses the sign on $i$:
\begin{equation}
F^{-1} |y\rangle = \frac{1}{\sqrt {2^n}} \sum_{x=0}^{2^n-1} e^{-2\pi i xy/{2^n}} |x\rangle .
\end{equation}

$\mathbf {Shor's\mbox{ }factoring\mbox{ }algorithm.\mbox{ }}$We want to find a factor of a number $N$,
where $2^{2n-2}<N^2<2^{2n}$.
Shor's factoring algorithm on a quantum computer runs in $O((log\mbox{ } N)^3)$ steps. We need a quantum
computer with two registers (which we shall refer to simply as left and right). The left register contains
$2n$ qubits, and the right register contains $log_{_2} N$ qubits. The values of the qubits
in both registers are initialized to $|0\rangle$:
\begin{equation}
|00 \cdots 0\rangle \otimes |00\cdots 0\rangle .
\end{equation}

\emph{Step 1:} Chose $m$, $2 \le m \le N-2$.  If $gcd(m,N) \ge 2$, we have found a proper factor of $N$.  Otherwise
proceed as follows, in Steps 2-5.

\emph{Step 2:} Do a Walsh transform $W_{2^{2n}}$ of the qubits in the left register to create a superposition of all states in the left register:
\begin{equation}
(W_{2^{2n}} \otimes \mathbf 1_{{log}_2 N}) (|00 \cdots 0\rangle \otimes |00 \cdots 0\rangle) =
|\psi_s\rangle \otimes |00 \cdots 0\rangle 
= \frac{1}{\sqrt {2^{2n}}} \sum_{x=0}^{2^{2n}-1} |x\rangle \otimes |00 \cdots 0\rangle .
\end{equation}

\emph{Step 3:} Apply the transform $f_m(|x\rangle \otimes |00 \cdots 0\rangle) \rightarrow |x\rangle \otimes |m^x
\mbox{ mod }N\rangle$:
\begin{equation}
 f_m(|\psi_s\rangle \otimes |00 \cdots 0\rangle) 
= \frac{1}{\sqrt {2^{2n}}} \sum_{x=0}^{2^{2n}-1} |x\rangle \otimes |m^x\mbox{ mod }N\rangle .
\end{equation}
Note that at this point, if we measured the right register, or allowed it to decohere, it would collapse into
a given value of $m^x$ mod $N$, such as $Z = m^z$ mod $N$.  Hence, in the left register, all amplitudes of states
would go to zero, except for those states $x$ such that $m^x$ mod $N = Z$.  If, for example, the order of $m$
was $5$, then the amplitudes of states would read something like:
\begin{equation}
\cdots, 0, 0, 0, c, 0, 0, 0, 0, c, 0, 0, 0, 0, c, 0, 0, 0, 0, c, 0, 0 \cdots 
\end{equation}
The amplitude would be non-zero on every $5th$ value.
The states were previously in an equal superposition with amplitude $\frac{1}{\sqrt {2^{2n}}}$, but the
surviving values would now have amplitude approximately $c = \frac{1}{\sqrt {\frac{2^{2n}}{5}}}$.
This is the idea, although (following Shor), we don't actually observe the right register at this point.
Instead we proceed to Step 4.

\emph{Step 4:} Do a quantum Fourier transform $F$ on the qubits in the left register:
\begin{equation}
(F\otimes \mathbf 1)(f_m(|\psi_s\rangle \otimes |00 \cdots 0\rangle) 
= \frac{1}{2^{2n}} \sum_{x=0}^{2^{2n}-1} \sum_{y=0}^{2^{2n}-1} e^{2\pi i xy/{2^{2n}}} |y\rangle \otimes |m^x\mbox{ mod }N\rangle .
\end{equation}

\emph{Step 5:} Observe the system registers.  This will give some concrete value of $w$ for $y$ and
$m^z$ mod $N$ for $m^x$ mod $N$:
\begin{equation}
(F\otimes \mathbf 1)(f_m(|\psi_s\rangle \otimes |00 \cdots 0\rangle) \rightarrow |w,m^z\mbox{ mod }N\rangle
\end{equation}
with probability equal to the square of the associated amplitude:
\begin{equation}
| \frac{1}{2^{2n}} \sum_{x: m^x = m^z \mbox{ mod } N} e^{2\pi i xw/{2^{2n}}}|^2.
\end{equation}
Thus with high probability, the observed $w$ will be near an integer multiple of $\frac{2^{2n}}{r}$.
This ends the quantum part of the calculation.  We now use the result to determine the period $r$.

First find the fraction that best approximates $\frac{w}{2^{2n}}$ with denominator $r' < N < 2^n$:
\begin{equation}
|\frac{w}{2^{2n}} - \frac{d'}{r'}| < \frac{1}{2^{2n+1}}.
\end{equation}
This may be done using continued fractions (see   \cite[chapter 12]{HW}).

Second try $r'$ in the role of $r$.  If $m^{r'} = 1$ mod $N$, we have, for even $r'$, $(m^{\frac{r'}{2}}-1)(m^{\frac{r'}{2}}+1) = 0\mbox{ mod } N$.
We then check $gcd(N,m^{\frac{r'}{2}}-1)$ and $gcd(N,m^{\frac{r'}{2}}+1)$ for a factor of N.  In the event $r'$ is
odd, or if $r'$ is even and we don't obtain a factor, we repeat the steps $O($log log $N)$ times using the same
value for $m$.  If that doesn't work, we change $m$ and start over. 

\subsection{The RSA game}
RSA is an encryption system widely used in banking and elsewhere.  Consider the ring of integers $Z_N$, where $N = pq$ for two distinct large primes $p$ and $q$. For encryption, RSA allows only the \emph{units} of $Z_N$ (i.e., eliminate all multiples of $p$ or $q$ from $Z_N$). The remaining set of integers, called $Z_N^*$, is an abelian group under multiplication, with order (Euler's totient function) $\phi = (p-1)(q-1) = (n+1)-(p+q)$. The RSA crypto system choses a relatively small odd integer $e$, and calculates $d = e^{-1}$ mod $\phi$. A message $M$ in $Z_n^*$ is then encrypted as $M^e$ mod $N$, and decrypted as $M^{ed} = M^{\phi+1} = M$ mod $N$.  The numbers $e$ and $N$ are publicly known, while
the decryption key $d$ is known only to the message recipient.

Alice challenges Bob to the following game.  She will create a public key $N$ and $e$, and encrypt a message
$M$.  The three components $(N,e,M^e)$ will be sent to Bob.  If Bob can decrypt the message, $M^e \rightarrow M$,
within (log $N)^3$ steps, Bob wins \$1,000.  Else he loses \$1,000.

Now RSA uses very large numbers $N$.  But we are going to use an extremely simple example in order to illustrate
the steps in Shor's algorithm. We assume that Alice sends Bob the triplet $(77,11,67)$. We first note that $77^2 = 5929$, and $2^{12} < 5929 < 2^{14}$.  The left quantum register will need 14 qubits, while the right register will require 7 qubits.  

\emph{Step 1:} Bob randomly chooses $m = 39$, where $2 \le 39 \le 75$.  The $gcd(39,77) = 1$, so Bob proceeds
to Step 2.

\emph{Step 2:} In the left qubit register, Bob creates a superposition of all numbers from 0 to $16383 =
2^{14} -1$.

\emph{Step 3:}  Bob applies the transform $f_m$ which associates to each $x$ in the superposition, the
value $39^x$ mod $77$.  Since $39^{30}$ mod $77 = 1$, we have $m^x = 1$ mod $77$, for $x \in S = \{30, 60,
90, 120, 150,\cdots,16380\}$.  That is $m = 39$ has period $r = 30$.  But Bob doesn't know this yet.

\emph{Step 4:}  Bob does a quantum Fourier transform on the left register, which contains the values
of $x$.  He then observes both registers and gets $w = 14,770$ for the left register state, and $Z = 53$
for the value of $39^z$ mod $77$ in the right register.

Bob now wants to find the fraction that best approximates $\frac{14770}{16384}$ with denominator less than
$77$.  This fraction is very close to $\frac{27}{30}$, so Bob tries $r' = 30$, or $\frac{r'}{2} = 15$.  He
gets $39^{15}-1$ mod $77 = 42$, $39^{15}+1$ mod $77= 44$, and $gcd(77,42) = 7$, $gcd(77,44) = 11$.  With these 
two factors in
hand, Bob calculates $\phi = (7-1)(11-1) = 60$.  Therefore for the decryption key $d$, he wants $d = e^{-1}$ mod
$60$, which gives $d = 11^{-1}$ mod $60 = 11$.  The decryption key is the same as the encryption key.  (This
is only a result of the trivially small modulus $N = 77$ we used.)  Bob now decrypts Alice's encrypted
message $(M^{e})^{d} = 67^{11}$ mod $77 = 23$.  Bob tells Alice the message $M = 23$ and collects his \$1,000.

\subsection{Nash equilibrium and prisoner's dilemma} 

We want to look at $2 \times 2$ games that are not zero sum, and the traditional game theoretic concept
of Nash equilibrium, and to extend it to quantum games.  Both Alice and Bob may gain
from a game, but may or may not do as well as some obtainable maximum.  We assume both try to maximize
utility, or \emph{expected utility} with mixed strategies or uncertain outcomes, and that utility can be 
assigned a cardinal number  \cite{PF}.

Non-zero sum games are traditionally presented in static form.  A matrix of payoffs corresponding to
moves is given, and some notion of \emph{equilibrium} is presented, without explaining how the players
got to that point.  But once they get there, they are expected to stay.  That's because they have
a \emph{dominant strategy} that indicates they are better off playing the corresponding move.

Let $s_A^i \in S_A$ be moves (including convex combinations of simple moves, if appropriate) available to Alice, and $s_B^j \in S_B$ be moves available to Bob.  Then a \emph{dominant strategy} for Alice is a move $s_A$ such that the payoff $\pi_A$ to Alice has the property
\begin{equation}
\pi_A(s_A,s_B^j) \geq \pi_A(s_A^i,s_B^j)
\end{equation}
for all $s_A^i \in S_A$, $s_B^j \in S_B$, provided such a move exists.  For an example, consider
Table VI.  Alice and Bob each have two possible moves, labeled C (cooperate) or D (defect).  The
values in parenthesis represent the payoffs $\pi$; the first number is the payoff to Alice, the second
number is the payoff to Bob.  Clearly for Alice $s_A = D$, because if Bob plays $C$, $\pi_A(D,C) = 5 > 3$,
\begin{table}[ht]
\begin{tabular}{|r|r|r|}
\hline
\ & Bob C & Bob D\\
\hline
Alice C & (3,3) & (0,5)\\
Alice D & (5,0) & (1,1)\\
\hline
\end{tabular}
\caption{Prisoner's Dilemma}
\end{table}
while if Bob plays $D$, $\pi_A(D,D) = 1 > 0$.  For similar reasons, $s_B = D$ also, so the game will be
in \emph{equilibrium} with $\{s_A,s_B\} = \{D,D\}$ and $\{\pi(s_A),\pi(s_B)\} = \{1,1\}$.  This outcome
is referred to as \emph{Prisoner's Dilemma} because clearly Bob and Alice would each be better off if
both played C, which would yield $\pi_A = \pi_B = 3$.

A \emph{Nash equilibrium} is a combination of moves $\{s_A,s_B\}$ such that neither party can increase his
or her payoff by unilaterally departing from the given equilibrium point:
\begin{eqnarray}
\pi_{A}(s_A,s_B) \geq \pi_{A}(s_A^i,s_B),\\
\pi_{B}(s_A,s_B) \geq \pi_{B}(s_A,s_B^j).
\end{eqnarray}
In Table VI, $\{D,D\}$, yielding payoffs $\{1,1\}$ is a Nash equilibrium, because if Alice switches to C, her
payoff goes from 1 to 0, and similarly for Bob.

A payoff point $\{\pi_A,\pi_B\}$ is \emph{jointly dominated} by a different point $\{\pi_A^*,\pi_B^*\}$ if
$\pi_A^* \ge \pi_A$ and $\pi_B^* \ge \pi_B$, and one of the inequalities is strict.  In Table VI, the point $\{1,1\}$ is jointly dominated by $\{3,3\}$. A pair of payoffs $\{\pi_A,\pi_B\}$ is \emph{Pareto optimal} if it is not jointly dominated by another point, and if neither party can increase his or her payoff without decreasing the payoff to the other party.  In Table VI, the point $\{3,3\}$ is Pareto optimal, because unilateral departure from it by either Alice or Bob decreases the payoff to the other party.  What about $\{1,1\}$?  Here, too, neither party can increase their payoff without decreasing the payoff to the other party (indeed, neither can unilaterally increase his payoff at all).  However, $\{1,1\}$ is jointly dominated by $\{3,3\}$, so it is not Pareto optimal. 

An \emph{evolutionarily stable strategy} (ESS) is a more restrictive notion than Nash equilibrium.  (That is,
strategies that are evolutionarily stable form a subset of Nash equilibria.)  Strategy $s_i$ is \emph{evolutionarily stable} against $s_j$ if $s_i$ performs better than $s_j$ against $s_i + (1-\eta) s_j$ for sufficiently small $\eta$.  
The notion is that of a population playing $s_i$ that is invaded by mutants playing $s_j$.  An ESS is then defined
as a strategy that is evolutionarily stable against all other strategies.  Note that an ESS holds for $\eta$
sufficiently small, say $\eta \in [0,\eta_0)$.  The value $\eta_0$ is called the \emph{invasion barrier}.  For
values of $\eta > \eta_0$, $s_i$ no longer performs better than $s_j$ against the combination, so members
of the population will switch to $s_j$.  We will return to this concept in the \emph{evolutionarily stable
strategy game} considered later. 

\subsection{Escaping prisoner's dilemma in a quantum game}
We now have enough background to tentatively define a quantum game. A \emph{quantum game $\Gamma$} is an
interaction between two or more players with the following elements: $\Gamma = \Gamma(\mathbf H, \Lambda,
\{s_i\}_j, \{\pi_i\}_j)$.  $\mathbf H$ is a Hilbert space, $\Lambda$ represents the initial state of the
game, $\{s_i\}_j$ is the set of moves of player $j$, while $\{\pi_i\}_j$ is a set of payoffs to player
$j$.  The object of the game is that of endogenously determining the strategies that maximize the payoffs 
to player $j$.  In the course of doing so, we may or may not determine an equilibrium to the game, and the
value $\overline{\pi}_j$ of the game to player $j$. 

We want, at this point, to give an introduction to the \emph{quantum} version of Prisoner's Dilemma, even
though final details will be deferred until later.
In the quantum version of prisoner's dilemma \cite{EWL}, each of Alice and Bob possesses a qubit and is able
to perform manipulations on his/her own qubit.  Each qubit lies in $\mathbf H_2$ which has as basis vectors 
$|C\rangle$ and $|D\rangle$, and the game lies in $\mathbf H_2 \otimes \mathbf H_2$ with basis vectors $|CC\rangle$,
$|CD\rangle$, $|DC\rangle$, and $|DD\rangle$.  Alice's qubit is the left-most qubit in each pair, while Bob's 
is the right-most.  The game is a simple quantum network.

The initial state $\Lambda$ of the game is
\begin{equation}
\Lambda = U|CC\rangle ,
\end{equation}
where $U$ is a unitary operator, known both to Alice and Bob, that operates on both qubits.
Alice and Bob have as strategic moves $s_A$, $s_B$,
\begin{eqnarray}
s_A = U_A\\
s_B = U_B
\end{eqnarray}
where $U_A$ and $U_B$ are unitary matrices that operate only on the respective player's qubit.  After
Alice and Bob have made their moves, the state of the game is
\begin{equation}
(U_A \otimes U_B) U |CC\rangle .
\end{equation}
Alice and Bob forward their qubits for final measurement.  The inverse of the unitary operator $U$ 
is now applied, to bring the game to the state:
\begin{equation}
U^\dagger (U_A \otimes U_B) U |CC\rangle .
\end{equation}
The measurement is then taken, and yields one of the four basis vectors of $\mathbf H_2 \otimes \mathbf H_2$. 
The associated payoff values to Alice and Bob are those previously given in Table VI.

How Alice and Bob escape prisoner's dilemma in this quantum game by selection of their respective
unitary matrices $U_A$, $U_B$ depends on their playing \emph{entanglement}-related strategies.  Therefore we 
will defer further discussion of the quantum prisoner's dilemma game until we have considered entanglement
in the next section. However, we wanted to make the point that \emph{a pure quantum strategy is
a unitary operator acting on the player's qubit}.

\subsection{Entanglement}

We have been considering vectors $|\psi\rangle$ in a Hilbert space $\mathbf H$.  The vector or state $|\psi\rangle$ is \emph{entangled} if it does not factor relative to a given tensor product decomposition of the Hilbert space, $\mathbf H = \mathbf H_1 \otimes \mathbf H_2$. For example, the state $|\psi_1\rangle = a|00\rangle + b |01\rangle$ can be decomposed into a tensor
product
\begin{equation}
|\psi_1\rangle = a|00\rangle + b |01\rangle = |0\rangle \otimes (a|0\rangle + b |1\rangle),
\end{equation}
so it is not entangled.  On the other hand, the state $|\psi_2\rangle = a|00\rangle + b |11\rangle$ cannot be decomposed into a tensor product, and is therefore entangled.  Entangled states act as a single whole without reference
to space or time.  Any operation performed on one entangled qubit instantly affects the states of the qubits
with which it is entangled.  Entanglement generates `spooky action at a distance'.

Instead of the orthonormal computational basis we have been using for Hilbert space, sometimes a different
orthonormal basis, called the \emph{Bell basis}, is used.  The Bell basis is a set of maximally entangled
states.  For two-qubits in $\mathbf H_4$, we can denote this entangled basis as
\begin{eqnarray}
|b_0\rangle = \frac{1}{\sqrt 2}(|00\rangle + |11\rangle)\\
|b_1\rangle = \frac{1}{\sqrt 2}(|01\rangle + |10\rangle)\\
|b_2\rangle = \frac{1}{\sqrt 2}(|00\rangle - |11\rangle)\\
|b_3\rangle = \frac{1}{\sqrt 2}(|01\rangle - |10\rangle) .
\end{eqnarray}
It is easy to transform the computational basis into the Bell basis by using a combination of a Hadamard
transformation $H$ and a c-NOT gate.
First apply the Hadamard transform to the left-most qubit. Then apply c-NOT (review equation 69) with the left qubit as the source and the right qubit as the target. Shorthand for this transformation is $ \neg (H \otimes \mathbf 1)$:
\begin{eqnarray}
\neg (H \otimes \mathbf 1) |00\rangle \rightarrow \neg \frac{1}{\sqrt 2}(|0\rangle+|1\rangle) |0\rangle \rightarrow |b_0\rangle\\  
\neg (H \otimes \mathbf 1) |01\rangle \rightarrow \neg \frac{1}{\sqrt 2}(|0\rangle+|1\rangle) |1\rangle \rightarrow |b_1\rangle\\  
\neg (H \otimes \mathbf 1) |10\rangle \rightarrow \neg \frac{1}{\sqrt 2}(|0\rangle-|1\rangle) |0\rangle \rightarrow |b_2\rangle\\  
\neg (H \otimes \mathbf 1) |11\rangle \rightarrow \neg \frac{1}{\sqrt 2}(|0\rangle-|1\rangle) |1\rangle \rightarrow |b_3\rangle .
\end{eqnarray}
We will now show how quantum entanglement can get players out of prisoner's dilemma.

\subsection{Return to the quantum Prisoner's Dilemma}

Let's return to the quantum version of Prisoner's Dilemma.  For consistency of notation, we map
$|C\rangle \rightarrow |0\rangle$ and $|D\rangle \rightarrow |1\rangle$.  When we left the final
state of the game, equation (123), it had the form
\begin{equation}
|\psi_f\rangle = U^\dagger(U_A \otimes U_B) U |00\rangle .
\end{equation}
When a measurement of the system is taken, it is projected into one of the four basis vectors
$|00\rangle$, $|01\rangle$, $|10\rangle$, $|11\rangle$, with associated probability, yielding as
expected payoff $\overline{\pi}_A$ to Alice  (refer to Table VI):
\begin{equation}
\overline{\pi}_A = 3 |\langle \psi_f|00\rangle|^2 + 0 |\langle \psi_f|01\rangle|^2 + 5 |\langle \psi_f|10\rangle|^2 
+ 1 |\langle \psi_f|11\rangle|^2 .
\end{equation}
The payoff probabilities depend on the final state of the game, which in turn depends on the unitary
matrix $U$ and the player moves $U_A$ and $U_B$.  Let's consider each of these in turn.

The purpose of the unitary matrix $U$ is to entangle Alice's and Bob's qubits.  Without this entanglement
the payoffs to Bob and Alice remain the same as in the classical game (namely, the Nash equilibrium
of (1,1)).  

Let's let our unitary matrix $U$ be (where $\otimes n$ simply means the tensor product $n$ times):
\begin{equation}
U = \frac{1}{\sqrt 2}(\mathbf 1^{\otimes 2} + i \sigma_x^{\otimes 2}) .
\end{equation}
The inverse is
\begin{equation}
U^\dagger = \frac{1}{\sqrt 2}(\mathbf 1^{\otimes 2} - i \sigma_x^{\otimes 2}) .
\end{equation}
Then, after the first application of $U$, the system state becomes:
\begin{equation}
U |00\rangle = \frac{1}{\sqrt 2}(|00\rangle + i |11\rangle) .
\end{equation}
Now let's first consider some traditional moves of Alice and Bob, either cooperate (apply matrix $U_A = 
U_B = \mathbf 1$) or defect (apply the spin-flip Pauli matrix $U_A = U_B = \sigma_x$):
\begin{eqnarray}
\mbox{both cooperate: } (\mathbf 1 \otimes \mathbf 1) U |00\rangle = \frac{1}{\sqrt 2} (|00\rangle + i|11\rangle)\\
\mbox{Alice defects: } (\sigma_x \otimes \mathbf 1) U |00\rangle = \frac{1}{\sqrt 2} (|10\rangle + i|01\rangle)\\
\mbox{Bob defects: } (\mathbf 1 \otimes \sigma_x) U |00\rangle = \frac{1}{\sqrt 2} (|01\rangle + i|10\rangle)\\
\mbox{both defect: } (\sigma_x \otimes \sigma_x) U |00\rangle = \frac{1}{\sqrt 2} (|11\rangle + i|00\rangle) .
\end{eqnarray}
Then when we apply the inverse of the unitary transformation $U$, namely $U^{-1} = U^\dagger$, we get
\begin{eqnarray}
\mbox{both cooperate: } U^\dagger \frac{1}{\sqrt 2}(|00\rangle + i|11\rangle) = |00\rangle\mbox{ with probability }1\\
\mbox{Alice defects: } U^\dagger \frac{1}{\sqrt 2}(|10\rangle + i|01\rangle) = |10\rangle\mbox{ with probability }1\\
\mbox{Bob defects: } U^\dagger \frac{1}{\sqrt 2}(|01\rangle + i|10\rangle) = |01\rangle\mbox{ with probability }1\\
\mbox{both defect: } U^\dagger \frac{1}{\sqrt 2}(|11\rangle + i|00\rangle) = |11\rangle\mbox{ with probability }1 .
\end{eqnarray}
These correspond to the four classical outcomes in Table VI, demonstrating that the classical game is
encompassed by the quantum prisoner's dilemma.  

Now let's consider some less traditional quantum moves by Alice and Bob. For example, suppose Alice plays
$\mathbf 1$ and Bob plays the Hadamard matrix $H$:  
\begin{equation}
(\mathbf 1 \otimes H) U |00\rangle = \frac{1}{2} |0\rangle(|0\rangle + |1\rangle) + \frac{i}{2} |1\rangle(|0\rangle - |1\rangle) = \frac{1}{2} [|00\rangle + |01\rangle + i |10\rangle - i |11\rangle].
\end{equation}
Then applying $U^\dagger$ to the last equation we get the final state as
\begin{equation}
U^\dagger (\mathbf 1 \otimes H) U |00\rangle = \frac{1}{\sqrt 2} (|01\rangle - i |11\rangle).
\end{equation}
Since $|\frac{1}{\sqrt 2}|^2 = \frac{1}{2}$ and $|\frac{-i}{\sqrt 2}|^2 = \frac{1}{2}$, a
measurement of the latter state will give Alice a payout of 0 or a payout of 1 with equal probability, so $\overline{\pi}_A = 0.5$, $\overline{\pi}_B = 3$.

Conversely, suppose Bob plays
$\mathbf 1$ and Alice plays the Hadamard matrix $H$:  
\begin{equation}
(H \otimes \mathbf 1) U |00\rangle = \frac{1}{2} [|00\rangle + |10\rangle + i |01\rangle - i |11\rangle].
\end{equation}
Then applying $U^\dagger$ to the last equation we get the final state of the reversed play as
\begin{equation}
U^\dagger (H \otimes \mathbf 1) U |00\rangle = \frac{1}{\sqrt 2} (|10\rangle - i |11\rangle).
\end{equation}
A measurement of the latter state will give Alice a payout of 5 or a payout of 1 with equal probability, so $\overline{\pi}_A = 3$, $\overline{\pi}_B = 0.5$.

We will summarize the remaining cases we want to consider:
\begin{eqnarray}
(H \otimes \sigma_x) U |00\rangle = \frac{1}{2} [|01\rangle + |11\rangle + i |00\rangle - i |10\rangle]\mbox{ } \\
(\sigma_x \otimes H) U |00\rangle = \frac{1}{2} [|10\rangle + |11\rangle + i |00\rangle - i |01\rangle]\mbox{ } \\
(H \otimes H) U |00\rangle = \frac{1}{\sqrt{2^3}} [|00\rangle + |10\rangle + |01\rangle + |11\rangle 
+ i |00\rangle - i |10\rangle - i |01\rangle + i |11\rangle],\mbox{ } \\
U^\dagger (H \otimes \sigma_x) U |00\rangle = \frac{1}{\sqrt 2} [|11\rangle - i |10\rangle],
\overline{\pi}_A = 3, \overline{\pi}_B = 0.5\mbox{ } \\
U^\dagger (\sigma_x \otimes H) U |00\rangle = \frac{1}{\sqrt 2} [|11\rangle - i |01\rangle],
\overline{\pi}_A = 0.5, \overline{\pi}_B = 3\mbox{ } \\
U^\dagger (H \otimes H) U |00\rangle = \frac{1}{2} [|00\rangle + |11\rangle - i |01\rangle - i |10\rangle], \overline{\pi}_A = \overline{\pi}_B = 2.25.\mbox{ }
\end{eqnarray}
Let `$\succ$' denote `is preferred to'.  Alice no longer has a preferred strategy.  While $\sigma_x \succ_A 
\mathbf 1$, if Bob plays $\sigma_x$ or $H$, then $H \succ_A \sigma_x$.  This is shown in Table VII.  In addition,
\begin{table}[ht]
\begin{tabular}{|r|r|r|r|}
\hline
\ & Bob $\mathbf 1$ & Bob $\sigma_x$ & Bob $H$\\
\hline
Alice $\mathbf 1$ & (3,3) & (0,5) & ($\frac{1}{2}$,3)\\
Alice $\sigma_x$ & (5,0) & (1,1) & ($\frac{1}{2}$,3)\\
Alice $H$ & (3,$\frac{1}{2}$) & (3,$\frac{1}{2}$) & (2$\frac{1}{4}$,2$\frac{1}{4}$)\\
\hline
\end{tabular}
\caption{Prisoner's Dilemma with allowed quantum moves of $\sigma_x$, $H$.}
\end{table}
The payoff state $(1,1)$ corresponding to $(\sigma_x,\sigma_x)$ is no longer a Nash equilibrium. However,
the outcome $(2\frac{1}{4},2\frac{1}{4})$ corresponding to $(H,H)$ is now a Nash equilibrium, although
it is not Pareto optimal.  Clearly the addition of quantum moves changes the game outcome.

To induce Pareto optimality, let's expand the set of allowed moves to be members of $S = \{\mathbf 1, \sigma_x,
H, \sigma_z\}$.  The result is shown in Table VIII.
The outcome $(2\frac{1}{4},2\frac{1}{4})$ is no longer a Nash equilibrium, but we have a new Nash equilibrium at $(3,3)$
corresponding to $(\sigma_z,\sigma_z)$.  The payoffs are equal to those of the non-equilibrium strategy 
point $(\mathbf 1,\mathbf 1)$, so it is not jointly dominated.  This Nash equilibrium is Pareto optimal.  
End of Prisoner's Dilemma.
\begin{table}[ht]
\begin{tabular}{|r|r|r|r|r|}
\hline
\ & Bob $\mathbf 1$ & Bob $\sigma_x$ & Bob $H$ & Bob $\sigma_z$\\
\hline
Alice $\mathbf 1$ & (3,3) & (0,5) & ($\frac{1}{2}$,3) & (1,1)\\
Alice $\sigma_x$ & (5,0) & (1,1) & ($\frac{1}{2}$,3) & (0,5)\\
Alice $H$ & (3,$\frac{1}{2}$) & (3,$\frac{1}{2}$) & (2$\frac{1}{4}$,2$\frac{1}{4}$) & (1$\frac{1}{2}$,4)\\
Alice $\sigma_z$ & (1,1) & (5,0) & (4,1$\frac{1}{2}$)& (3,3)\\
\hline
\end{tabular}
\caption{Prisoner's Dilemma with allowed quantum moves of $\sigma_x$, $H$, $\sigma_z$. The outcome $(3,3)$ corresponding
to moves $(\sigma_z,\sigma_z)$ is not only a Nash equilibrium, it is also Pareto optimal.}
\end{table}

What is the meaning of the unitary matrix $U$ that is applied at the beginning and end of the game?  That remains
to be determined.  Sometimes it is ascribed to a third player, a referee or a co-ordinator.  But there are
other interpretations.  Perhaps the best is that `it acts as a collaborator to the players and serves to
maximize the payoff at the Nash equilibria' \cite{CT}.  An Invisible Hand in prisoner's dilemma?  More work is
needed.

\subsection{Battle of the sexes game: a quantum game with entanglement}
The so-called `battle of the sexes' game is not really a battle: it's a love fest with conflicting values.
Alice and Bob want to spend an evening together, and if they spend it apart, their respective payoffs are
$\{\gamma,\gamma\}$. As usual, Alice's payoff is listed first and Bob's payoff second. Alice prefers to spend the evening at the Opera (O), while Bob prefers to spend the evening watching TV (T).  The payoffs for both at the Opera are $\{\alpha,\beta\}$, while for both watching TV, the payoffs are $\{\beta,\alpha\}$. It is assumed $\alpha > \beta > \gamma$.  Alice
and Bob are both at work at their respective jobs, and are not able to communicate (no cellphones). Each
plans to show up either at the Opera or at Bob's house for TV, in hopes of meeting the other at that place.
The moves for each are thus members of the set $\{O,T\}$.  The game is shown in Table IX.

Inspection of the Table shows two Nash equilibria in moves: $(O,O)$ and $(T,T)$. A unilateral departure of either
player from one of these equilibria results in a smaller payoff.  However $\cdots$, there is a Nash equilibrium in
each row for Alice, and in each column for Bob.  So how does either player decide what to do? 
\begin{table}[ht]
\begin{tabular}{|r|r|r|}
\hline
\ & Bob O & Bob T\\
\hline
Alice O & ($\alpha,\beta$) & ($\gamma,\gamma$)\\
Alice T & ($\gamma,\gamma$) & ($\beta,\alpha$)\\
\hline
\end{tabular}
\caption{Battle of the Sexes $(\alpha > \beta > \gamma$)}
\end{table}
In addition, there is a third hidden Nash equilibrium in mixed strategies resulting from Alice playing $O$ with probability $p$
and $T$ with probability $1-p$, while Bob plays $O$ with probability $q$ and $T$ with probability $1-q$, where
$p$ and $q$ are neither $0$ nor $1$.  Calculation shows $p = \frac{\alpha-\gamma}{\alpha+\beta-2\gamma}$, while
$q = \frac{\beta-\gamma}{\alpha+\beta-2\gamma}$.  These probabilities give the expected payoffs to Alice and
Bob as
\begin{equation}
\overline{\pi}_A(p,q) = \overline{\pi}_B(p,q) = \frac{\alpha \beta-\gamma^2}{\alpha+\beta-2\gamma}.
\end{equation}
In the corner Nash equilibria shown in Table IX, one of Alice or Bob receives a payoff of $\alpha$ and the
other a payoff of $\beta$.  But $\alpha > \beta > \overline{\pi}_A(p,q)$.  So both Alice and Bob are worse off in the
third Nash equilibrium.

To find this third Nash equilibrium, we first write Alice's expected payoff given the assumed probabilities
of each move of Alice and Bob:
\begin{equation}
\overline{\pi}_A = pq\alpha + p(1-q)\gamma + (1-p)q\gamma+(1-p)(1-q)\beta .
\end{equation}
Then, maximizing over $p$,
\begin{equation}
\frac{\partial \overline{\pi}_A}{\partial p} = q\alpha + (1-q)\gamma - q\gamma - (1-q)\beta = 0.
\end{equation}
Solving the latter equation for $q$ results in $q = \frac{\beta-\gamma}{\alpha+\beta-2\gamma}$.  A similar
calculation maximizing Bob's expected payoff yields $p$.  

How do quantum strategies change things?  Let's map $|O\rangle \rightarrow |0\rangle$ and $|T\rangle \rightarrow
|1\rangle$,and then entangle states by applying our unitary matrix $U$, 
\begin{equation}
U = \frac{1}{\sqrt 2}(\mathbf 1^{\otimes 2} + i \sigma_x^{\otimes 2}),
\end{equation}
to an initial state $|00\rangle$.  Then, after the first application of $U$, the system state becomes:
\begin{equation}
U |00\rangle = \frac{1}{\sqrt 2}(|00\rangle + i |11\rangle),
\end{equation}
as before. Both Alice and Bob know $U$ and the initial state $|00\rangle$.

We again allow Alice and Bob to make moves from the strategy set $S = \{\mathbf 1, \sigma_x,
H, \sigma_z\}$ on their individual qubits. And then we apply $U^\dagger$ to the result. The final 
states are those calculated previously in Prisoner's Dilemma, but the expected payoffs are different, 
as shown in the following Table X.

\begin{table}[ht]
\begin{tabular}{|r|r|r|r|r|}
\hline
\ & Bob $\mathbf 1$ & Bob $\sigma_x$ & Bob $H$ & Bob $\sigma_z$\\
\hline
Alice $\mathbf 1$ & ($\alpha,\beta$) & ($\gamma,\gamma$) & ($\frac{\beta +\gamma}{2},\frac{\alpha +\gamma}{2}$) & ($\beta,\alpha$)\\
Alice $\sigma_x$ & ($\gamma,\gamma$) & ($\beta,\alpha$) & ($\frac{\beta +\gamma}{2},\frac{\alpha +\gamma}{2}$) & ($\gamma,\gamma$)\\
Alice $H$ & ($\frac{\beta +\gamma}{2},\frac{\alpha +\gamma}{2}$) & ($\frac{\beta +\gamma}{2},\frac{\alpha +\gamma}{2}$) & ($\frac{\alpha +\beta +2\gamma}{4},\frac{\alpha +\beta +2\gamma}{4}$) & ($\frac{\alpha +\gamma}{2},\frac{\beta +\gamma}{2}$)\\
Alice $\sigma_z$ & ($\beta,\alpha$) & ($\gamma,\gamma$) & ($\frac{\alpha +\gamma}{2},\frac{\beta +\gamma}{2}$)& ($\alpha,\beta$)\\
\hline
\end{tabular}
\caption{Battle of the Sexes Game with quantum moves. The Nash equilibrium is $(\beta,\alpha)$ corresponding to
$(\sigma_x,\sigma_x)$.  Alice and Bob spend the evening watching TV.}
\end{table}
The upper left-hand entries show the classical game is contained in the quantum game.  The only Nash equilibrium
in the Table is $(\beta,\alpha)$ corresponding to $(\sigma_x,\sigma_x)$.  Alice and Bob spend an evening
watching television together, with Alice having a payoff of $\beta$ less than Bob's payoff of $\alpha$.  At
$(\sigma_x,\sigma_x)$ neither Alice nor Bob can unilaterally increase his or her payoff, and since this set
of payoffs is not jointly dominated by another set of payoffs, it is also Pareto optimal.  Television rules!

It remains to consider mixed strategies.  It is clear the four corner payoffs in the Table are the extreme
points of a convex set.  So we only need consider
consider convex combinations of $\mathbf 1$ and $\sigma_z$.  Alice's expected payoff takes the form
\begin{equation}
\overline{\pi}_A = pq\alpha + p(1-q)\beta +(1-p)q\beta+(1-p)(1-q)\alpha .
\end{equation}
Maximizing over $p$,
\begin{equation}
\frac{\partial \overline{\pi}_A}{\partial p} = q\alpha + (1-q)\beta - q\beta - (1-q)\alpha = 0.
\end{equation}
Solving for $q$ gives $q = \frac{1}{2}$.  Similarly, $p = \frac{1}{2}$.  The mixed strategies
$(\frac{1}{2}\mathbf 1 +\frac{1}{2} \sigma_z,\frac{1}{2}\mathbf 1+\frac{1}{2} \sigma_z)$ yield
payoffs of $(\frac{\alpha+\beta}{2},\frac{\alpha+\beta}{2})$.  At last equality between Bob and Alice! This
Nash equilibrium is also Pareto optimal, as it is not jointly dominated by either $(\alpha, \beta)$ or $(\beta,
\alpha)$.

\subsection{Newcomb's Game: a game against a Superior Being}
Alice plays the following game against a Superior Being (SB).  The SB may be thought of as God, a superior
intelligence from another planet, or as a supercomputer that is very good at predicting Alice's thought
processes \cite{SJB}.  There are two boxes $B_1$ and $B_2$.  $B_1$ contains \$1000.  $B_2$ contains either \$1,000,000
or \$0, depending on which amount SB put in the box.  Alice may choose to take either both boxes or only $B_2$.
If the SB has predicted that Alice will choose both boxes, then SB puts \$0 in $B_2$, while if the SB has
predicted Alice will take only box $B_2$, then SB puts \$1,000,000 in $B_2$.  The game is depicted in Table
XI.
\begin{table}[ht]
\begin{tabular}{|r|r|r|}
\hline
\ & SB predicts Alice will take only box $B_2$ & SB predicts Alice will take both boxes\\
\hline
Alice takes only box $B_2$ & \$1,000,000 & \$0\\
Alice takes both boxes  & \$1,001,000 & \$1000\\
\hline
\end{tabular}
\caption{Newcomb's Game.}
\end{table}
Alice clearly has a dominant strategy, which is to take both boxes, as each payoff in the second row
is greater than the corresponding payoff in the first row.  On the other hand, the dominant strategy
conflicts with expected utility theory (here utility is taken to be linear in the payoffs).  Suppose
the predictive accuracy of SB is $p$.  Then according to expected ultility theory, Alice will be indifferent 
between taking both boxes or only
$B_2$ if
\begin{equation}
p\mbox{ } \$1,000,000 + (1-p)\mbox{ } \$0 = (1-p)\mbox{ } \$1,001,000 + p\mbox{ } \$1000 .
\end{equation}
For $p>.5005$ Alice would prefer the strategy of only taking box $B_2$, conflicting with the dominant
strategy. There are various ways to resolve
this dilemma \cite{SJB}.  For example, if SB is omniscient (p=1), then the Table has only two entries,
\$1000 and \$1,000,000. So automaton Alice will choose whichever SB has predicted, and the paradox is resolved.  

But here we are interested in the quantum game \cite{PS2}.  SB surely knows the universe is based on
quantum physics, not on classical physics, which is only the biased view of beings who are approximately
two meters high.  The quantum Newcomb's game takes place in the Hilbert space $\mathbf H_1 \otimes
\mathbf H_2$, which we will take to be a 2-qubit space, with the left qubit denoting Alice's actions,
and the right qubit denoting the actions of the SB.  For SB, $|0\rangle$ represents the placement of
\$1,000,000 in box $B_2$, while $|1\rangle$ represents the placement of \$0 in $B_2$.  For Alice,
$|0\rangle$ represents taking $B_2$ only, while $|1\rangle$ represents taking both boxes. The basis vectors 
of $\mathbf H_1 \otimes \mathbf H_2$ are $|00\rangle$, $|01\rangle$, $|10\rangle$, $|11\rangle$, corresponding 
to the payoff states in Table XI.

The initial state of the game is $\Lambda =  |00\rangle$ if SB puts \$1,000,000 in box $B_2$, or
$\Lambda =  |11\rangle$ if SB puts nothing in $B_2$.  The course of the game is as follow.

\emph{Step 1:} SB makes its choice, $|0\rangle$ or $|1\rangle$.  Once made this choice cannot be altered.

\emph{Step 2:} SB applies the Hadamard matrix $H$ to Alice's qubit; that is, the operator $H \otimes \mathbf 1$
to the initial state $\Lambda$.

\emph{Step 3:} Alice applies the spin flip operator $\sigma_x \otimes \mathbf 1$ with probability $w$ or
the identity matrix $\mathbf 1 \otimes \mathbf 1$ with probability $1-w$ to the current state of the game.  
(These operate only on her own qubit.)

\emph{Step 4:} The SB applies $H \otimes \mathbf 1$ to the current state of the game, and the payoff to Alice
is determined.  

If the SB has chosen $|0\rangle$, then the sequence of steps in the game is as follow:
\begin{eqnarray}
(H \otimes \mathbf 1) |00\rangle \rightarrow \frac{1}{\sqrt 2}(|00\rangle + |10\rangle)\\
 w(\sigma_x \otimes \mathbf 1)(H \otimes \mathbf 1) |00\rangle \rightarrow \frac{w}{\sqrt 2}(|00\rangle + |10\rangle)\\
\Rightarrow (w(\sigma_x \otimes \mathbf 1)+(1-w)(\mathbf 1 \otimes \mathbf 1))(H \otimes \mathbf 1) |00\rangle \rightarrow
 \frac{1}{\sqrt 2}(|00\rangle + |10\rangle) \\
 (H \otimes \mathbf 1) (w(\sigma_x \otimes \mathbf 1)+(1-w)(\mathbf 1 \otimes \mathbf 1))(H \otimes \mathbf 1) |00\rangle \rightarrow |00\rangle .
 \end{eqnarray}
 Thus Alice takes only box $B_0$ and receives \$1,000,000.  The SB has correctly predicted Alice's move.
 
If the SB has chosen $|1\rangle$, then the sequence of steps in the game is as follow:
\begin{eqnarray}
(H \otimes \mathbf 1) |11\rangle \rightarrow \frac{1}{\sqrt 2}(|01\rangle - |11\rangle)\\
 w(\sigma_x \otimes \mathbf 1)(H \otimes \mathbf 1) |11\rangle \rightarrow \frac{w}{\sqrt 2}(|11\rangle - |01\rangle)\\
\Rightarrow (w(\sigma_x \otimes \mathbf 1)+(1-w)(\mathbf 1 \otimes \mathbf 1))(H \otimes \mathbf 1) |11\rangle \rightarrow
 \frac{1-2w}{\sqrt 2}(|01\rangle - |11\rangle) \\
 (H \otimes \mathbf 1) (w(\sigma_x \otimes \mathbf 1)+(1-w)(\mathbf 1 \otimes \mathbf 1))(H \otimes \mathbf 1) |11\rangle \rightarrow (1-2w)|11\rangle .
 \end{eqnarray}
The final value is maximized when $w=0$. Thus Alice takes both boxes and receives \$1,000.  The SB has again perfectly
predicted Alice's move.  The SB did not require omiscience to achieve this result, only a knowledge of quantum mechanics. By applying the Hadamard matrix (the quantum Fourier transform) to the initial state of the game, the SB
induced Alice to behave in a way so as to confirm the SB's prediction. 

\subsection{Evolutionarily stable strategy game}
It seems that quantum games are played about us every day at a molecular level. Gogonea and Merz \cite{GMerz}
indicate games are being played at the quantum mechanical level in protein folding. Turner and Chao \cite{TC} studied
the evolution of competitive interactions among viruses in an RNA phage, and found the fitness of the phage
generates a payoff matrix conforming to the two-person prisoner's dilemma game.  We want to briefly touch
on some game theory aspects of biology.

The concept of 
\emph{evolutionarily stable strategy} (ESS), which we previously defined in connection with the concept of Nash equilibrium, was introduced into game theory \cite{EGTSE} to deal with some problems in population biology and 
with the fact there may be multiple Nash equilibria. In \emph{Evolution and the Theory of Games} \cite{MSETG} Maynard
Smith noted that `game theory is more readily applied to biology than to the field of economic behaviour for which
it was originally designed'.     

Consider a population of $N$ members who are randomly matched in pairs to play a symmetric bimatrix 
(i.e., $2 \times 2$) game.  By \emph{symmetric} is meant the following.  Let $S$ be the set of player moves, and let $s_i$, $s_j$ be moves that are available to both Alice and Bob.  Then Alice's expected payoff when she 
plays $s_i$ and Bob plays $s_j$ is the same as Bob's expected payoff if he plays $s_i$ and Alice plays $s_j$:
\begin{equation}
\overline{\pi}_A (s_i,s_j) = \overline{\pi}_B (s_j,s_i) .
\end{equation}
That is, Alice's payoff matrix $\Pi_A$ is the transpose of Bob's payoff matrix: $\Pi_A = \Pi_B^T$.
This defines the symmetry of the game.  The game becomes \emph{evolutionary} if over time moves $s_i$ with
higher payoffs gradually replace those $s_j$ with lower payoffs.  In such a game, Maynard Smith and Price
\cite{MSP} showed that a population which adopts an ESS can withstand a small invading group.

But what if the current population, in equilibrium while playing classical moves, is invaded by a population
playing quantum moves? This is the problem considered by Iqbal and Toor \cite{IT}.

Suppose the proportion of the population playing the move $s_i$ in a symmetric bimatrix game is $p_i$,
while the proportion playing the move $s_j$ is $p_j$.  Define the \emph{fitness} $w$ of moves $s_i$ and $s_j$
as follows:
\begin{eqnarray}
w(s_i) = p_i \overline{\pi}(s_i,s_i) + p_j \overline{\pi}(s_i,s_j)\\
w(s_j) = p_i \overline{\pi}(s_j,s_i) + p_j \overline{\pi}(s_j,s_j).
\end{eqnarray}
The first equation says the fitness of move $s_i$ is a weighted average of the payoff to playing
$s_i$ against an opponent also playing $s_i$ and of the payoff to playing $s_i$ against an opponent playing
$s_j$.  The respective weights are the proportions of the population playing $s_i$ and $s_j$.  The second
equation is really the same as the first with indexes switched.

For our \emph{quantum evolutionarily stable strategy game} we will assume that the symmetric bimatrix game
played between the two population groups is the Prisoner's Dilemma game. The payoff matrix for this game
is that previously given in Table VI. Note that the payoff matrix of
one player is the transpose of the payoff matrix of the other player, which is required for symmetry.  Note
also that the unitary matrix $U = \frac{1}{\sqrt 2}(\mathbf 1^{\otimes 2} + i \sigma_x^{\otimes 2})$ used
in the quantum Prisoner's Dilemma game is also symmetric between the two players.  For classical moves,
the payoff state $\{s_A,s_B\} = \{D,D\}$ and $\{\pi(s_A),\pi(s_B)\} = \{1,1\}$, which is a Nash equilibrium, 
is also an evolutionarily stable strategy.  Consider, however, the effect of an invading force of mutants
playing quantum moves.  For ease of reference, we will reproduce Table VIII here as Table XII. We will label 
$\{\mathbf 1, \sigma_x\}$ as classical moves, and $\{H,\sigma_z\}$ as mutant moves.
\begin{table}[ht]
\begin{tabular}{|r|r|r|r|r|}
\hline
\ & Classical $\mathbf 1$ & Classical $\sigma_x$ & Mutant $H$ & Mutant $\sigma_z$\\
\hline
Classical $\mathbf 1$ & (3,3) & (0,5) & ($\frac{1}{2}$,3) & (1,1)\\
Classical $\sigma_x$ & (5,0) & (1,1) & ($\frac{1}{2}$,3) & (0,5)\\
Mutant $H$ & (3,$\frac{1}{2}$) & (3,$\frac{1}{2}$) & (2$\frac{1}{4}$,2$\frac{1}{4}$) & (1$\frac{1}{2}$,4)\\
Mutant $\sigma_z$ & (1,1) & (5,0) & (4,1$\frac{1}{2}$)& (3,3)\\
\hline
\end{tabular}
\caption{Population playing classical moves of $\mathbf 1$, $\sigma_x$, is invaded by mutants play the quantum move $H$; a later invasion of mutants plays $\sigma_z$ and wipes out the previous mutants.}
\end{table}

We see that $\sigma_x$ is not evolutionarily stable against $H$.  Members playing $\sigma_x$ will die out and
the population will soon be comprised of mutants playing $H$.  The new ESS will yield the payoff 2$\frac{1}{4}$
to either mutant party.  If this new population is now invaded by different mutants playing $\sigma_z$, then
$H$ is no longer an ESS.  Members playing $H$ will die out, and the population will soon be comprised of mutants
playing $\sigma_z$.  These mutants will enjoy a payoff of 3, and will appear fat and happy when contrasted
with the original population.

\subsection{Card game: a quantum game without entanglement}  
The following game doesn't use entanglement, but is heuristic for its mathematical setup, and is good preparation
for more complicated games that follow.  Bob and Alice play the following card game  \cite{JXHMXR}.  There are three cards, otherwise identical, except for the following
markings:  the first card has a circle on each side; the second card has a dot on each side; the third card
has a circle on one side and a dot on the other.  Alice puts the three cards in a black box and shakes it to
randomize the three cards.  Bob is allowed to blindly draw one card from the box.  If it has the same mark on
each side, Alice wins $+1$ from Bob.  If the card has different marks on each side, Bob wins $+1$ from Alice.
Of course, two of the cards having the same mark on each side, Alice has expected payoff $\overline{\pi}_A =
\frac{2}{3} (1) + \frac{1}{3} (-1) = \frac{1}{3}$, while Bob has expected payoff $\overline{\pi}_B = \frac{1}{3} (1)
+ \frac{2}{3} (-1) = - \frac{1}{3}$.  The game is unfair to Bob.

One way to make the game fair, in a classical sense, would be to allow Bob to look in the black box and see the
upper faces of the three cards before drawing one of them.  Then if Bob saw two circles facing up among the three
cards, he would randomly draw one of those two cards, while if he saw two dots facing up, he would radomly draw one of the latter two cards. Since one of the two cards with identical upside marks must have different markings on each side, this would give Bob an expected payoff $\overline{\pi}_B = 0$.  The game would now be fair.  However, we are not going 
to let Bob do this.  In fact, it's a black box so that he \emph{can't} look inside, but he can stick his hand in
and pull one card out.

Instead, to create the quantum equivalent of looking at the upper faces of all three cards, we are going to 
1) allow Bob to make a single \emph{query} to the black box or qubit database $|r\rangle$; and 2), allow Bob to withdraw 
from the game once he sees the upper face of the card he draws.  This setup is highly artificial, and it is
doubtful we are even describing the same game, but this quantized version of the Card Game will allow us to
make several heuristic points.

To describe the quantum game setup, let the card state be $|0\rangle$ if the card has a circle up, and $|1\rangle$
if a card has a dot up.  The three-card state can be written as
\begin{equation}
|r\rangle = |r_0 r_1 r_2\rangle
\end{equation}
where $r_k \in \{0,1\}$.

As part of Bob's query, we will require the following unitary matrix $U_k$:
\begin{equation}
U_k = \left( \begin{array}{cc}
1 & 0 \\
0 & e^{i\pi r_k} \end{array} \right).
\end{equation}
Note that if $r_k = 0$, then $U_k = \mathbf 1$, while if $r_k = 1$, then $U_k = \sigma_z$.  Now we
apply the Hadamard matrix $H$ to $U_k$ to form $H U_k H$ and obtain:
\begin{equation}
H U_k H = \frac{1}{2} \left( \begin{array}{cc}
1 & 1 \\
1 & - 1 \end{array} \right)\left( \begin{array}{cc}
1 & 0 \\
0 & e^{i\pi r_k} \end{array} \right) \left( \begin{array}{cc}
1 & 1 \\
1 & -1 \end{array} \right)
= \frac{1}{2}
\left( \begin{array}{cc}
1 + e^{i\pi r_k} & 1 - e^{i\pi r_k} \\
1 - e^{i\pi r_k} & 1 + e^{i\pi r_k} \end{array} \right).
\end{equation}
Thus, applying this transformation to the state $|0\rangle$, we get
\begin{equation}
H U_k H |0\rangle = \frac{1}{2} 
\left( \begin{array}{cc}
1 + e^{i\pi r_k} & 1 - e^{i\pi r_k} \\
1 - e^{i\pi r_k} & 1 + e^{i\pi r_k} \end{array} \right)
\left( \begin{array}{c}
1 \\
0 \end{array} \right)
= \frac{1}{2}
\left( \begin{array}{c}
1 + e^{i\pi r_k} \\
1 - e^{i\pi r_k} \end{array} \right)
= 
\frac{1 + e^{i\pi r_k}}{2} |0\rangle + \frac{1 - e^{i\pi r_k}}{2} |1\rangle.
\end{equation}
Note that if $r_k = 0$, $H U_k H |0\rangle = |0\rangle$, while if $r_k = 1$, $H U_k H |0\rangle = |1\rangle$.
Thus,
\begin{equation}
H U_k H |0\rangle = |r_k\rangle .
\end{equation}
So now let's assume that Bob has a query machine that depends on state $|r\rangle$ in the black box.  The machine
has three inputs and gives three outputs.  To determine the upside marks of the three cards, Bob inputs $|000\rangle$
to obtain:
\begin{equation}
(H U_k H \otimes H U_k H \otimes H U_k H) |000\rangle = |r_0 r_1 r_2\rangle.
\end{equation}
So after Bob's query, he knows the upside marks of the three cards: either some element of the set $S_0 = 
\{\mbox{ 3-qubit permuations of } \{|0\rangle, |0\rangle, |1\rangle\}\}$ or some element of the set $S_1 = 
\{\mbox{ 3-qubit permuations of } \{|0\rangle, |1\rangle,|1\rangle\}\}$. If $S_0$ descibes the
state of the black box, then Bob knows the winning card has a circle on the upside face.  If $S_1$ describes the
state of the black box, then Bob know the winning card has a dot on the upwise face.  So now Bob draws his card,
and gets to look at the upside face only.
If the drawn card has a circle on the upside face, and the black box $\in S_0$, then Bob has an equal chance of
winning.  But if the black box $\in S_1$, then Bob refuses to play because he knows the drawn card is a losing card.
A similar analysis applies when the drawn card has a dot on the upside face. 

So a query to the database shows Bob whether there are two circles or two dots showing face up in the black
box, and thus when he draws his card he knows that if it matches the two upside marks, then he has a 50-50
chance of winning, while if the drawn card doesn't matched the two upside marks, the card is definitely a loser
and he should exercise his option to withdraw from the game.

With respect to entanglement, the operators $H$ and $U_k$ form simple linear combinations of qubits, while the
quantum query machine is a tensor product of these operations.  Hence there is no entanglement of states in this game.
Du \emph{et. al.} note that that the general rule appears to be that entanglement is required in static
quantum games to make a difference from classical outcomes, but not in dynamic games.  The key is the ability
of the player to affect the state of others' qubits.  This can be done through entanglement or through the time
steps of a dynamic game.

\subsection{Quantum teleportation and pseudo-telepathy}

Alice and Bob are seven light-years apart and share an entangled pair of qubits, say $|b_0\rangle = \frac{1}{\sqrt 2}(|00\rangle + |11\rangle)$.  
If Alice measures her qubit and finds it is in the state $|0\rangle$, then Bob's qubit is guaranteed to
be in the state $|0\rangle$ also. If Alice finds by measurement her qubit is in the state $|1\rangle$, then
Bob's qubit will also be found in the state $|1\rangle$.  That is, \emph{Alice's measurement affects the state
of Bob's qubit}.  As far as we know, this transmission of influence through the Bohr channel takes place 
instantaneously.  It is not affected by distance or limited by the speed of light.  It is spooky action at 
a distance.  It is also the basis for quantum teleportation.

$\mathbf{Teleportation\mbox{ }}$.  The quantum teleportation protocol \cite{BBCJPW}, by contrast, does not take place instantaneously, since it uses a classical channel as well as a Bohr (EPR) channel.  On the other hand, a quantum state disappears in one place and reappears in another: hence it is teleported.  The traditional teleportation protocol works like this.  Alice has an unknown quantum state $|\psi\rangle$ she wants to transmit to Bob.  She will do this in two pieces:  she will use an entangled Bohr channel, and an  additional classical channel to transmit some classical bits.  Alice and Bob have made previous arrangement  to share an entangled pair of particles, this time say in the Bell state $|b_3\rangle$:
\begin{equation}
|b_3\rangle = \frac{1}{\sqrt 2}(|01\rangle - |10\rangle) .
\end{equation}
The unknown state Alice is trying to transmit may be written in terms of unknown amplitudes $a$, $b$, $|a|^2 +
|b|^2 = 1$, as
\begin{equation}
|\psi\rangle = a |0\rangle + b |1\rangle .
\end{equation}
We may write the initial state of the 3-qubit system as:
\begin{eqnarray}
|\psi\rangle \otimes |b_3\rangle
= (a |0\rangle + b |1\rangle) \otimes (\frac{1}{\sqrt 2}(|01\rangle - |10\rangle))\\
= \frac{a}{\sqrt 2}|001\rangle - \frac{a}{\sqrt 2}|010\rangle + \frac{b}{\sqrt 2}|101\rangle - \frac{b}{\sqrt 2}|110\rangle .
\end{eqnarray}
We want to rewrite this state in terms of the Bell basis, for reasons that will become apparent.  To do
this, we take the inner product of $|\psi\rangle \otimes |b_3\rangle$ with each of the Bell vectors in
order to find the multiplier on each Bell state. Note that we take the inner product with the 
\emph{two left-most} qubits in equation (184).  These qubits are under the control of Alice.
\begin{eqnarray}
\langle b_0 |(|\psi\rangle \otimes |b_3\rangle) \rangle = +\frac{a}{2} |1\rangle - \frac{b}{2} |0\rangle \\
\langle b_1 |(|\psi\rangle \otimes |b_3\rangle) \rangle = -\frac{a}{2} |0\rangle + \frac{b}{2} |1\rangle \\
\langle b_2 |(|\psi\rangle \otimes |b_3\rangle) \rangle = +\frac{a}{2} |1\rangle + \frac{b}{2} |0\rangle \\
\langle b_3 |(|\psi\rangle \otimes |b_3\rangle) \rangle = -\frac{a}{2} |0\rangle - \frac{b}{2} |1\rangle .
\end{eqnarray}
Using these residual state multipliers, we can then write the state $|\psi\rangle \otimes |b_3\rangle$ in terms 
of the Bell basis:
\begin{equation}
|\psi\rangle \otimes |b_3\rangle = \frac{1}{2}\lbrack\left(\begin{array}{c}
-b \\
+a \end{array} \right) |b_0\rangle + \left(\begin{array}{c}
-a \\
+b \end{array} \right) |b_1\rangle + \left(\begin{array}{c}
+b \\
+a \end{array} \right) |b_2\rangle + \left(\begin{array}{c}
-a \\
-b \end{array} \right) |b_3\rangle\rbrack .
\end{equation}
Now let's rewrite the last equation in terms of $2 \times 2$ matrices:
\begin{eqnarray}
|\psi\rangle \otimes |b_3\rangle = 
\frac{1}{2}\lbrack \left( \begin{array}{cc}
0 & -1 \\
1 & 0 \end{array} \right)\left(\begin{array}{c}
a \\
b \end{array} \right) |b_0\rangle + \left( \begin{array}{cc}
-1 & 0 \\
0 & 1 \end{array} \right)\left(\begin{array}{c}
a \\
b \end{array} \right) |b_1\rangle + \\ 
\left( \begin{array}{cc}
0 & 1 \\
1 & 0 \end{array} \right)\left(\begin{array}{c}
a \\
b \end{array} \right) |b_2\rangle + \left( \begin{array}{cc}
-1 & 0 \\
0 & -1 \end{array} \right)\left(\begin{array}{c}
a \\
b \end{array} \right) |b_3\rangle\rbrack .
\end{eqnarray}
We can rewrite this again in terms of the Pauli spin matrices:
\begin{equation}
|\psi\rangle \otimes |b_3\rangle = 
\frac{1}{2}\lbrack -i\sigma_y \left(\begin{array}{c}
a \\
b \end{array} \right) |b_0\rangle - \sigma_z \left(\begin{array}{c}
a \\
b \end{array} \right) |b_1\rangle + \sigma_x 
\left(\begin{array}{c}
a \\
b \end{array} \right) |b_2\rangle - \mathbf 1 
\left(\begin{array}{c}
a \\
b \end{array} \right) |b_3\rangle\rbrack .
\end{equation}
Now, to teleport her qubit to Bob, Alice must couple the unknown state $|\psi\rangle$ with her member of the
entangled qubit pair. To do this she makes a joint (von Neumann) measurement of these two qubits, which
comprise the two left-most qubits of $|\psi\rangle \otimes |b_3\rangle$.  Alice's measurement projects 
her two qubits into one of the four Bell states.  This destroys the unknown state $|\psi\rangle$.  But not to worry.  Alice's measurement also leaves Bob's qubit in one of the following four states:
\begin{eqnarray}
|\psi\rangle \otimes |b_3\rangle \rightarrow |b_0\rangle \Longrightarrow \mbox{ Bob's qubit }= -i\sigma_y\left(\begin{array}{c}
a \\
b \end{array} \right)\\
|\psi\rangle \otimes |b_3\rangle \rightarrow |b_1\rangle \Longrightarrow \mbox{ Bob's qubit }= -\sigma_z\left(\begin{array}{c}
a \\
b \end{array} \right)\\
|\psi\rangle \otimes |b_3\rangle \rightarrow |b_2\rangle \Longrightarrow \mbox{ Bob's qubit } =
\sigma_x\left(\begin{array}{c}
a \\
b \end{array} \right)\\
|\psi\rangle \otimes |b_3\rangle \rightarrow |b_3\rangle \Longrightarrow \mbox{ Bob's qubit } = 
-\mathbf 1 \left(\begin{array}{c}
a \\
b \end{array} \right) .
\end{eqnarray}
Alice then, through a classical channel, transmits to Bob the results of her measurement: i.e., the Bell
state she obtained.  Then Bob applies the corresponding spin operator (which is its own inverse) to his qubit to recover the state
$|\psi\rangle = \left(\begin{array}{c}
a \\
b \end{array} \right)$: $i\sigma_y$ for $|b_0\rangle$, $-\sigma_z$ for $|b_1\rangle$, $\sigma_x$ for $|b_2\rangle$,
or $-\mathbf 1$ for $|b_3\rangle$. 
(Actually, the overall signs [signs that multiply both $a$ and $b$ equally] don't matter, since 
$-|\psi\rangle$ is the same state as $|\psi\rangle$. So, for example, multiplication by $\sigma_z$ or
by $\mathbf 1$ is sufficient.)

To summarize, Alice and Bob share an entangled state $|\theta\rangle$ of two qubits.  Alice wishes to teleport an unknown
state $|\psi\rangle$ to Bob.  To do this, she first performs a measurement of $|\psi\rangle \otimes |\theta\rangle$ in
the Bell basis on her two qubits (the unknown state, and her qubit in the entangled state).  She transmits the information of which Bell state she obtained to Bob.  Bob applies the corresponding Pauli spin operator to his qubit
and recovers the unknown state $|\psi\rangle$.

$\mathbf{Pseudo-telepathy\mbox{ }}$. `Entanglement is perhaps the most non-classical manifestation of quantum
mechanics.  Among its many interesting applications to information processing, it can be harnessed to \emph{reduce}
the amount of communication required to proces a variety of distributed computational tasks.  Can it be used
to \emph{eliminate} communication altogether?  Even though it cannot serve to signal information between remote
parties, there are distributed tasks that can be performed without any need for communication, provided the parties
share prior entanglement: this is the realm of \emph{pseudo-telepathy}.' \cite{BBT}

Consider the following \emph{Pseudo-Telepathy Game} $\Gamma_N$ between $N$ players.  Since there are more than two players, we can't call them Alice and Bob, so we'll let them all be subscript Alices:  $A_1, A_2, \cdots, A_N$. There are also two functions $f$ and $g$, each of which take $N$-qubit inputs. The game has the following steps.

\emph{Step 1}:  The players mingle, discuss strategy, share random variables (in the classical
setting) or entanglement (in the quantum setting).  

\emph{Step 2}:  The players separate and are not allowed to engage in any form of communication.  Each player $A_i$
is given a single qubit input $x_i$ and requested to produce the single qubit output $y_i$.  The players \emph{win} 
$+1$ if
\begin{equation}
f(x_1, x_2, \cdots, x_N) = g(y_1, y_2, \cdots, y_N) .
\end{equation}
else they lose this amount.  
The functions $f$ and $g$ are defined as followings.  Players are guaranteed that the sum of the qubits they are
given is an even number: $\sum_i x_i$ is even.  (Think of what this means.  If $\sum_i x_i$ is even, then it
is divisible by 2.  Thus $\frac{1}{2} \sum_i x_i$ is a whole number that is either odd or even.  If odd, then $\frac{1}{2} \sum_i x_i\mbox{ mod } 2 = 1$.  If even, then $\frac{1}{2} \sum_i x_i \mbox{ mod } 2 = 0$.  But 
the latter case means $\frac{1}{2} \sum_i x_i \mbox{ mod }2$ is also divisible by two, so that the original sum $\sum_i x_i$ was divisible by 4.)  The players are asked to produce an even sum of output bits $\sum_i y_i$ if and only if the sum of the input bits $\sum_i x_i$ is divisible by 4.  Thus the criterion for the $N$-players to win is:
\begin{equation}
\sum_i y_i \mbox{ mod } 2 = \frac{1}{2} \sum_i x_i \mbox{ mod } 2 .
\end{equation}
The left-hand side of this equation is $g$ and the right-hand side $f$.  A win depends solely on the global state of
the $N$ qubits, even though each player controls only $1$ qubit, and is not allowed to communicate with the other
players.  Note that the expected payoff to the players if any player $i$ randomizes the submission of $y_i$ is $0$, as mod $2$ produces only two outcomes. This is a very nice game, because it highlights
the issue of cooperation between players, and because the game is scalable to any number $N$ of players.

Now, the amazing thing is that if the players are allowed to share prior entanglement, as in Step 1, then they
always win $\Gamma_N$.  To see how they do this, we need as components the Bell states $|b_0\rangle$ and $|b_2\rangle$,
the Hadamard transform $H$, and the unitary or rotation matrix introduced in the Card Game, except here
we will define it as:
\begin{equation}
U_{\frac{\pi}{2}} = \left( \begin{array}{cc}
1 & 0 \\
0 & e^{i\frac{\pi}{2}} \end{array} \right)
=
\left( \begin{array}{cc}
1 & 0 \\
0 & i \end{array} \right),
\end{equation}
remembering that $cos(\frac{\pi}{2}) + i \mbox{ }sin(\frac{\pi}{2}) = i$. Note that $U_{\frac{\pi}{2}} |0\rangle =
|0\rangle$ but $U_{\frac{\pi}{2}} |1\rangle = i\mbox{ }|1\rangle$.

Since $N$ players share the entangled Bell states, the latter will have to be $N$-qubit Bell states.
Let's write our $N$-qubit Bell states in the following simplified form:
\begin{eqnarray}
|b_0^N\rangle = \frac{1}{\sqrt 2}(|0^N \rangle + |1^N \rangle)\\
|b_2^N\rangle = \frac{1}{\sqrt 2}(|0^N \rangle - |1^N \rangle).
\end{eqnarray}
The first $N$-qubit state, $|b_0^N\rangle$ is the entangled state that all players agree to share.  The second state
may evolve in the course of play.  

Consider now the effect of the unitary matrix operating on a single qubit of $|b_0^N\rangle$:
\begin{equation}
U_{\frac{\pi}{2}} |b_0^N\rangle = \frac{1}{\sqrt 2}(|0^N \rangle + i\mbox{ }|1^N \rangle).
\end{equation}
The powers of $i$ are $i,\mbox{ } i^2 = -1,\mbox{ } i^3 = -i,\mbox{ } i^4 = 1$.  So if $U_{\frac{\pi}{2}}$ is applied to two
qubits, the sign on $|1^N\rangle$ becomes $-1$, and thus $|b_0^N\rangle \rightarrow |b_2^N\rangle$.  If
applied to four qubits, the sign is unchanged, so $|b_0^N\rangle \rightarrow |b_0^N\rangle$.  So if
$m$ players apply $U_{\frac{\pi}{2}}$ to their individual qubits, the initial state $|b_0^N\rangle$ will
remain unchanged if $m = 0\mbox{ mod }4$.  If $m = 2\mbox{ mod }4$, then $|b_0^N\rangle \rightarrow |b_2^N\rangle$.

If each player applies the Hadamard matrix to his qubit when the entangled state is $|b_0^N\rangle$, the
result is a superposition of all states \emph{with an even number of 1 bits}:
\begin{equation}
(H \otimes^N) |b_0^N\rangle = \frac{1}{\sqrt{2^{N-1}}} \sum_{even\mbox{ }bit\mbox{ } y}^{2^N-1} |y\rangle .
\end{equation}
Note that this does \emph{not} mean the states $|y\rangle$ in the summation are even numbers.  For example, $|101\rangle = |5\rangle$ is an
odd number, but has an even number of 1 bits, while $|100\rangle = |4\rangle$ is an even number, but has an
odd number of 1 bits.  To see that the $N$-fold Hadamard transform (the Walsh transform) turns Bell state $|b_0^N\rangle$
into a superposition of even-bit numbers (meaning an even number of 1 bits), consider Table XIII, which is an analog of Table V.
\begin{table}[ht]
\begin{tabular}{|r|r|r|r|}
\hline
$|b\rangle$ & $|y\rangle$ & $b\cdot y$ & $(-1)^{b \cdot y}$\\
\hline
$|111\rangle$ & $|000\rangle$ & $0$ & $1$\\
$|111\rangle$ & $|001\rangle$ & $1$ & $-1$\\
$|111\rangle$ & $|010\rangle$ & $1$ & $-1$\\
$|111\rangle$ & $|011\rangle$ & $0$ & $1$\\
$|111\rangle$ & $|100\rangle$ & $1$ & $-1$\\
$|111\rangle$ & $|101\rangle$ & $0$ & $1$\\
$|111\rangle$ & $|110\rangle$ & $0$ & $1$\\
$|111\rangle$ & $|111\rangle$ & $1$ & $-1$\\
\hline
\end{tabular}
\caption{Walsh transform with intitial qubit $|111\rangle$}
\end{table}
Note that the minus signs appear on the numbers with an odd number of 1 bits.  So if we apply $(H\otimes H\otimes H)$ to
$\frac{1}{\sqrt 2}(|000\rangle + |111\rangle)$, we get $
\frac{1}{\sqrt {2^4}} (|0\rangle + |1\rangle + |2\rangle + |3\rangle + |4\rangle + |5\rangle + |6\rangle + |7\rangle + |0\rangle - |1\rangle - |2\rangle + |3\rangle - |4\rangle + |5\rangle + |6\rangle - |7\rangle) = 
\frac{2}{\sqrt {2^4}} (|0\rangle + |3\rangle + |5\rangle + |6\rangle$), a superposition of numbers all of which have an even number of 1 bits.

If the state has evolved to the state $|b_2^N\rangle$ due to player action, and each player applies the
Hadamard matrix to his qubit, then the result is a superposition of all odd bit states (meaning states with
an odd number of 1 bits):
\begin{equation}
(H \otimes^N) |b_2^N\rangle = \frac{1}{\sqrt{2^{N-1}}} \sum_{odd\mbox{ }bit\mbox{ } y}^{2^N-1} |y\rangle .
\end{equation}
 
So here, then, are the steps each player takes with respect to his or her qubit in the game $\Gamma_N$:

\emph{Player Step 2a:} If a player receives qubit $x_i =1$, the player applies $U_{\frac{\pi}{2}}$ to his or 
her qubit in the entangled Bell state $|b_0^N\rangle$.  Otherwise the player does nothing.  \emph{Consequence:} 
Because the sum of bits $\sum_i x_i$ is even, an
even number of players will perform this step.  If $\sum_i x_i$ is divisible by 4, then the Bell state
$|b_0^N\rangle$ is left unchanged.  But if $\sum_i x_i = 2 \mbox{ mod } 4$ then $|b_0^N\rangle \rightarrow |b_2^N\rangle$.

\emph{Player Step 2b:} Each player applies the Hadamard matrix $H$ to his or her qubit. \emph{Consequence:}  
If the entangled
state is still in the state $|b_0^N\rangle$ from Step 2a, then this present step transforms the entangled state into a superposition of all \emph{even} bit states.  But if the entangled state has been transformed into $|b_2^N\rangle$, then this step transforms the entangled state into a superposition of all \emph{odd} bit states.

\emph{Player Step 2c:}  Each player now measures his qubit in the computational basis ($|0\rangle$ vs. $|1\rangle$)
to produce $y_i$.  

If $\sum_i x_i$ was divisible by 4, the entangled qubit is in a superposition of even bit states, so will be
projected under the measurement  into a number with an even number of 1 bits.  The players win, because
$\sum_i y_i\mbox{ mod } 2 = 0$.  If $\sum_i x_i = 2\mbox{ mod }4$, then the entangled qubit is in a superposition 
of odd bit states, so will be projected under the measurement into a number with an odd number of 1 bits.  The players win again, because $\sum_i y_i\mbox{ mod } 2 = 1$.

The players have demonstrated pseudo-telepathy by acting as though each knew what the other was doing, even
though there was no communication between players.  This was made possible by the shared entangled state $|b_0^N\rangle$
acting as a quantum invisible hand.

We may characterize this pseudo-telepathy game in terms of traditional $N$-person game theory as follows.  No player can secure any value by himself, so the value of a one-person coalition $\{i\}$ is $0$: $v\{i\} = 0$.  The value of the coalition of all players is $1$: $v(N) = 1$.  Such a game is said to
be in $(0,1)$-\emph{normalization}. Let $S$ be a subset of the set of players $N$.  If for all $S \subset N$ either
$v(S) = 0$ or $v(S) = 1$, a game is said to be \emph{simple}.  Thus the pseudo-telepathy game is also simple;
indeed $v(S) = 0$ for all $S$ save $S = N$.  Finally, a game is said to be constant sum if $v(S) + v(N-S) = v(N)$.
The pseudo-telepathy game is \emph{not} constant sum, as $v(S) + v(N-S) = 0$ for $S \ne N$, but $v(N) = 1$.

The set of imputations for this game is the set of probability vectors $P = \{p_1, p_2, \cdots, p_N\}$.  This fulfills
the requirement that $\sum_{i \in N} p_i = v(N) = 1$, and also the requirement that $p_i \ge v(\{i\}) = 0$, for all
$i \in N$.  None of these allocation vectors is dominated by another, for $S \subset N$. Thus the \emph{core} of this game is the convex set of probability vectors $P$.   

\subsection{Quantum secret sharing}
The IRA has some secret information they want to preserve among their members, but are fearful that some
of them may be MI5 informants, and that others may be arrested and reveal what they know under interrogation.
So they need a secure way to embed the secret among themselves.  A $(k,n)$ \emph{threshold} scheme \cite{CGL} is one
in which any $k \le n$ members can reconstruct a secret, but $k-1$ members cannot find \emph{any} information
about the secret at all.  

Let's first, however, consider a simple example where two parties must cooperate to discover a secret quantum
state \cite{HBB}.  Alice, Bob, and Gerald share the following entangled state (the left qubit is Alice's, the
right qubit is Gerald's):
\begin{equation}
|\psi\rangle = \frac{1}{\sqrt 2}(|000\rangle + |111\rangle).
\end{equation}
First note we can rewrite this in terms of a different basis.  Let
\begin{eqnarray}
|x^+\rangle = \frac{1}{\sqrt 2}(|0\rangle + |1\rangle)\\
|x^-\rangle = \frac{1}{\sqrt 2}(|0\rangle - |1\rangle) .
\end{eqnarray}
This implies the reciprocal relations
\begin{eqnarray}
|0\rangle = \frac{1}{\sqrt 2}(|x^+\rangle + |x^-\rangle)\\
|1\rangle = \frac{1}{\sqrt 2}(|x^+\rangle - |x^-\rangle) .
\end{eqnarray}
So the original state in terms of the new basis would be
\begin{equation}
|\psi\rangle = \frac{1}{2\sqrt 2} [(|x^+ x^+\rangle + |x^- x^-\rangle)(|0\rangle + |1\rangle)
                                 + (|x^+ x^-\rangle + |x^- x^+\rangle)(|0\rangle - |1\rangle)] .
\end{equation}
Alice wishes to send a secret qubit $|\phi_{secret}\rangle = a |0\rangle + b|1\rangle$ to Bob and Gerald in such a way that Bob and Gerald must
cooperate in order to learn the secret.  She essentially does this through the teleportation protocol,
but we will also need the definitions of $(|x^+\rangle,|x^-\rangle)$ for part of the procedure.  Alice combines the secret qubit $|\phi_{secret}\rangle$ with the shared state $|\psi\rangle$ to form the overall state
\begin{equation}
|\phi_{secret}\rangle \otimes |\psi\rangle = \frac{1}{\sqrt 2}(a |0000\rangle + b |1000\rangle
+ a |0111\rangle + b |1111\rangle) .
\end{equation}
Alice now rewrites this in terms of the Bell basis.  The multipliers on the Bell states are:
\begin{eqnarray}
\langle b_0 |(|\phi_{secret}\rangle \otimes |\psi\rangle) \rangle = \frac{a}{2} |00\rangle + \frac{b}{2} |11\rangle \\
\langle b_1 |(|\phi_{secret}\rangle \otimes |\psi\rangle) \rangle = \frac{a}{2} |11\rangle + \frac{b}{2} |00\rangle \\
\langle b_2 |(|\phi_{secret}\rangle \otimes |\psi\rangle) \rangle = \frac{a}{2} |00\rangle - \frac{b}{2} |11\rangle \\
\langle b_3 |(|\phi_{secret}\rangle \otimes |\psi\rangle) \rangle = \frac{a}{2} |11\rangle - \frac{b}{2} |00\rangle .
\end{eqnarray}
Alice now measures her two qubits in the Bell basis, sends the result to Gerald, and tells Bob to measure
his qubit in the $(|x^+\rangle,|x^-\rangle)$ basis.  After Alice's Bell measurement, the qubits of Bob and
Gerald will be in one of the following states:
\begin{eqnarray}
|b_0\rangle \rightarrow a |00\rangle + b |11\rangle \\
|b_1\rangle \rightarrow a |11\rangle + b |00\rangle \\
|b_2\rangle \rightarrow a |00\rangle - b |11\rangle \\
|b_3\rangle \rightarrow a |11\rangle - b |00\rangle .
\end{eqnarray}
If Bob gets $|x^+\rangle$ upon his measurement, then Gerald's qubit becomes
\begin{eqnarray}
a |00\rangle + b |11\rangle \rightarrow a |0\rangle + b |1\rangle\\
a |11\rangle + b |00\rangle \rightarrow a |1\rangle + b |0\rangle\\
a |00\rangle - b |11\rangle \rightarrow a |0\rangle - b |1\rangle\\
a |11\rangle - b |00\rangle \rightarrow a |1\rangle - b |0\rangle
\end{eqnarray}
while if Bob gets $|x^-\rangle$, Gerard's qubit becomes
\begin{eqnarray}
a |00\rangle + b |11\rangle \rightarrow a |0\rangle - b |1\rangle\\
a |11\rangle + b |00\rangle \rightarrow -a |1\rangle + b |0\rangle\\
a |00\rangle - b |11\rangle \rightarrow a |0\rangle + b |1\rangle\\
a |11\rangle - b |00\rangle \rightarrow -a |1\rangle - b |0\rangle .
\end{eqnarray}
To reconstruct Alice's qubit, Gerald needs to know what measurement Bob obtained, so that Gerald can apply
the appropriate Paul spin matrix to his final qubit state.  Thus Gerald and Bob together can reconstruct
Alice's qubit, but neither can do so alone.  The appropriate Pauli spin matrices to be applied to Gerald's final
state are:
\begin{table}[ht]
\begin{tabular}{|r|r|r|}
\hline
Bell$\backslash$ Bob & $|x^+\rangle$ & $|x^-\rangle$\\
\hline
$|b_0\rangle$ & $\mathbf 1$ & $\sigma_z$\\
$|b_1\rangle$ & $\sigma_x$ & $\sigma_x \sigma_z$\\
$|b_2\rangle$ & $\sigma_z$ & $\mathbf 1$\\
$|b_3\rangle$ & $\sigma_z \sigma_x$ & $-\sigma_x$\\
\hline
\end{tabular}
\caption{Pauli spin matrix to be applied to Gerald's final qubit state}
\end{table}

Now that we have seen the close relation of quantum secret sharing to teleportation, at least in one example,
let's return to the $(k,n)$ threshold notion, and consider an example of a $(2,3)$ threshold scheme.  This scheme works
by splitting up a state among three parties in such a way that any two can reconstruct the original state.
We begin with an unknown secret state that is not a qubit, but rather a \emph{qutrit}.  A qutrit is a ternary
`trit' that can take values in the three-dimensional Hilbert space spanned by $(|0\rangle, |1\rangle,
|2\rangle)$.  We've simply added one more dimension to a qubit.  Note that for this example, tensor products
expand by powers of 3, so 3 qutrits occupy a Hilbert space of dimension 27: $\mathbf {H_{27} = H_3 \otimes
H_3 \otimes H_3}$. 

We have an secret state $|\phi_{secret}\rangle = \alpha |0\rangle + \beta |1\rangle + \gamma |2\rangle$.  We
have an encoding transformation that maps this 1-qutrit state into a mixed 3-qutrit state:
\begin{equation}
|\phi_{secret}\rangle \rightarrow \alpha (|000\rangle +|111\rangle +|222\rangle) + \beta (|012\rangle +|120\rangle
+|201\rangle) + \gamma (|021\rangle +|102\rangle +|210\rangle).
\end{equation}
Now we can split this mixed 3-qutrit state between Alice, Bob, and Gerald.  The left qutrit belongs
to Alice, and the right qutrit to Gerald.  Given their qutrits, no one has any idea about the original
state, because the state they posses has an equal mixture of $|0\rangle$, $|1\rangle$, and $|2\rangle$.
However, any two people can reconstruct the secret state $|\phi_{secret}\rangle$.  For example, Alice and
Bob get together.  Alice adds her qutrit to Bob's modulo 3, then Bob adds his (new) qutrit to Alice's.  The result
is the state
\begin{equation}
(\alpha |0\rangle + \beta |1\rangle + \gamma |2\rangle) (|00\rangle + |12\rangle + |21\rangle).
\end{equation}
To see this, let's consider just the multipliers on $\alpha$.  When Alice and Bob get together, they
have
\begin{equation}
\alpha(|000\rangle + |111\rangle + |222\rangle) + \cdots.
\end{equation}
Adding Alice's qutrit to Bob's modulo 3 we get
\begin{equation}
\alpha(|000\rangle + |111\rangle + |222\rangle) + \cdots \rightarrow \alpha(|000\rangle + |121\rangle + |212\rangle)+ \cdots.
\end{equation}
Then adding Bob's (new) qutrit to Alice's we get
\begin{eqnarray}
\alpha(|000\rangle + |121\rangle + |212\rangle) + \cdots \rightarrow \alpha(|000\rangle + |021\rangle + |012\rangle)+ \cdots\\
= (\alpha |0\rangle + \cdots) (|00\rangle + |12\rangle + |21\rangle) .
\end{eqnarray}
Alice's qutrit is now identical with the secret state $|\phi_{secret}\rangle$, which has been disentangled from 
the other qutrits.  By a similar process Gerald and Bob could recover the secret state, or Alice and Gerald.

\subsection{The density matrix and quantum state estimation}
The `No Cloning Theorem' forbids a quantum copier of the following sort: the copier takes one quantum state as
input and outputs two systems of the same kind.  The no cloning theorem got its name after Nick Herbert proposed
a faster-than-light communication device, published in \emph{Foundations of Physics} in 1982 \cite{NH82}.  This generated
widespread attention and a flaw in the argument was soon found: the device required quantum cloning, and there
were problems with producing identical copies of a quantum state.  (Further background is found in \cite{AP2004}.) 

However, that is not the whole story.  Preparing virtually identical copies is no problem, if we don't try
to do it in a single measurement.  By statistical procedures the input state can be determined to any degree 
of accuracy.  For example, for the unknown state $|\psi\rangle$,
\begin{equation}
|\psi\rangle = a |0\rangle + b |1\rangle
\end{equation}
repeated measurement of $n$ such prepared states in the computational basis will yield $|0\rangle$ $n_a$ times 
and $|1\rangle$ $n_b$ times,
where $n_a + n_b = n$.  Then clearly
\begin{eqnarray}
\frac{n_a}{n} \simeq |a|^2 = |\langle \psi|0\rangle|^2\\
\frac{n_b}{n} \simeq |b|^2 = |\langle \psi|1\rangle|^2 .
\end{eqnarray}
That is, the $n$ measurements will yield $(x_1,x_2,\cdots,x_n)$, where each $x_i$ is either $0$ or $1$.  This
corresponds to a set of Bernoulli trials whose Likelihood Function is
\begin{equation}
L(p) = \prod_{i=1}^n p^{x_i}q^{1-x_i} = p^{\sum x_i} q^{n-\sum x_i} .
\end{equation}
where p is the probability of $1$ and $q = 1-p$ is the probability of $0$.  Maximizing $L(p)$ yields the
estimate for $p$ as
\begin{equation}
\hat{p} = \frac{1}{n} \sum x_i = \frac{n_b}{n}.
\end{equation}

This leads to the statistically-based \emph{density matrix} $\rho$:
\begin{equation}
\rho = \left (\begin{array}{cc}
\frac{n_a}{n} & 0 \\
0 & \frac{n_b}{n} \end{array} \right) =
 \frac{n_a}{n} \left (\begin{array}{cc}
1 & 0 \\
0 & 0 \end{array} \right) +
 \frac{n_b}{n} \left (\begin{array}{cc}
0 & 0 \\
0 & 1 \end{array} \right) =
 \frac{n_a}{n} |0\rangle \langle 0| + \frac{n_b}{n} |1\rangle \langle 1|.
\end{equation}
From the statistical point of view, the quantum state is a mathematical encoding of all data that can be collected this way.

Before proceeding further we need to explain the differences between \emph{pure} states and \emph{mixed} states.
If a quantum state $|\psi\rangle$ is a convex combination of other quantum states, it is said to be in a
\emph{mixed} state.  Note that mixture involves classical probabilities or combinations, not amplitudes. But if a state
$|\psi\rangle$ cannot be expressed as a convex combination of other states, it is said to be in a
\emph{pure} state.  Pure states are the extreme points of a convex set of states.

For a pure state $|\phi\rangle$, the ket-bra $|\phi\rangle\langle\phi|$ is called a \emph{projection
operator}.  It projects $|\phi\rangle$ onto itself ($|\phi\rangle\langle\phi|\phi\rangle = |\phi\rangle$), and any state 
$|\theta\rangle$ orthogonal to $|\phi\rangle$ is projected onto $0$
($|\phi\rangle\langle\phi|\theta\rangle = 0$).  For a pure state $\phi$, the density matrix is simply
$\rho = |\phi\rangle\langle\phi|$.
For a mixed state, where the system will be found in one of the extreme points $|\phi_j\rangle$ with probability $p_j$, the \emph{density matrix} $\rho$ is defined as the sum of the projectors weighted with the respective probabilities:
\begin{equation}
\rho = \sum_j p_j |\phi_j\rangle\langle\phi_j| .
\end{equation}
Since the probabilities are non-negative and sum to one, this means $\rho$ is a positive semidefinite Hermitian
operator (the eigenvalues are non-negative) and the trace of $\rho$ (the sum of the diagonal elements of the matrix,
i.e. the sum of its eigenvalues) is equal to one.

For example, let the pure state $|\psi\rangle$ be $|\psi\rangle = a |0\rangle + b|1\rangle$, where $a$ and $b$ are
complex numbers with respective complex conjugates $a^*$ and $b^*$.  Then the density matrix $\rho$ for
$|\psi\rangle$ is
\begin{equation}
\rho = |\psi\rangle \langle \psi| = 
\left( \begin{array}{cc}
aa* & ab* \\
ba* & bb* \end{array} \right).
\end{equation}
For $a = \sqrt{\frac{2}{3}}$, $b = \sqrt{\frac{1}{3}}$, this becomes
\begin{equation}
\rho = |\psi\rangle \langle \psi| = 
\left( \begin{array}{cc}
\frac{2}{3} & \frac{\sqrt 2}{3} \\
\frac{\sqrt 2}{3} & \frac{1}{3} \end{array} \right).
\end{equation}
A measurement of $|\psi\rangle$ in the computational basis will yield $|0\rangle$ with probability 
$\frac{2}{3}$ or $|1\rangle$ with
probability $\frac{1}{3}$.  These probabilities are found in the trace of $\rho$. We may rewrite $\rho$
as $\rho = \frac{2}{3} |0\rangle \langle 0| + \frac{1}{3} |1\rangle \langle 1|$, losing any information in
the off-diagonal elements. (This is what happens, as we shall see, during cloning.) Note that \emph{after}
the measurement, then either $|\psi\rangle = |0\rangle$ with probability 1, or $|\psi\rangle = |1\rangle$
with probability 1.

As another example, suppose $\frac{3}{4}$ of the states in an ensemble of states are prepared in the state
$|\psi_1 \rangle = .8 |0\rangle + .6 |1\rangle$, while $\frac{1}{4}$ are prepared in the state
$|\psi_2 \rangle = .6 |0\rangle - .8 i |1\rangle$.  Then the density matrix for this mixed ensemble, using
equation (240), is
\begin{equation}
\rho = .75 |\psi_1 \rangle \langle \psi_1| + .25 |\psi_2 \rangle \langle \psi_2| =
\left( \begin{array}{cc}
.57 & .36 + 12 i \\
.36 - 12 i & .43 \end{array} \right).
\end{equation}
A particle drawn from this ensemble and measured in the $(|0\rangle,|1\rangle)$ basis will be found
in state $|0\rangle$ with probability .57 or in state $|1\rangle$ with probability .43.  But if we
wanted to use $\rho$ to find the probabilities for a \emph{different} basis, we would need the off
diagonal elements as well as the trace.  To see this, suppose we draw a particle from the same
ensemble and take a measurement in the orthonormal basis $(|\phi_1\rangle,|\phi_2\rangle)$, where
$|\phi_1\rangle = .6 |0\rangle + .8 |1\rangle$ and $|\phi_2\rangle = .8 |0\rangle - .6 |1\rangle$.  Note that $\langle\phi_1|\phi_2\rangle = 0$ and $|\langle\phi_1|\phi_1\rangle|^2 =  |\langle\phi_2|\phi_2\rangle|^2 = 1$.  
Then $\rho$ gives as the probabilities $P$ of observing $|\phi_1\rangle$ and $|\phi_2\rangle$ as
\begin{eqnarray}
P(|\phi_1\rangle) = (.6,\mbox{ }.8) \rho \left( \begin{array}{c}
.6 \\
.8 \end{array} \right) = .826\\
P(|\phi_2\rangle) = (.8,\mbox{ }-.6) \rho \left( \begin{array}{c}
.8 \\
-.6 \end{array} \right) = .174   .
\end{eqnarray}

Suppose we choose an observable $\aleph$, such as the spin state of an electron. Then in the von Neumann formulation
of quantum measurement, each observable is associated with a Hermitian operator $A$, with $A |\psi_j\rangle = a_j |\psi_j\rangle$, where $|\psi_j\rangle$ are the eigenvectors of $A$, and 
$a_j$ are the eigenvalues.  Thus, using the same basis for $\rho$ and $A$, namely the eigenvectors of $A$, we have
\begin{equation}
A \rho = \sum_j p_j A |\psi_j\rangle\langle\psi_j| = \sum_j p_j a_j |\psi_j\rangle\langle\psi_j|.
\end{equation}
Now the expected value of $A$, $\overline{A}$, is simply
\begin{equation}
\overline{A} = \sum_j p_j\mbox{ } a_j.
\end{equation}
Thus the latter may be represented as
\begin{equation}
\overline{A} = \mbox{ trace }(A\rho).
\end{equation}

There are many approaches to \emph{quantum state estimation} via the density matrix $\rho$. The problem of state
estimation is closely related to the problem of cloning, and is connected to issues of entanglement.
The \emph{maximum likelihood} approach considered earlier is probably the best.  For the heuristic purposes
of this essay a \emph{Bayesian} framework \cite{MSred} is revealing.  We might start with the principle of indifference, or insufficient reason, and make the initial assumption that the density matrix has the fully mixed form (for a system in $\mathbf H_2$):
\begin{equation}
\rho = \frac{1}{2} \mathbf 1 = \left( \begin{array}{cc}
\frac{1}{2} & 0 \\
0 & \frac{1}{2} \end{array} \right).
\end{equation}
This corresponds to an ensemble, half of which are in an up state and half of which are in a down state:
\begin{equation}
\rho = \frac{1}{2} |u\rangle \langle u| + \frac{1}{2} |d\rangle \langle d|
= \frac{1}{2} \left( \begin{array}{c}
1 \\
0 \end{array} \right) (1\mbox{ }0) + \frac{1}{2} \left( \begin{array}{c}
0 \\
1 \end{array} \right) (0\mbox{ }1) = \frac{1}{2}\left( \begin{array}{cc}
1 & 0 \\
0 & 0 \end{array} \right) + \frac{1}{2}\left( \begin{array}{cc}
0 & 0 \\
0 & 1 \end{array} \right) =  \frac{1}{2} \mathbf 1 .
\end{equation}

Or we may start with the general form of the density matrix, which can be written in terms of the Pauli spin 
matrices and real numbers $r_x$, $r_y$, and $r_z$ as follows:
\begin{eqnarray}
\rho = \frac{1}{2} (\mathbf 1 + \mathbf r \cdot \sigma)\\
= \frac{1}{2} (\mathbf 1 + r_x\mbox{ } \sigma_x + r_y\mbox{ } \sigma_y + r_z\mbox{ } \sigma_z)\\
= \frac{1}{2} \left( \begin{array}{cc}
1+r_z & r_x-i r_y \\
r_x+ i r_y & 1-r_z \end{array} \right).
\end{eqnarray}
Here we require that the determinant of $\rho$ be non-negative, $\mathbf{det}\mbox{ }\rho \ge 0$, which implies 
$\frac{1}{4} [1-(r_x^2 + r_y^2 + r_z^2)] \ge 0$, or that 
$\mathbf r^2 = r_x^2 + r_y^2 + r_z^2 \leq 1$, so that each density matrix may be associated with a ball of radius $1$, called a \emph{Bloch sphere}.  Points on the surface of the ball correspond to pure states, while interior points correspond to mixed states.

If we assume this form of the density matrix $\rho$ and then measure spin in the $z$ direction, obtaining a
series of $n$ results $u$ and $d$ with frequencies $n_u$ and $n_d$, then the likelihood is
\begin{equation}
L(n_u) = [\frac{1}{2}(1+r_z)^{\frac{n_u}{n}}] [\frac{1}{2} (1-r_z)^{\frac{n - n_u}{n}}] .
\end{equation}

Now consider the following \emph{State Discrimination Game} $\Gamma_{sd}$. There are $N$ states, members of the
set $S = \{|\psi_j\rangle,\mbox{ } j=0,1,\cdots,N-1\}$.  Each of these states is represented by a density
matrix $\rho_j = \eta_j |\psi_j\rangle \langle\psi_j|$.  Alice prepares a state $\rho_k$, unknown to Bob, and 
forwards it to Bob, along with the information that the associated $|\psi_k\rangle$ is a member of $S$.  
She also tells him the probabilities $\eta_j$ of each state in $S$.

The $\eta_j$ are called \emph{prior probabilities}. This, of course, immediately suggests a Bayesian framework,
so let's consider a Bayesian strategy called \emph{quantum hypothesis testing} \cite{AC}. Because there are $N$ 
states, Bob will follow a procedure that gives him $N$ outcomes, which we will label $a_j$.  If Bob obtains outcome
$a_m$ he will assume that the state he was sent was $\rho_m$.  There is error probability $p_E$ that $\rho_m \ne
\rho_k$ and probability $1-p_E = p_D$ that $\rho_m = \rho_k$.

To complete the game description, we need to define the \emph{channel matrix} $[h(a_m|\rho_k)]$ which expresses the
probabilities that Bob will find $a_m$ given that $\rho_k$ was sent, and the \emph{cost matrix} $[c_{mk}]$ which
assigns a cost to making the hypothesis $a_m$ when $\rho_k$ was sent. No matter what $\rho_k$ was sent, Bob's
measurement will yield one of the $a_m$.  This gives rise to the completeness condition that
\begin{equation}
\sum_{m=1}^N h(a_m|\rho_k) = 1.
\end{equation}
Then the total error probability is
\begin{equation}
p_E = 1 - \sum_{k=1}^N \eta_k h(a_k|\rho_k) .
\end{equation}
The average amount $c_B$ Bob will pay Alice is given by the Bayesian cost matrix
\begin{equation}
c_B = \sum_{mk} \eta_k c_{mk} h(a_m|\rho_k) .
\end{equation}
Bob's goal is to minimize $c_B$.  The only thing Bob controls are the elements in the channel matrix $h$.  
Thus Bob's problem is
\begin{equation}
min_{\mbox{ }\mathbf h}\mbox{ } \sum_{mk} \eta_k c_{mk} h(a_m|\rho_k).
\end{equation}
This puts quantum state \emph{discrimination} (finding a state in a given set of states) in the context of game
theory.  If we set the diagonal elements of the cost matrix equal to 0 (Bob pays nothing for being correct)
and the other elements equal to a constant $c$ (all errors cost the same) then, comparing equations (256) and
(257), Bob's problem reduces to
\begin{equation}
min_{\mbox{ }\mathbf h}\mbox{ } p_E .
\end{equation}

The number of states here is finite.  By contrast, in quantum state \emph{estimation} the set of states is
infinite.  Since a quantum state itself is not observable, quantum state estimation means estimating the
density matrix $\rho$ of the quantum state, as we have already seen. This, too, can be put in the context of 
game theory.

In the \emph{State Estimation Game} \cite{LJ} Alice chooses an arbitrary pure state $|\psi\rangle \in \mathbf H_d$ and sends $|\psi\rangle^{\otimes N}$ to Bob
and $|\psi\rangle$ to a referee. After receiving the $N$ states from Alice, Bob performs a measurement on them
and then sends a pure state $|\phi\rangle$ to the referee.  After receiving the two states from Bob and Alice, the referee compares them according to some criterion (see cloning, below), then awards a payoff to Alice if the
two states are not sufficiently close, or to Bob if they are. Of course Bob's task is to construct the best
quantum state measurement he can given the $N$ states received from Alice.

\subsection{Quantum cloning}
In econometrics one tries, by some procedure, to produce an estimate $\hat{a}$ of some unknown parameter $a$.
This can be considered an attempt, by our estimation procedure, to \emph{clone} the parameter $a$.  We don't
expect to achieve a perfect clone, but only a best estimate that lies within an interval of uncertainty. Which
brings us to the cloning of quantum states. The object of an \emph{optimal cloning device} \cite{RFW} is to prepare 
near copies as close to the original as possible. 

Optimal cloning can be formulated in terms of a quantum game, the \emph{Cloning Game}, played between Alice and 
Clare, the cloning queen. This game will have $N$ input systems and $M$ output systems.  We start with Alice, who has a pure state described by a density matrix $\rho$ in 2-dimensional Hilbert space $\mathbf H_2$.  She is going to run her state preparing procedure $N$ times, giving rise to a composite system in Hilbert space $\mathbf H_2\otimes^N$:
\begin{equation}
\mathbf 1_2\otimes^N \rho = \rho\otimes^N.
\end{equation}
Alice then ships $\rho\otimes^N$ off to Clare.  Clare uses a cloning device $T_m$ of her choice to produce
$M$ output systems $T_m \rho\otimes^N$.  Next, Alice produces $M$ copies of her original system, 
$\rho\otimes^M$.  The outcome of the game depends on
\begin{equation}
T_m \rho\otimes^N\mbox{ vs. }\rho\otimes^M .
\end{equation}
Since $T_m$ maps density matrices to density matrices, it is restricted to being a linear completely positive
trace preserving map.

One way of assigning payoffs to this game would be to base them on the norm difference
\begin{equation}
||T_m \rho\otimes^N\mbox{ - }\rho\otimes^M|| .
\end{equation}
Another way would be to use the \emph{fidelity}, based on $\mathbf{trace} (\rho\otimes^M T_m \rho\otimes^N)$.  This would be $1$ if the cloning machine were perfect.  The fidelity could depend on the input density matrix $\rho$.  Define $F(T)$ by
\begin{equation}
F(T) = \mbox{inf}_\rho \mbox{ trace }(\rho\otimes^M T_m \rho\otimes^N) < 1 .
\end{equation}
Then Clare's job is to maximize $F(T)$.  This makes the Cloning Game a maximin problem. A cloner is called
`universal' if the fidelity of the output clones is independent of the input state. The maximal
fidelity of cloning for a universal cloner is $\frac{5}{6}$, which can be achieved by unitary evolution
or by a teleportation scheme \cite{BDEFMS}.

A \emph{universal quantum cloner} of 1 qubit $\rightarrow$ 2 qubits is a quantum machine that takes as input
an unknown quantum state $|\psi\rangle$ and generates as output two qubits in a state that may be described
by a density matrix of the form $\rho = \eta |\psi\rangle \langle \psi | + (1-\eta) \frac{1}{2} \mathbf 1$. 
The parameter $\eta$ describes the shrinking of the original Bloch vector $\mathbf r$ corresponding to the
density operator $|\psi\rangle \langle\psi|$. For example, if $|\psi\rangle \langle\psi| = 
\frac{1}{2}(\mathbf 1 + \mathbf r \cdot \sigma)$, then $\rho = \frac{1}{2}(\mathbf 1 + \eta \mathbf r \cdot \sigma)$.  Then the optimal cloner involves maximizing the fidelity by maximizing $\eta < 1$:
\begin{equation}
\mbox{max}_\eta\mbox{ } F = \langle \psi|\rho|\psi\rangle = \frac{1}{2}(1+\eta) .
\end{equation}
A Bloch vector shrinkage of $\eta = \frac{2}{3}$ corresponds to the maximal fidelity of $\frac{5}{6}$.

The cloning process goes like this.  Let $|B\rangle$ denote the initial state of blank copies (the destination of the clones) plus any auxillary qubits (`ancilla') needed in the process.  The qubit $|\psi\rangle$ to be cloned  is encoded in the basis $(|0\rangle,|1\rangle)$. Then the universal quantum cloning machine (UQCM) transformation $T_{UQCM}$ performs the following transformations on the basis vectors or states:
\begin{eqnarray}
T_{UQCM} |0\rangle |B\rangle \rightarrow \sqrt{\frac{2}{3}} |0\rangle |0\rangle |A_\perp \rangle +
\sqrt{\frac{1}{6}} (|01\rangle + |10\rangle) |A\rangle \\
T_{UQCM} |1\rangle |B\rangle \rightarrow \sqrt{\frac{2}{3}} |1\rangle |1\rangle |A \rangle +
\sqrt{\frac{1}{6}} (|01\rangle + |10\rangle) |A_\perp \rangle .
\end{eqnarray}
Here $A$ and $A_\perp$ represent two possible orthogonal final states for the ancilla qubits.  Note that
this implies for the input state $|\psi\rangle$, the output
\begin{equation}
T_{UQCM} |\psi\rangle |B\rangle \rightarrow
\end{equation}
\begin{equation}
(\sqrt{\frac{2}{3}} |0\rangle |0\rangle |A_\perp \rangle +
\sqrt{\frac{1}{6}} (|01\rangle + |10\rangle) |A\rangle, \mbox{ }
\sqrt{\frac{2}{3}} |1\rangle |1\rangle |A \rangle +
\sqrt{\frac{1}{6}} (|01\rangle + |10\rangle) |A_\perp \rangle) \left( \begin{array}{c}
a \\
b \end{array} \right) .
\end{equation}

The next step is
to \emph{trace} over the ancilla qubits, which yields a two-qubit mixed state.  Then another trace is
performed with respect to each individual qubit, giving two copies of the same mixed one-qubit state,
which has a fidelity of $\frac{5}{6}$ when compared to the original state.

\subsection{Conclusion}
At this point the reader has enough background to start doing quantum game thory.  Of course, there is much
more to be said, as the references will indicate.  The reader is referred especially to the notes on quantum
computation \cite{EHI} \cite{ZM} \cite{JPres}.

This essay has demonstrated that traditional game theory is a subset of quantum game theory, and the latter
has a much richer structure and a broader set of outcomes.  That is all the justification required for doing
\emph{quantum} game theory.  Nothing is given up, and more is obtained by switching to the latter.
Therefore the study of traditional game theory is neither an evolutionarily stable strategy nor a Nash equilibrium,
and will be relegated to the dust-bin of extinct species and nonequilibrium payoffs.  That being said, can the current state of quantum game theory survive an invasion of mutants?  I hope those invading mutants will be mathematical economists coming to fix what's wrong with quantum mechanics.  Indeed, Lambertini \cite{LL} argues that
mathematical economics and quantum mechanics are isomorphic.

A quantum game $\Gamma = \Gamma(\mathbf H, \Lambda, U, \{s_i\}_j, \{\pi_i\}_j)$, where $\mathbf H$ is a Hilbert
space; $\Lambda$ is the initial state of the game; $U$ is a unitary matrix applied to all the player's qubits at
the beginning and end of the game; $\{s_i\}_j$ are the set of moves of player $j$, including convex combinations;
and $\{\pi_i\}_j$ are the set of payoffs to player $j$.  The purpose of the game is to endogenously determine
the strategies that maximize player $j$'s expected payoff.  Generally, a pure quantum move $s_i$ is a unitary matrix
applied to the player's individual qubit.

In the course of this essay, we have seen the Spin Flip game, the Guess a Number games I and II, the RSA
game, Prisoner's dilemma, Battle of the sexes, Newcomb's game, Evolutionarily stable strategy game, Coin flip game, Pseudo-telepathy game, and game theoretic aspects of Teleportation, Secret sharing, State estimation, and Quantum cloning. In the Spin Flip game, Bob was able to exploit quantum superposition via the Hadamard transform $H$ to always win the game, though to be sure this outcome was also dependent on the sequence of player moves. The key to Guess a Number Game I was use of the Grover search algorithm to rotate a state vector in Hilbert space to the approximate
location of the unknown number.  This search was speeded up from $N$ moves to $\sqrt N$ moves by the use of
superposition and calls to the $f_a$ oracle.  In the Guess a Number game II, the Bernstein-Vazirani oracle
was used to create the Walsh transform $W_{2^n}$ of the unknown number after a single call to the oracle. In the
RSA game, Shor's factoring algorithm was used to project a superimposed state of integers into, with high
probability, a number that is near an integer multiple of $\frac{2^{2n}}{r}$ for the given composite RSA prime $N = pq$,
where $r$ is the order of the tested element. The probability was controlled by use of the quantum Fourier transform.

In the Prisoner's dilemma game, we saw that the addition of quantum moves $H$ and $\sigma_z$ to $\mathbf 1$ and $\sigma_x$ added to the traditional game outcomes, and indeed attained a Pareto optimal point as a Nash equilibrium.  In the Battle
of the sexes game, the same quantum moves produced a unique Nash and Pareto optimal equilibrium in pure strategies;
and equality between Alice and Bob, also a Nash equilibrium and Pareto optimal, in mixed strategies.  Newcomb's
paradox was resolved by the Superior Being's ability to perfectly predict (control) Alice's choice through the use of
superposition, which replaced omniscience on the part of the Superior Being, and the incentive to cheat on the
part of Alice.  These games also show, through the use of the unitary matrix U, the partial irrelevance of the
categories `cooperative' and `noncooperative'.  If players' qubits are entangled in the game, there are hidden channels of communication (an invisible hand) when a player simply focuses on maximizing his or her own expected utility. In the Evolutionarily stable strategy game, invading mutants playing quantum moves were able to wipe out existing species playing only classical moves. The Coin flip game demonstrated the use of a quantum oracle, in a game without entanglement, to turn an unfair game into a fair one.  

In the Pseudo-telepathy game, communication among players was not necessary in order for them to conspire to win the game, as long as they shared a quantum entangled state. The game could be won with certainty with an implied coalition
of all $N$ players, while any proper subset of $N$ had expected payoff of $0$. We also saw that \emph{N-dimensional probability space} was the \emph{core} of the pseudo-telepathy game. Does this mean quantum entanglement gives rise to quantum probability? We saw that qubit states are unobservable, and under measurement are projected onto the measurement basis, typically $0$ or $1$, and hence destroyed.  This creates opportunity as well as difficulties.  Measurement in
the Bell basis is at the heart of the teleportation protocol.  And while quantum states can only be cloned with a
certain fidelity, they can be used for secret sharing and secure communication.  The problems of quantum state
discrimination using maximum likelihood in a Bayesian framework, or quantum state estimation using the same in
connection with the Bloch sphere representation of the density matrix, are not concepts fundamentally foreign to economists.  

Piotrowski and Sladkowski \cite{PS3} have stated what they called the \emph{Quantum anthropic principle}: 
Even if at earlier stages of civilization markets were governed by classical laws, the incomparable efficiency of quantum algorithms in conveying comparative advantage should result in market evolution such that quantum behaviors 
will prevail over classical ones.  Since nature already plays quantum games, it would appear that humans do so also
using their personal quantum computers (human brains). Thus, while speculative, Gottfried Mayer's comment
in \emph{Complexity Digest} is not so far fetched:
`It might be that while observing the due ceremonial of everyday market transactions we are in fact
observing capital flows resulting from quantum games eluding classical description.  If human decisions
can be traced to microsopic quantum events one would expect that nature would have taken advantage
of quantum computation in evolving complex brains.  In that sense one could indeed say that quantum
computers are playing their market games according to quantum rules.' \cite{GJM}


\begin{thebibliography}{99}

\bibitem{JSB}
Bell\, J.S., `On the Einstein Podolsky Paradox', \emph{Physics}, 1(3), 1964, 195-200.

\bibitem{BBCJPW}
Bennett\, Charles H., Gilles Brassard, Claude Cr\'{e}peau, Richard Jozsa, Asher Peres, William K. Wootters,
`Teleporting an unknown quantum state via dual classical and EPR channels', http://www.enricozimuel.net/documenti/BBC+93.ps .

\bibitem{BV}
Bernstein\, E. and U. Vazirani, `Quantum complexity theory', in \emph{Proceedings of the 25th Annual ACM
Symposium on the Theory of Computing}, San Diego, Calif., 16-18 May 1993, New York:ACM, 1993, 11-20,
http://www.cs.berkeley.edu/$\sim$vazirani/pubs/bv.ps

\bibitem{SJB}
Brams\, Steven J., \emph{Superior Beings: If they exist, how would we know?  Game theoretic implications
of omniscence, omnipotence, immortality, and incomprehensibility}, New York:Springer-Verlag, 1983.

\bibitem{BBT}
Brassard\, Gilles, Anne Broadbent, Alain Tapp, `Recasting Mermin's multi-player game into the framework
of pseudo-telepathy', arXiv: quant-ph/0408052 v1 6 Aug 2004. 

\bibitem{SLB}
Braunstein\, Samual L., `Quantum Computation', http://www-users.cs.york.ac.uk/$\sim$schmuel/comp
/comp\_best.pdf .

\bibitem{BK}
Braunstein\, Samuel L. and H. J. Kimble, `Teleportation of continuous quantum variables', \emph{Physical
Review Letters} 80, 4, 26 January 1998, http://www-users.cs.york.ac.uk/$\sim$schmuel/papers/bk98.pdf

\bibitem{BDEFMS}
Bru\ss\, Dagmar, David P. DiVincenzo, Artur Kert, Christopher A. Fuchs, Chiara Macchiavello, John A.
Smolin, `Optimal universal and state-dependent quantum cloning', arXiv: quant-ph/9705038 v3 6 Dec 1997.

\bibitem{AC}
Chefles\, Anthony, `Quantum state discrimination', arXiv: quant-ph/0010114 v1 31 Oct 2000.

\bibitem{CT}
Cheon\, Taksu and Izumi Tsutsui, `Classical and quantum contents of solvable game theory on Hilbert
space,' arXiv; quant-ph/0503233 v1 31 Mar 2005

\bibitem{CGL}
Cleve\, Richard, Daniel Gottesman, Hoi-Kwong Lo, `How to share a quantum secret', December 1998,
http://www.hpl.hp.com/techreports/98/HPL-98-205.pdf 

\bibitem{DeSc}
Debreu\, G. and H. E. Scarf, `A limit theorem on the core of an economy', \emph{International Economic
Review}, 4, 1963, 235-246.

\bibitem{DD}
Deutsch\, D., `Quantum Theory, the Church-Turing principle and the universal quantum computer', \emph{Proc.
Roy. Lond.} A400, 1985, 97-117.

\bibitem{DD89}
Deutsch, D., `Quantum computational networks,' Proceedings of the Royal Society of London, A425, 1989,
73-90.

\bibitem{DD2002}
Deutsch, D., `It from Qubit', Sept. 2002, http://www.qubit.org/people/david/Articles/ItFromQubit.pdf

\bibitem{DJ}
Deutsch\, D. and R. Jozsa, `Rapid solution of problems by quantum computation,' \emph{Proceedings Royal Society
London}, A400, 1992, 73-90.

\bibitem{JXHMXR}
Du\, Jianfeng, Xiaodong Xu, Hui Li, Mingjun Shi, Xianyi Zhou, Rongdian Han, `Quantum strategy without
entanglement', arXiv: quant-ph/0011078 v1 19 Nov 2000.

\bibitem{EPR}
Einstein\, A., B. Podolsky, N. Rosen, `Can quantum mechanical description of physical reality be considered complete?', 
\emph{Phys. Rev.} 47, 1935, 777-780.

\bibitem{EW}
Eisert\, Jens and Martin Wilkens, `Quantum Games,' arXiv:quant-ph/0004076 v1 19 Apr 2000.

\bibitem{EWL}
Eisert\, Jens, Martin Wilkens, and Maciej Lewenstein, `Quantum games and quantum strategies', arXiv:
quant-ph/9806088 v3 29 Sept 1999.

\bibitem{EHI}
Ekert\, Artur, Patrick Hayden and Hitoshi Inamori, \emph{Basic concepts in quantum computation}, arXiv:
quant-ph/0011013 v1 2 Nov 2000,

\bibitem{RF}
Feynman\, Richard P., `Simulating Physics with Computers,' \emph{International Journal of Theoretical Physcis}, 
21, 1982, 467.

\bibitem{PF}
Fishburn\, Peter C., `Expected utility theories: a review note', in R. Henn and O. Moeschlin, eds., \emph{Mathematical Economics and Game Theory: Essays in honor of Oskar Morgenstern}, Lecture Notes in Economics and Mathematical
Systems, 141, Berlin:Springer-Verlag, 1977.

\bibitem{DGale}
Gale\, David, \emph{The Theory of Linear Economic Models}, New York: McGraw-Hill, 1960.

\bibitem{NG2005}
Gisin\, Nicolas, `How come the correlations?' http://arxiv.org/ftp/quant-ph/papers/0503/0503007.pdf

\bibitem{GMerz}
Gogonea\, V. and K. M. Merz, `Fully quantum mechanical description of proteins in solution – combining linear scaling
quantum mechanical methodologies with the Poisson-Boltzmann equation', \emph{J. Phys. Chem. A}, 103 (1999) 5171–5188.

\bibitem{DGott}
Gottesman\, Daniel, `The Heisenberg representation of quantum computers', arXiv: quant-ph/9807006 v1
1 July 1998.

\bibitem{LG}
Grover\,Lov K., `A fast quantum mechanical algorithm for database search', arXiv: quant-ph/9605043.

\bibitem{HW}
Hardy\, G. H. and E. M. Wright, \emph{An Introduction to the Theory of Numbers}, Fifth edition, Oxford:Clarendon
Press 1979.

\bibitem{NH82}
Herbert,\ N. `FLASH--a superluminal communicator based upon a new type of quantum measurement",
\emph{Found. Phys.} 12, 1982, 1171.

\bibitem{HBB}
Hillary\, Mark, Vladimir Buzek, and Andre Berthiaume, `Quantum secret sharing', \emph{Physical Review A},
vol 59, no 3, March 1999, 1829-1834, http://www.quniverse.sk/buzek/mypapers/99pra1829.pdf

\bibitem{HM}
Hunziker\, Markus and David A. Meyer, `Quantum algorithms for highly structured search problems,' 
http://www3.baylor.edu/$\sim$Markus\_Hunziker/HunzikerMeyer2002.pdf .

\bibitem{IT}
Iqbal\, A. and A.H. Toor, `Evolutionary stable strategies in quantum games', arXiv: quant-ph/0007100 v3
11 Dec 2000.

\bibitem{JR}
Jaroszkiewicz\,George and Jason Ridgway-Taylor, `Quantum Computational Representation of the Bosonic
Oscillator', arXiv:quant-ph/0502166 v1 25 Feb 2005

\bibitem{Jammer}
Jammer\,Max, \emph{The Philosophy of Quantum Mechanics}, New York: Wiley, 1974.

\bibitem{JJ}
Johnson\, Joseph F., `The problem of quantum measurement', arXiv quant-ph/0502124 v1 21 Feb 2005.

\bibitem{LL}
Lambertini, Luca, `Quantum mechaics and mathematical economics are isomorphic,' 29 Feb 2000,
http://www.dse.unibo.it/wp/370.pdf

\bibitem{LJ}
Lee\, Chiu Fan and Neil F. Johnston, `Game theoretic discussion of quantum state estimation and cloning',
arXiv: quant-ph/0207139 v2 29 Nov 2002.

\bibitem{SL}
Lomonaco, Jr.\,Samuel J., `A lecture on Grover's quantum search algorithm', arXiv:quant-ph/0010040 v2 18 Oct 2000.

\bibitem{LR}
Luce\, R. Duncan and Howard Raiffa, \emph{Games and Decisions}, New York: Wiley, 1957.

\bibitem{MW}
Marinatto\, Luca and Tullio Weber, `A quantum approach to static games of complete information',
arXiv: quant-ph/0004081 v2 27 June 2000 .

\bibitem{GJM}
Mayer, Gottfried J., Editor's Note to \emph{Complexity Digest}, 27, 2 July 2001.

\bibitem{MSP}
Maynard Smith\, J. and G.R. Price, `The logic of animal conflict', \emph{Nature}, 246, 1973, 15-18.

\bibitem{MSETG}
Maynard Smith\, J., \emph{Evolution and the Theory of Games}, Cambridge: Cambridge University Press, 1982.

\bibitem{ZM}
Meglicki, Zdzislaw, `Introduction to quantum computing', February 5, 2002,
http://beige.ucs.indiana.edu/M743/M743.pdf .

\bibitem{DM}
Meyer\, David A., `Quantum Games and Quantum Algorithms', arXiv:quant-ph/0004092 v2, 3 May 2000.

\bibitem{MOR}
Milman\, P. H. Ollivier, and J. M. Raimond, `Universal quantum cloning in cavity QED',
http://www.imperial.ac.uk/physics/qgates/papers/ENS\_QG04.pdf, 23 Jan 2003.

\bibitem{NT}
Nawaz\, Ahmad and A. H. Toor, `Dilemma and Quantum Battle of the Sexes', arXchiv:quant-ph/0110096 v3,
26 Mar 2004.

\bibitem{vN28}
Neumann\, John von, `Zur Theorie der Gesellschaftspiele' \emph{Mathematische Annalen}, 1928.
100:295-320.

\bibitem{vN32}
Neumann\,John von, \emph{Mathematische Grundlagen der Quantenmechanik}, Berlin: Springer-Verlag, 1932.

\bibitem{vN37}
Neumann\, John von, `A Model of General Economic Equilibrium' (`\"{U}ber ein \"{o}konomisches
Gleichungssystem und eine Verallgemeinerung des Brouwerschen Fixpunktsatzes') in K. Menger, ed., \emph{Ergebnisse eines mathematischen Kolloquiums, 1935-36}, 1937.

\bibitem{vN56}
Neumann, John von, `Probabilistic logics and the synthesis of reliable organisms from unreliable components',
\emph{Automata Studies}, Princeton University Press, 1956, 329-378.

\bibitem{vNM}
Neumann\, John von and Oscar Morgenstern, \emph{The Theory of Games and Economic Behavior}, New York: Wiley, 1944.

\bibitem{OO}
Ore\, Oystein, \emph{Number Theory and Its History}, New York: Dover (reprint of New York: McGraw-Hill, 1948), 1988.

\bibitem{RP}
Penrose, Roger, \emph{The Emperor's New Mind}, Oxford: Oxford University Press, 1989.

\bibitem{AP2004}
Peres\, Asher, `How the no-cloning theorem got its name,' arXiv: quantum-ph/0205076 v1 14 May 2002.

\bibitem{PS1}
Piotrowski\, Edward W. and Jan Sladkowski, `An invitation to quantum game theory', arXiv: quant-ph/0211191 v1
28 Nov 2002.

\bibitem{PS2}
Piotrowski\, Edward W. and Jan Sladkowski, `Quantum solution to the Newcomb's paradox', arXiv: quant-ph/0202074 v1
13 Feb 2002.

\bibitem{PS3}
Piotrowski\, Edward W. and Jan Sladkowski, `Trading by quantum rules--quantum anthropic principle', 
http://alpha.uwb.edu.pl/ep/RePEc/sla/eakjkl/9.pdf .


\bibitem{SP}
Pirandola\, Stefano, `A quantum teleportation game', arXiv: quant-ph/0407248 v3 17 Nov 2004.

\bibitem{JPres}
Preskill\, John, `Lecture notes for Physics 229: quantum information and computation', California Institute
of Technology, September 1998, http://www.theory.caltech.edu/people/preskill/ph229/\#lecture

\bibitem{PS}
Shor\, P. W., `Algorithms for quantum computation: discrete logarithms and factoring', in \emph{Proc. 35th Annual
Symposium on the Foundations of Computer Science}, edited by S. Goldwasser, Los Alamitos, Calif.:IEEE Computer
Society Press, 1994, 124-134,  http://www.ennui.net/~quantum/papers/9508027.pdf .

\bibitem{MSred}
Srednick\, Mark, `Subjective and objective probabilities in quantum mechanics,' arXiv: quant-ph/0501009
v2 14 Jan 2005.

\bibitem{EGTSE}
Stanford Encyclopedia of Philsophy, `Evolutionary game theory', 
http://plato.stanford.edu/entries/game-evolutionary/ .

\bibitem{HS1}
Stapp\, Henry, `Why classical mechanics cannot naturally accomodate consciousness, but quantum meachanics can,'
http://psyche.cs.monash.edu.au/v2/psyche-2-05-stapp.html .

\bibitem{HS2}
Stapp\, Henry, \emph{The Mindful Universe}, http://www-physics.lbl.gov/~stapp/MUA.pdf

\bibitem{TC}
Turner\, P.E. and L. Chao, `Prisoner's dilemma in an RNA virus,' \emph{Nature}, 398(6726), April 1, 1999,
441-3.

\bibitem{SU}
Ulam, S.M., \emph{Adventures of a Mathematician}, New York:Charles Scribner's Sons, 1976.

\bibitem{RFW}
Werner\, R. F., `Optimal cloning of pure states', arXiv: quant-ph/9804001 v1 1 April 1998.

\bibitem{CZ}
Zalka\,Chris, `Grover's quantum searching algorithm is optimal', arXiv:quant-ph/9711070 v2, 2 Dec 1999.

\end{thebibliography}
\end{document}